\documentclass[prc,preprint,nofootinbib,superscriptaddress,showpacs]{revtex4}
\usepackage[dvips]{graphicx}
\DeclareGraphicsExtensions{.ps,.eps,.pdf,.jpg}
\usepackage{amssymb, latexsym, amsmath, amsfonts}
\usepackage{tabularx}
\usepackage{url}
\usepackage{color}
\allowdisplaybreaks

\usepackage{graphics}
\usepackage{graphicx}
\usepackage{color}
\usepackage{slashed}
\usepackage{multirow}
\usepackage{umoline}
\usepackage{footnote}
\usepackage{array}

\allowdisplaybreaks

\newcommand{\pt}{\partial}

\newcommand{\mc}{\mathcal}
\newcommand{\mcL}{\mathcal{L}}
\newcommand{\mcT}{\mathcal{T}}
\newcommand{\dmcL}{\delta\mathcal{L}}
\newcommand{\dmcT}{\delta\mathcal{T}}

\newcommand{\cds}{\cdots}
\newcommand{\vds}{\vdots}

\newcommand{\rij}{\mathbf{r}_{i} -\mathbf{r}_{j}}

\newcommand{\mbsig}{\boldsymbol{\sigma}}

\newcommand{\bnabla}{\mathbf{\nabla}}
\newcommand{\la}{\langle}
\newcommand{\ra}{\rangle}
\newcommand{\mbr}{\mathbf{r}}
\newcommand{\taun}{\tau_{n}}
\newcommand{\taup}{\tau_{p}}

\newcommand{\unifi}{u n_{i} f_{i}}
\newcommand{\su}{s(u)}

\newcommand{\mus}{\mu_{s}}
\newcommand{\nun}{\nu_{n}}
\newcommand{\rn}{r_{N}}
\newcommand{\uni}{u n_{i}}
\newcommand{\unixi}{u n_{i} x_{i}}
\newcommand{\nii}{n_{i}}
\newcommand{\cu}{c(u)}
\newcommand{\xii}{x_{i}}

\newcommand{\muni}{\mu_{ni}}
\newcommand{\muno}{\mu_{no}}

\newcommand{\cald}{\mathcal{D}}
\newcommand{\caldp}{\mathcal{D}^{\prime}}
\newcommand{\nno}{n_{no}}
\newcommand{\foo}{f_{o}}

\def\mr{\multirow{2}{*}}

\begin{document}



\title{Nuclear Equation of State and Neutron Star Cooling}

\author{Yeunhwan Lim}
\email{ylim@tamu.edu}
\affiliation{
Cyclotron Institute, Texas A\&M University College Station, TX 77843, USA}

\author{Chang Ho Hyun}
\email{hch@daegu.ac.kr}
\affiliation{
Department of Physics Education, Daegu University
Gyeongsan 38453, Republic of Korea }

\author{Chang-Hwan Lee}
\email{clee@pusan.ac.kr}
\affiliation{
Department of Physics, Pusan National University,
Busan 46241, Republic of Korea}


\begin{abstract}

We investigate the cooling of neutron stars with relativistic and non-relativistic models 
of dense nuclear matter.
We focus on the effects of uncertainties originated from the
nuclear models, the composition of elements in the envelope region,
and the formation of superfluidity in the core and the crust of neutron stars.
Discovery of $2 M_\odot$ neutron stars PSR J1614-2230 and PSR J0343+0432
has triggered the revival of stiff nuclear equation of state at high densities.
In the mean time, observation of a neutron star in Cassiopeia A for more than 10 years
has provided us with very accurate data for the thermal evolution of neutron stars.
Both mass and temperature of neutron stars depend critically on the equation of state
of nuclear matter, so we first search for nuclear models that satisfy the constraints from
mass and temperature simultaneously within a reasonable range.
With selected models, we explore the effects of element composition in the evenlope region,
and the existence of superfluidity in the core and the crust of neutron stars.
Due to uncertainty in the composition of particles in the envelope region 
we obtain a range of cooling curves that can cover substantial region of observation data.

\end{abstract}

\maketitle
\newpage


\section{Introduction}\label{Sec:int}

A neutron star (NS) is born as a result of the core collapsing supernova explosion
if the initial mass of the main sequence progenitor is around 8 to 20 solar mass ($M_{\odot}$).
The resulting central density of a neutron star is expected to reach up to several times of the nuclear
saturation density ($n_0 \simeq 0.16$ fm$^{-3}$). Hence, a neutron star is one of the best
astrophysical laboratories to study the physics of the extremely dense nuclear matter. 
Mass distribution of neutron stars may depend on the binary evolution in addition
to the neutron star equations of states (EoSs) \cite{Lee14}.
Recent observations of $\sim 2 M_{\odot}$ neutron stars 
PSR J1614$-$2230 and PSR J0348$+$0432 \cite{demorest2010, Anto13}
ruled out many soft EoSs with which the maximum mass of neutron star becomes less than 
$2M_\odot$ .
Also recent anlayses on the mass and radius of neutron stars from low-mass X-ray binaries
\cite{slb2010} provide constraints to the EoS of nuclear matter. 

In recent works \cite{kaon2014, hyperon2014}, the behavior of nuclear EoS was
investigated at high densities by calculating the mass and radius of neutron stars with several
Skyrme force models and exotic matter.
In those works, it was confirmed that the models 
consistent with high-mass neutron star observations \cite{demorest2010, Anto13}
are also consistent with the mass-radius zone in Ref.~\cite{slb2010}.
Moreover, the conclusion doesn't change
even if exotic degrees of freedom such as kaon condensation \cite{kaon2014} or hyperons \cite{hyperon2014}
are included.
Nuclear EoS is also one of the key ingredients 
that determine the thermal evolution of neutron stars.
Available nuclear models predict very diverse mass-radius relations, so it is well expected that the
cooling behavior will be sensitive to nuclear models.
Therefore, cooling of neutron stars is expected to provide multi-test grounds for the nuclear models.
Any reasonable model should satisfy both the empirical mass-radius relation and the observed temperature.

\begin{figure}
\centerline{
\includegraphics[scale=0.4]{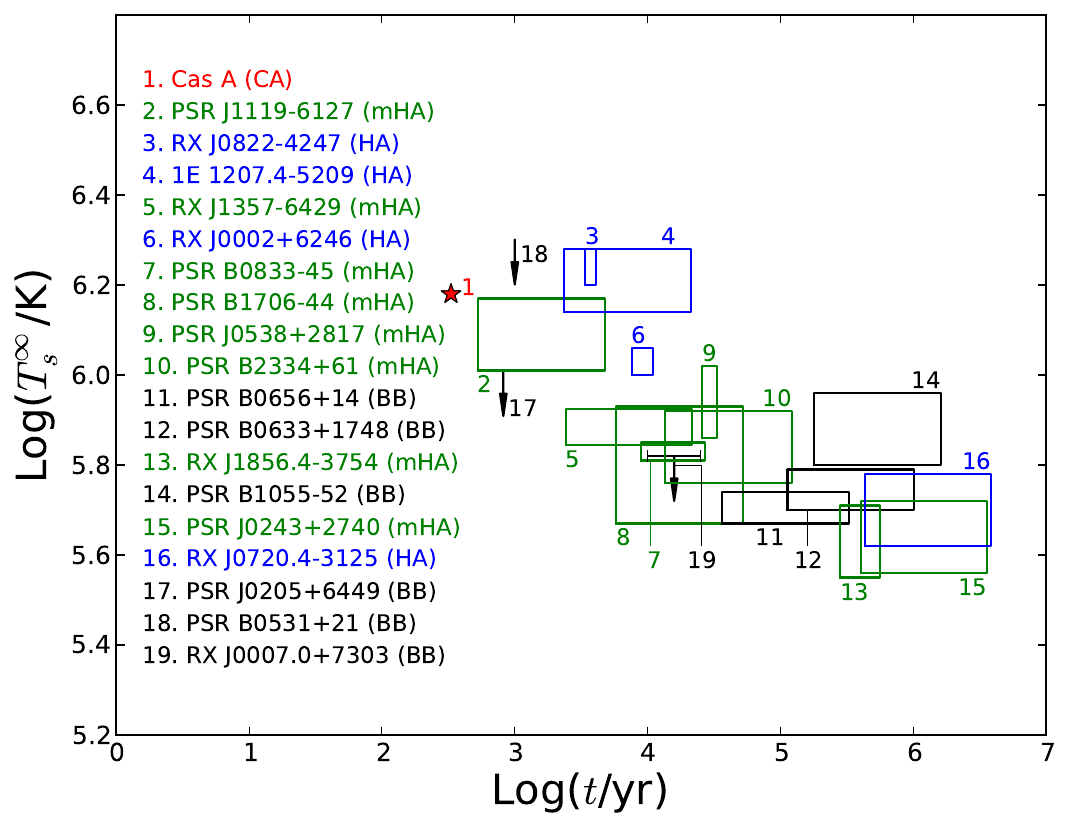} 
\includegraphics[scale=0.4]{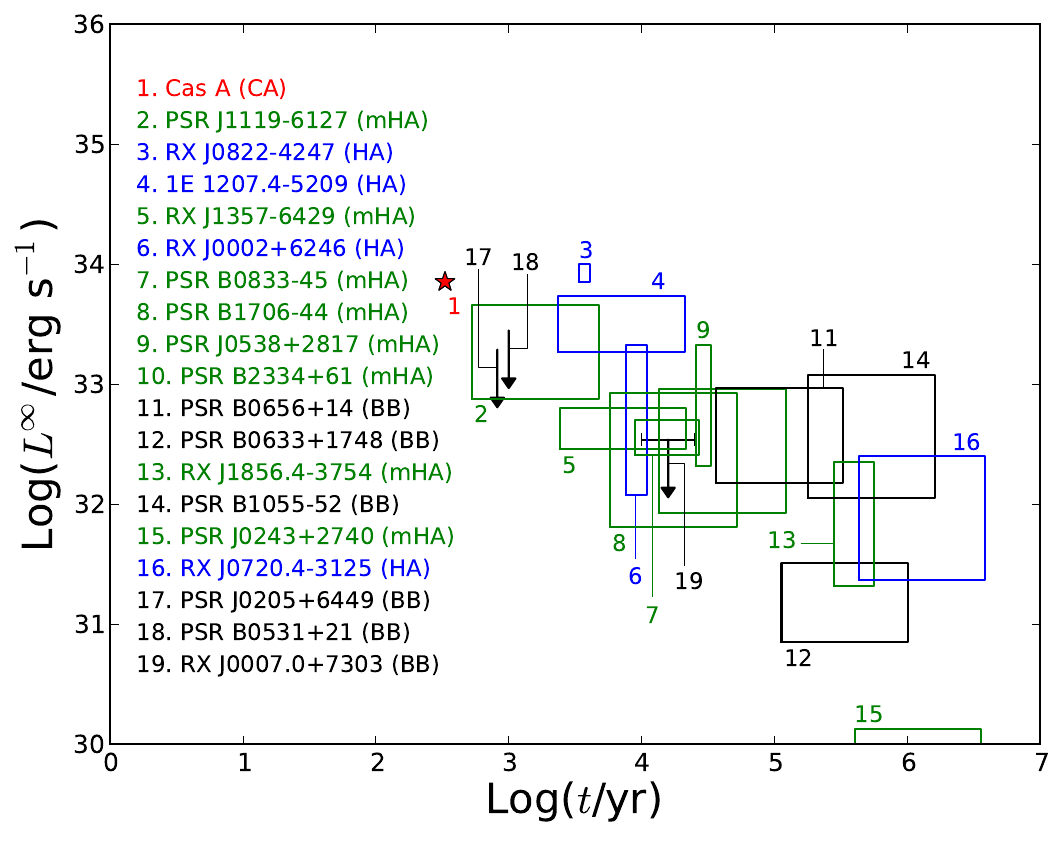} } 
\caption{(Color online) The effective temperature and the photon luminosity for observed neutron stars.
Left panel : the effective temperature at infinity. 
Right panel : photon luminosity seen by the observer at infinity.
Note that four models are used to link the effective 
temperature and the luminosity as in Table~\ref{tab:nsobs}.
Errors of the data are taken from the model that gives the largest uncertainties.
}
\label{fig:obsns}
\end{figure}

In Table~\ref{tab:nsobs} and Figure~\ref{fig:obsns}, 
we summarize the surface temperatures ($T_s^\infty$) and the photon luminosities 
($L^\infty$) of 19 isolated neutron stars.
Data of number 1 denote those of a neutron star in Cassiopeia A (Cas A), which have been
accumulated for the last decade.
The age of Cas A is very well defined, so the data provide crucial information to the thermal evolution
of young neutron stars.
For the stars with ages less than $10^4$ years, temperatures are around $10^6$~K, and similar to each other.
The temperature drops rather rapidly during $10^4 \sim 10^5$ years, and changes slowly later.
As a whole the data show slow-quick-slow cooling pattern.

In this work, we simulate the cooling of neutron stars with various EoSs obtained from
both relativistic and non-relativistic models. 
We note that the direct Urca process is a good indicator whether a specific nuclear
model is suitable for neutron star cooling simulation. 
Once the direct Urca process is turned on, the energy loss due to neutrino emssion is so fast that 
any other effects, e.g. superfluidity or existence of exotic states of matter 
cannot slow down the temperature drop.
As a result, the appearance of direct Urca process in the cooling curve is crucial to investigate 
the inner structure of neutron stars.
Thus, the study of cooling curve of neutron star can provide hints about
the inner structure of neutron stars and the EoS of dense nuclear matter.

In addition to the nuclear EoS, physical conditions such as the composition of elements
in the envelope, and existence of superfluidity in the core play crucial roles in
determining the cooling curve of neutron stars.
In general, the surface temperature depends sensitively on the elements in the envelope.
On the other hand, superfluidity directly determines the cooling rate.
If nucleons form cooper pairs and transit to a superfluid state,
the rate of neutrino emission is suppressed exponentially.
This may lead to a very slow cooling rate.
However, below a critical temperature, creation and destruction of cooper pairs
ignite a fast cooling mechanism, and this can give an abrupt decrease of temperature.
Recent literature succeeds to reproduce the cooling curve of Cas A in terms of 
pair breaking and formation (PBF) \cite{dpls2011,Shternin:2010qi}.
In this work, we incorporate PBF to various nuclear models and explore the extent to
which PBF can reconcile with observation data.

Structure of the paper is organized as follows. 
In Section \ref{sec:EoS} we summarize the equation of state of nuclear matter which we use in the study 
of thermal evolution of neutron stars. 
We do not consider exotic matter or quark matter in
the core of neutron stars but assume that the core is composed of
nucleons (protons and neutrons) and leptons (electrons and muons) 
in the form of uniform matter.
In Section~\ref{sec:cool} we present the
basic ingredients for the simulation of the neutron star cooling. 
In Section~\ref{sec:stancool}, we present the cooling curves 
for the standard cooling mechanisms with various nuclear models.
The effect of the envelope elements to the surface temperature is also presented.
In Section~\ref{sec:super}, we discuss the effect of superfluidity to the cooling process.
We give conclusions from
neutron star cooling curves combined with various EoSs
in Section~\ref{sec:con}. 

\begin{table}[tbp]
\renewcommand{\arraystretch}{0.8}
\begin{tabular}{clcccccc}
\hline 
No. & Source & Log$(t_{sd}/\text{yr})$ & 
Log$(t_{kin}/\text{yr}) $ & Log$(T_{s}^{\infty}/\text{K})$ &
Log$(L^{\infty}/\text{erg s}^{-1})$ &
  Model  & Ref. \\
\hline
1        &   Cas A        & 2.518$^{+0.007}_{-0.007}$ & - & $6.18^{+0.01}_{-0.01}$  & 33.83 - 33.88& CA & \cite{cas2010, cas2013} \\
2        & PSR J1119-6127 & 3.20 & - & $6.08^{+0.09}_{-0.07}$ & 32.88 - 33.66 & mHA & \cite{shk2008}\\  
\mr{3}  & \mr{RX J0822-4247$^\dagger$} &  \mr{3.90} & \mr{$3.57^{+0.04}_{-0.04} $} &
                            $6.24^{+0.04}_{-0.04}$ & 33.85 - 34.00 & HA & \mr{\cite{ztp1999,plps2004}}  \\ 
        &   &   &   & $6.65^{+0.04}_{-0.04}$ & 33.60 - 33.90 & BB &  \\
\mr{4}  &  \mr{1E 1207.4-5209} & \mr{$5.53^{+0.44}_{-0.19}$}  & \mr{$3.85^{+0.48}_{-0.48}$} & 
                          $6.21^{+0.07}_{-0.07}$ & 33.27 - 33.74 & HA & \mr{\cite{zps2004,plps2004}} \\ 
       &   &   &   & $6.48^{+0.01}_{-0.01}$ & 33.27 - 33.74 & BB &  \\  
\mr{5} &  \mr{PSR J1357-6429} & \mr{3.86} & \mr{-} & 
                           $5.88^{+0.04}_{-0.04}$ &  32.46 - 32.80 & mHA & \mr{\cite{zavlin2007}} \\ 
     &     &   &   & $6.23^{+0.05}_{-0.05}$ & 32.35 - 32.76 & BB &   \\ 
\mr{6} &  \mr{RX J0002+6246} & \mr{-} & \mr{$3.96^{+0.08}_{-0.08} $} & $6.03^{+0.03}_{-0.03}$ &
     33.08 - 33.33 & HA & \mr{\cite{pzst2002,plps2004}}  \\ 
  &     &    &   & $6.15^{+0.11}_{-0.11}$ &
     32.18 - 32.81 & BB &   \\  
\mr{7} &  \mr{PSR B0833-45$^\dagger$} & \mr{4.05} & \mr{4.26$^{+0.17}_{-0.31}$} & 
                          $5.83^{+0.02}_{-0.02}$ & 32.41 - 32.70 & mHA & \mr{\cite{pzsbg2001,plps2004}} \\ 
    &     &   &   & $6.18^{+0.02}_{-0.02}$ & 32.04 - 32.32 & BB &  \\    
\mr{8} &   \mr{PSR B1706-44} & \mr{4.24} & \mr{- }& $5.80^{+0.13}_{-0.13}$ & 
    31.81 - 32.93 & mHA & \mr{\cite{gowan2004,plps2004}} \\ 
  &     &   &    & $6.22^{+0.04}_{-0.04}$ &     32.48 - 33.08 & BB &   \\  
9 &  PSR J0538+2817 & $4.47^{+ 0.05}_{- 0.06}$ & - & $5.94^{+0.08}_{-0.08}$ &
    32.32 - 33.33 & mHA & \cite{zp2004} \\  
 %
10  &  PSR B2334+61 & 4.61 & - & $5.84^{+0.08}_{-0.08}$ &
      31.93 - 32.96 & mHA & \cite{zavlin2007a} \\ 
  %
11 &  PSR B0656+14 & 5.04 & - & $5.71^{+0.03}_{-0.04}$ &
    32.18 - 32.97 & BB & \cite{pmc1996,plps2004} \\    
12 &  PSR B0633+1748$^\dagger$ & 5.53 & - & $5.75^{+0.04}_{-0.05}$ &
     30.85 - 31.51 & BB & \cite{hw1997,plps2004} \\  
 %
 13 &  RX J1856.4-3754 & - & 5.70$^{+0.05}_{-0.25}$ & $5.63^{+0.08}_{-0.08} $ &
    31.32 - 32.35 & mHA & \cite{hkcap2007, plps2004} \\  
 %
14 &  PSR B1055-52 & 5.73 & - & $5.88^{+0.08}_{-0.08} $ &
    32.05 - 33.08 & BB & \cite{pz2003, plps2004} \\  
%
15 &  PSR J0243+2740 & 6.08 & - & $5.64^{+0.08}_{-0.08}$ &
      29.10 - 30.13 & mHA & \cite{zavlin2007a} \\   
%
%
16 &  RX J0720.4-3125 & 6.11 & - & $5.70^{+0.08}_{-0.08}$ &
    31.37 - 32.40 & HA & \cite{mzh2003} \\   
%
%
17 & PSR J0205+6449$^{\dagger\dagger}$ &  - & 2.91 & $< 6.01$ & $< 33.29$ & BB & \cite{shsm2004a} \\    
18 & PSR B0531+21$^{\dagger\dagger}$ & - & 3.0 & $< 6.30$ & $ < 34.45 $ & BB & \cite{weiss2004}\\    
19 & RX J0007.0+7303$^{\dagger\dagger}$ & - & 4.0-4.4 & $ < 5.82$ & $ < 32.54 $ & BB & \cite{hgchr2004}\\ 
\hline
\end{tabular}
\caption{Thermal emission from isolated neutron stars.
Temperatures were obtained using four models; 
carbon atmosphere (CA), hydrogen atmosphere (HA),
magnetized hydrogen atmosphere (mHA), and black-body (BB) models. 
$t_{sd}$ is the age of neutron star which is obtained from $t_{sd}=P/2\dot{P}$,
and $t_{kin}$ is the age from the kinematic information between its transverse
velocity and supernova remnant.
For sources of no.~3 $\sim$ 8, two different models were used to 
link the effective temperature and the photon luminosity.
Sources of no. 17 $\sim$ 19 have the limited observational data, thus have only upper limit.
Part of this table is adopted from Ref. \cite{plps2004, ozel2013}. \\
$^\dagger$ Alternative names:  Puppis A (PSR J0822-4247), Vela (PSR B0833-45), Geminga (PSR B0633+1748). 
$^{\dagger\dagger}$ PSR J0205+6449 is a pulsar in supernova remnant 3C 58, PSR B0531+21 is in SN 1054 in Crab Nebula, and RX J0007.0+7303 is in the CTA1.}\label{tab:nsobs}
\end{table}

\section{Neutron Star Equation of State}\label{sec:EoS}

Nuclear matter properties beyond the nuclear saturation density are not yet understood clearly, 
and many nuclear models give quite different masses and radii of neutron stars.
 In order to investigate the properties of neutron star matter, we first consider both 
relativistic and non-relativistic models for the neutron star core, 
which are consistent with $2.0 M_{\odot}$ neutron stars
 \cite{demorest2010, Anto13}. We consider the crust of neutron star 
 separately because heavy nuclei can exist in the crust. 
 The properties of neutron star crust are important 
 in understanding neutron star properties in low-mass X-ray binaries.

\subsection{Non-relativistic nuclear force model}

For the non-relativistic nuclear force model,
we use Skyrme force models to obtain the EoS for nuclear matter \cite{Ben03}. 
In the Skyrme force model, the interaction between two nucleons has the form of
\begin{eqnarray}\label{eq:skyint}
\hat{v}_{\rm SF}(\mathbf{r}_i,\mathbf{r}_j) 
&=& t_0(1+x_0\hat{P}_\sigma)\delta(\rij) +\frac{t_1}{2}(1+x_1\hat{P}_\sigma) \Bigl[ \delta(\rij)\hat{\bf k}^2  + \hat{\bf k}^{\dagger 2}\delta(\rij) \Bigr]  \nonumber \\ 
&& +  t_2(1+ x_2\hat{P}_\sigma)\hat{\bf k}^\dagger \cdot \delta (\rij)\hat{\bf k} + \frac{1}{6}t_3(1+x_3 \hat{P}_\sigma
)n^{\alpha}\delta(\rij) \nonumber \\ 
&& +  iW_0 \hat{\bf k}^\dagger \delta(\rij)\times \hat{\bf k} \cdot (\hat\mbsig_i+\hat\mbsig_j) \,,
\end{eqnarray}
where $\hat{P}_\sigma = \frac 12 (1+\hat\mbsig_i\cdot \hat\mbsig_j)$ is the spin-exchange operator, 
$t_i$, $x_i$ and $\alpha$ are the parameters of the interactions, 
and $\hat{\bf k}$ is defined as 
\begin{equation}
\hat{\bf k} =\frac{1}{2i}(\bnabla_{i}-\bnabla_j).
\end{equation}
Note that the interaction contains terms up to quadratic in derivatives, $t_3$-term 
is added to account for many body effects beyond quadratic order in density $n$, 
and $W_0$-term gives the spin-orbit interaction which is 
important to explain the nuclear structure. 

\begin{figure}
\centerline{
\includegraphics[scale=0.37]{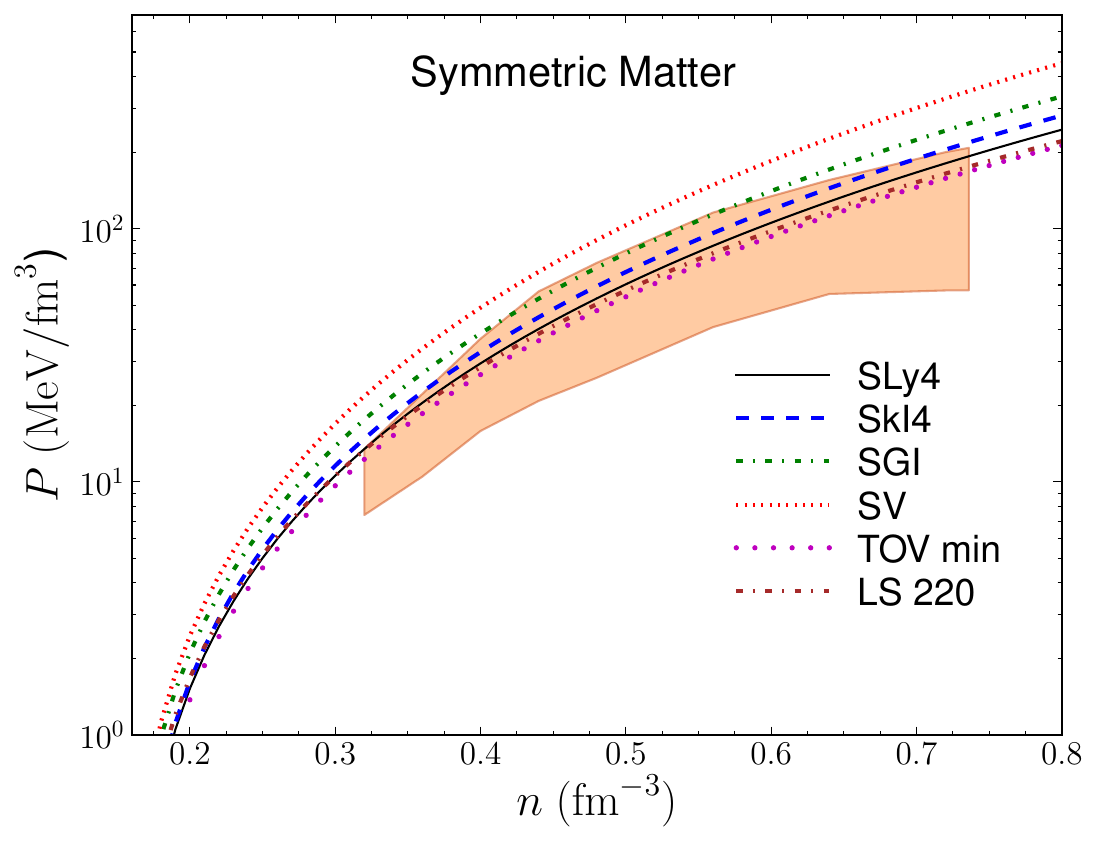} 
\includegraphics[scale=0.37]{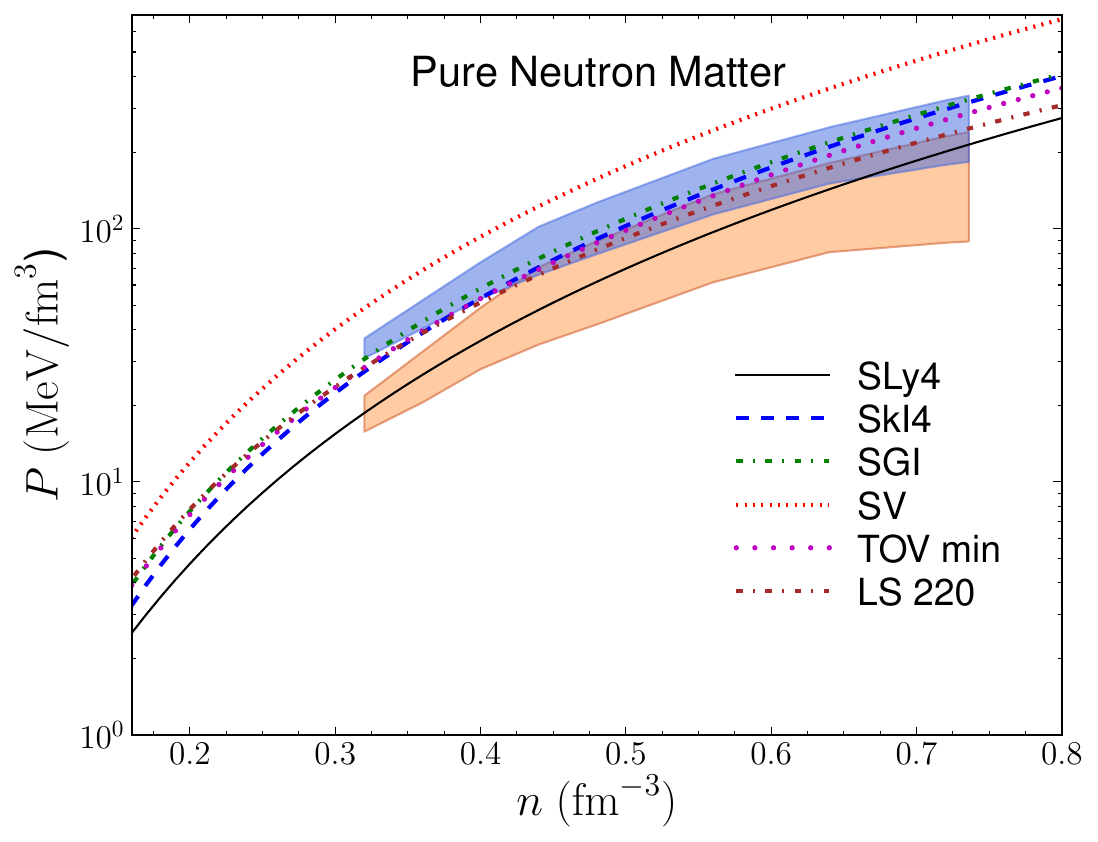}
}

\centerline{
\includegraphics[scale=0.37]{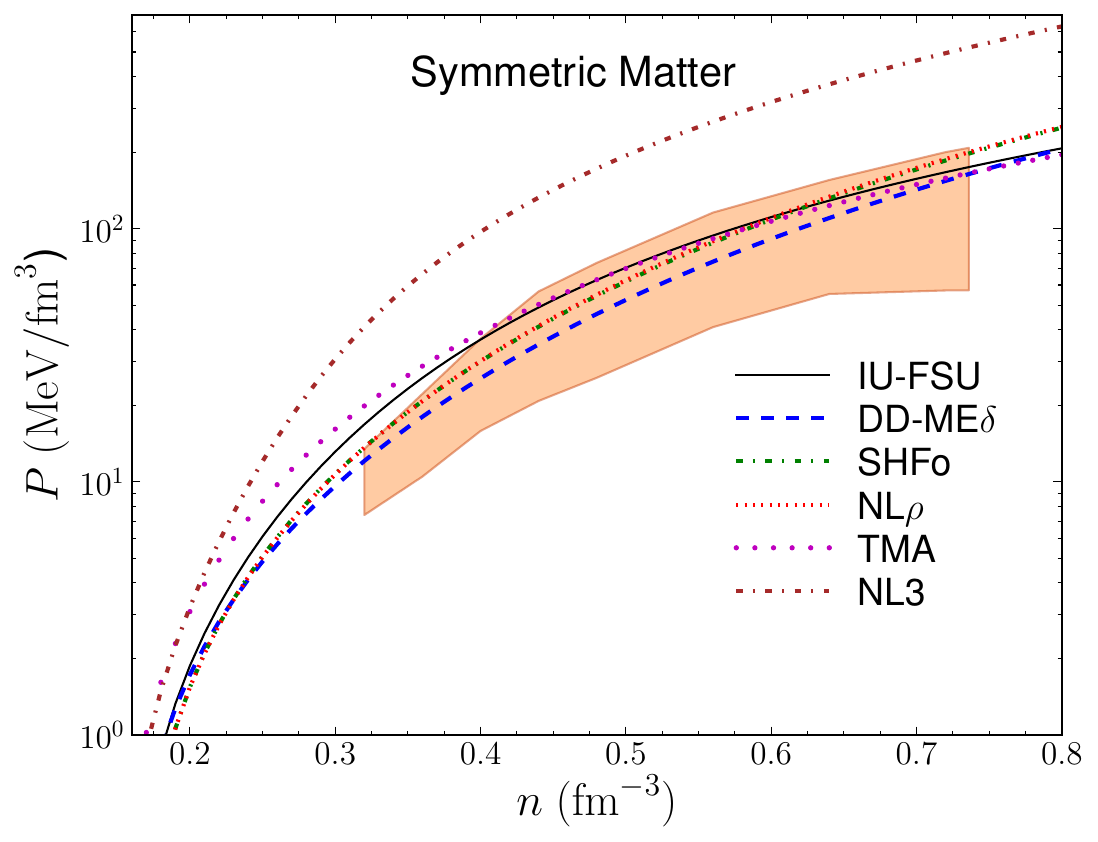} 
\includegraphics[scale=0.37]{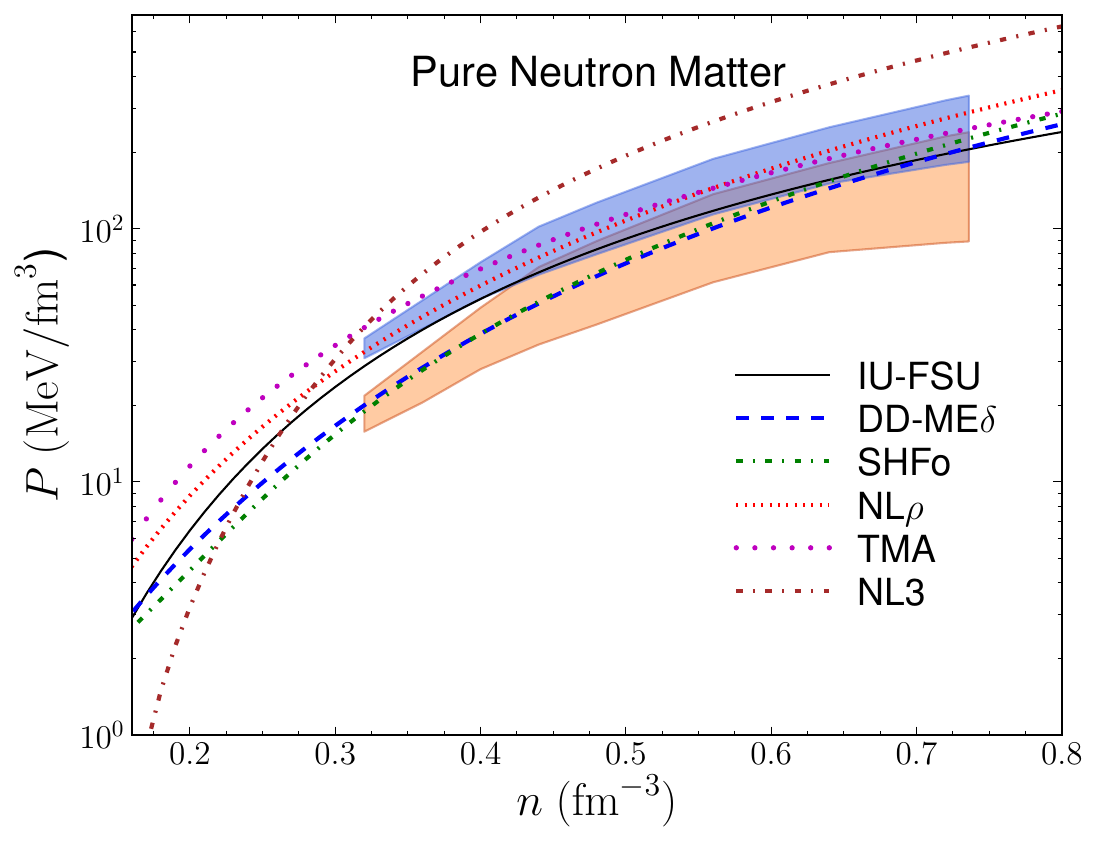}
}
\caption{(Color online) Pressure of symmetric nuclear matter and pure neutron matter from 
both non-relativistic Skyrme force models (upper panels)
and relativistic mean field models (lower panels).  
The shaded area is the result from the analysis of the flow experiment, Ref. \cite{dll2002}.
For the pure neutron matter (right panels), the upper (lower) shaded area in each plot represents the stiff (soft) equation of state.}
\label{fig:pres}
\end{figure}

\begin{figure} 
\centerline{
\includegraphics[scale=0.4]{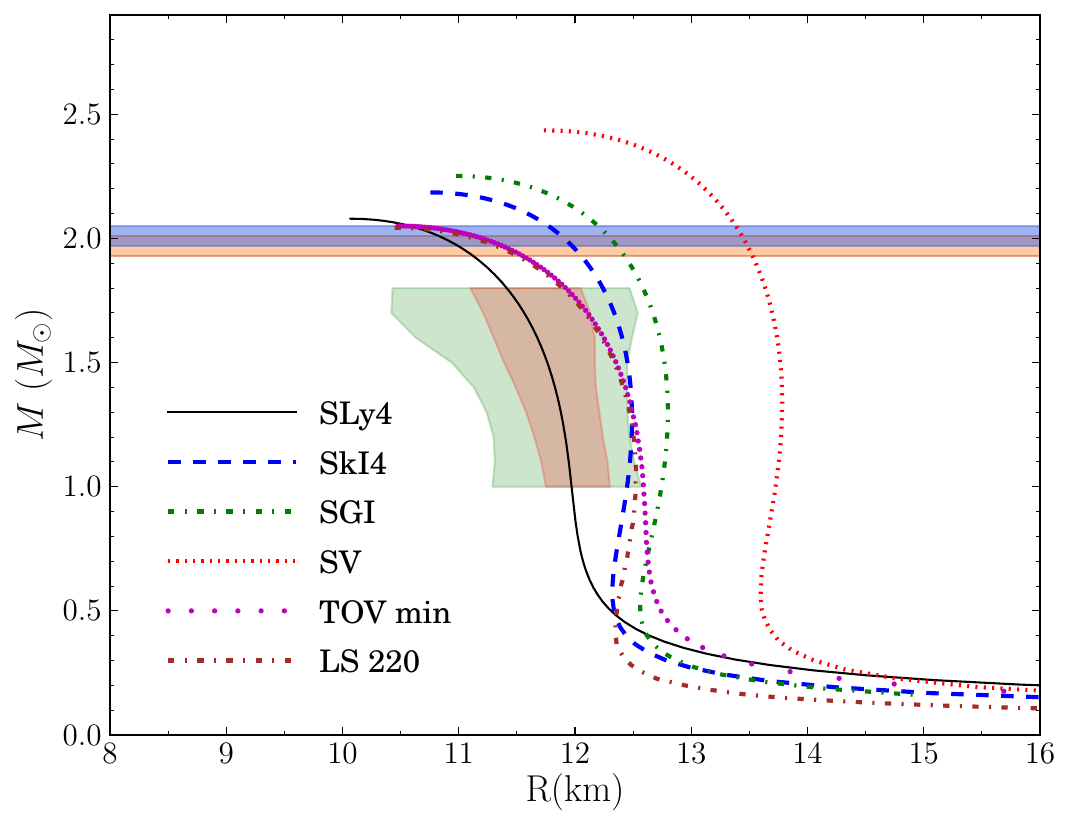} 
\includegraphics[scale=0.4]{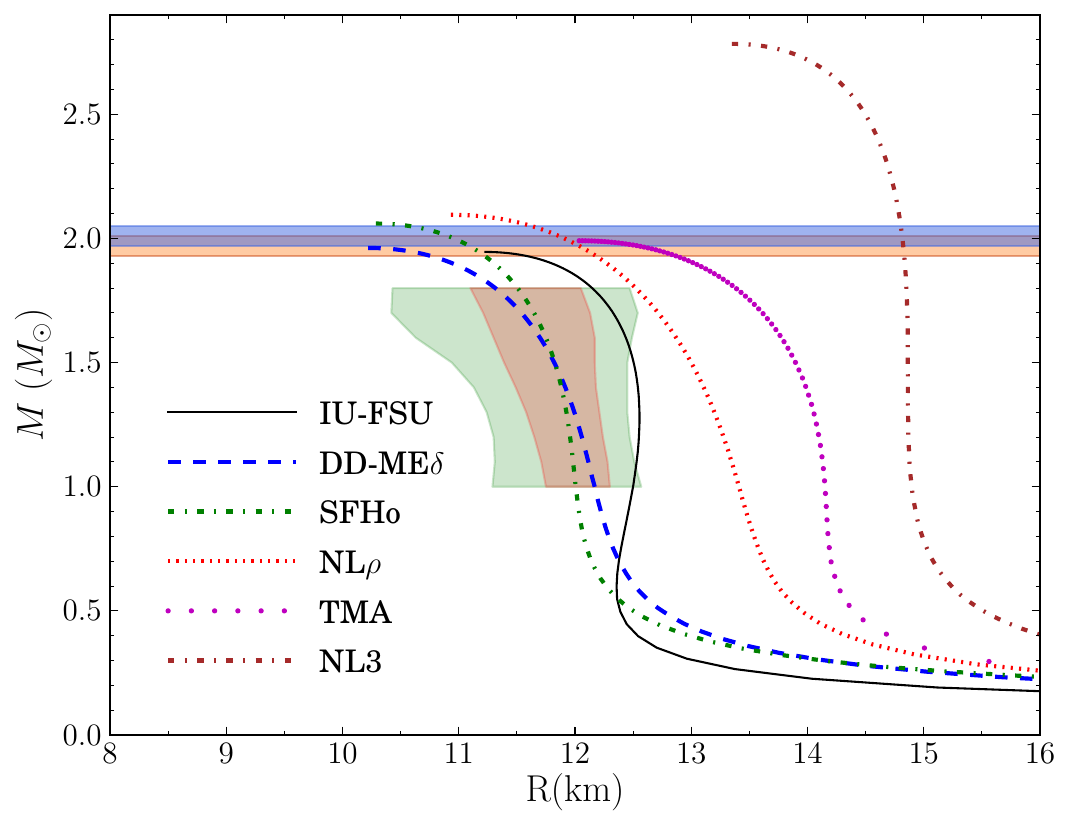}
}
\caption{(Color online) Neutron star's mass and radius relation from
non-relativistic Skyrme force models (left) and RMF
models (right). Thick horizontal lines indicate the masses of 
PSR J1614$-$2230 and PSR J0348$+$0432 \cite{demorest2010, Anto13}. 
The shaded area is the most probable mass and radius, $1 \sigma$ and
$2 \sigma$ region, from the analysis of Steiner \textit{et al}. \cite{slb2010}.}
\label{fig:nsmr}
\end{figure}

At the Hartree-Fock level,
the total energy can be expressed as
\begin{equation}
E = \sum_{ij} \la i \vert \hat{t} \vert j\ra \rho_{ji}
+ \frac{1}{2}\sum_{ijkl}\bar{v}_{ijkl}\rho_{ki}\rho_{lj},
\end{equation}
where $\hat{t}$ is the kinetic energy operator and
\begin{equation}
\bar{v}_{ijkl} =\la ij \vert \hat{v}(1-\hat{P}_{\sigma} \hat{P}_{\mbr} \hat{P}_{\tau}) \vert kl \ra.
\end{equation}
Here $\hat{P}_{\bf r} $ is the parity operator and
 $\hat{P}_\tau = \frac 12 (1+\hat\tau_i\cdot \hat\tau_j)$
 is the iso-spin exchange operator.
Total energy is obtained as 
\begin{equation}
E = \int d^3r\, \mathcal{E} 
= \int d^3r\, (\mathcal{E}_{B} + \mathcal{E}_{C} + \mathcal{E}_{g}
+ \mathcal{E}_{J}),
\end{equation}
where $\mc{E}_{B}$ is the bulk part contribution, 
$\mc{E}_{C}$ is the Coulomb contribution,
$\mc{E}_{g}$ is the contribution from the density gradient term,
and 
$\mc{E}_{J}$ is the contribution from the  spin-orbit term. 
For a uniform matter in the neutron star core,
$\mc{E}_{B}$ is dominant. Hence  the energy density can be approximated as \cite{sple2005}
\begin{eqnarray}
\mathcal{E} \simeq \mathcal{E}_B & = &
\frac{\hbar^2}{2 m_n}\tau_{n} + \frac{\hbar^2}{2 m_p}\tau_{p}
+ n (\tau_n + \tau_p) 
\biggl[ \frac{t_1}{4}\Bigl(1+\frac{x_1}{2}\Bigr)
+ \frac{t_2}{4}\Bigl(1+\frac{x_2}{2}\Bigr)\biggr]  \nonumber \\
&& \quad + (\taun n_n +\taup n_p)
\biggl[ \frac{t_2}{4}\Bigl(\frac{1}{2}+ x_{2}\Bigr)
- \frac{t_1}{4}\Bigl(\frac{1}{2} + x_{1}\Bigr)\biggr]  \nonumber \\
&& \quad + \frac{t_0}{2}
\biggl[\Bigl(1+ \frac{x_0}{2}\Bigr) n^2
- \Bigl(\frac{1}{2} + x_{0}\Bigr)( n_n^2+ n_p^2)\biggr] \nonumber \\
&& \quad + \frac{t_3}{12}
\biggl[\Bigl(1+ \frac{x_3}{2}\Bigr)n^2
- \Bigl(\frac{1}{2} + x_{3}\Bigr)(n_n^2+ n_p^2)\biggr]
n^{\alpha}\, ,
\label{eq:hamilN}
\end{eqnarray}
where $m_n$ and $m_p$ are neutron and proton masses,
$n_n$ and $n_p$ are the number densities of neutrons and protons,
the total baryon number density $n=n_n+n_p$, and $\tau_n$ and $\tau_p$
are kinetic energy densities of neutrons and protons, respectively. 
The pressure can be obtained by taking a density derivative of energy per baryon,
\begin{equation}
P = n^2 \frac{\pt (\mc{E}/ n)}{\pt  n}\,.
\end{equation}
In the upper part of Table~\ref{tab1}, we summarize the basic properties of 
nuclear matter for Skyrme force models which are used in this work.
In the upper panels of Figure~\ref{fig:pres}, we show the EoS of Skyrme force models 
for symmetric nuclear matter and pure neutron matter, respectively.
In the left panel of Figure~\ref{fig:nsmr}, masses and radii of neutron stars are 
summarized for the Skyrme force models.

\begin{table} 
\begin{center}
\begin{tabular}{lccccccc}
\hline
Model & $n_0$ (fm$^{-3}$)  & $B$ (MeV)  & 
 $S_v$ (MeV)  &  $L$ (MeV) &  $K$ (MeV) & 
$m^\star_N/m_N$ &   Ref \\  
 \hline  
SLy4                & 0.160 & 16.0 & 32.0 & 45.9 & 230 & 0.694 &   \cite{cbhmf1998} \\  
SkI4                 & 0.160 & 16.0 & 29.5 & 60.4 & 248 & 0.649 &   \cite{rf1994} \\  
SGI                  & 0.155 & 15.9 & 28.3 & 63.9 & 262 & 0.608 &   \cite{gs1981}\\  
SV                   & 0.155 & 16.1 & 32.8 & 96.1 & 306 & 0.383 &   \cite{bfgq1975} \\  
TOV-min          & 0.161 & 15.9 & 32.3 & 76.2 & 222 & 0.934 &   \cite{ehnrg2013} \\  
LS220              & 0.155 & 16.0 & 28.6 & 73.1 & 220 & 1.000 &   \cite{ls1991} \\  
\hline
IU-FSU              & 0.155 & 16.4 & 31.3 & 47.2 & 231 & 0.669 &   \cite{fhps2010} \\ 
DD-ME$\delta$ & 0.152 & 16.1 & 32.4 & 52.9 & 219 & 0.668 &   \cite{rvcrs2011} \\ 
SFHo                 & 0.158 & 16.2 &  31.6 & 47.1 & 245 &  0.810 &   \cite{shf2013} \\ 
NL$\rho$            & 0.160 & 16.1 &  30.4  &  84.6  & 241 &   0.800 &   \cite{lgbct2002} \\ 
TMA                   & 0.147 & 16.0 & 30.7 & 90.1 & 318 & 0.691 &   \cite{tma1995} \\ 
NL3                    & 0.148 & 16.2  & 37.3 & 118  & 272 & 0.655 &  \cite{lkr1997} \\ 
\hline
\end{tabular}
\end{center}
\caption{Nuclear matter properties at the saturation density ($n_0$). 
Upper 6 models correspond to non-relativistic Skyrme force models and lower 6 models correspond to RMF models.  
$B$ is the binding energy of the symmetric nuclear matter, $S_v$ is the symmetry energy, 
$L$ is the slope of the symmetry energy, $K$ is the compression modulus, 
and $m_N^\star$ is Landau effective nucleon mass (effective chemical potential). 
}
\label{tab1}
\end{table}

\subsection{Relativistic mean field model}

Relativistic mean field (RMF) models have been very successful in explaining 
finite nuclei properties such as binding energy, density profile, root mean square radius, etc. 
RMF models are typically described by the Lagrangian density \cite{muller96},
\begin{eqnarray}
\mathcal{L} & = & \bar{\psi}
\left[
i\slashed{\pt} - g_{\omega}\slashed{\omega}
-\frac{1}{2}g_{\rho} \vec{\tau}\cdot \vec{\slashed{b}}
+ g_\delta \vec{\delta}\cdot \vec{\tau} - m_N
+ g_{\sigma}\sigma - \frac{1}{2}e(1+\tau_{3})\slashed{A}
\right]\psi   -\frac{1}{4}F_{\mu\nu}F^{\mu\nu} \nonumber\\
&& + \frac{1}{2}\pt_{\mu}\sigma \pt^{\mu}\sigma
-\frac{1}{2}m_{\sigma}^2\sigma^2 
-\frac{1}{4}\Omega_{\mu\nu}\Omega^{\mu\nu} 
+ \frac{1}{2}m_{\omega}^2 \omega^{\mu}\omega_{\mu}
-\frac{1}{4}\vec{R}_{\mu\nu}\vec{R}^{\mu\nu}
+ \frac{1}{2}m_{\rho}^2 \vec{b}^{ \mu}\cdot \vec{b}_{\mu}  \nonumber\\
 && + \frac{1}{2}\pt_\mu \vec{\delta}\cdot\pt^\mu \vec{\delta}
 -\frac{1}{2}m_\delta^2 \,\vec{\delta}^2
  -V_{\rm eff} (\sigma, \omega^\mu\omega_\mu,  \vec{b}^{ \mu}\cdot \vec{b}_{\mu} ),
 \vphantom{\frac 12} 
\label{eq:rmflag}
 \end{eqnarray}
where $\sigma$ is the scalar field, $\omega^\mu$ is the vector-isoscalar field, 
$\vec b^\mu$ is the vector-isovector field,
$\vec{A}$ is the photon field, $\vec{\delta}$ is the scalar-isovector field,
$F_{\mu\nu} = \pt_{\mu}A_{\nu} - \pt_{\nu}A_{\mu}$, 
$\Omega_{\mu\nu} = \pt_{\mu}\omega_{\nu} - \pt_{\nu}\omega_{\mu}$,
$\vec{R}_{\mu\nu} = \pt_{\mu}\vec{b}_{\nu} - \pt_{\nu}\vec{b}_{\mu}$, and
$V_{\rm eff}$ is the general effective potential for meson fields.
\footnote{In some literature, meson mass terms are also included in 
the effective potential. However, in this work, mass terms are explicitly
specified and the effective potentials have only higher order interaction 
terms beyond mass terms.}
The equations of motion for meson fields can be obtained using the Euler-Lagrange equation
\begin{equation}
\pt_{\mu}\left( \frac{\pt \mc{L}}{\pt(\pt_{\mu} \phi)}\right)
- \frac{\pt \mc{L}}{\pt \phi} =0,
\end{equation}
where $\phi = \sigma, \omega^\mu, b_i^\mu, \delta_i$.
By taking expectation value of each field, one can define the scaled meson mean fields
$ \Phi \equiv  g_\sigma \la \sigma \ra$, $W \equiv g_\omega \la \omega^{0}\ra$,
$R \equiv g_\rho \la  b_{3}^{0}\ra$,
and $D \equiv g_\delta \la \delta_3 \ra$.
Then the equations of motion for the uniform nuclear matter become
\begin{eqnarray}
 n_s &=&  \frac{1}{c_\sigma^2} \Phi + \frac{\partial V_{\rm eff}(\Phi,W,R)}{\partial \Phi},  \\  
 n &=& \frac{1}{c_\omega^2} W - \frac{\partial V_{\rm eff}(\Phi,W,R)}{\partial W},   \\
\frac 12  n_3 
&=& \frac{1}{c_\rho^2} R -  \frac{\partial V_{\rm eff}(\Phi,W,R)}{\partial R},  
\end{eqnarray}
where the scaled coupling $c_i$'s are defined as $c_i \equiv g_i/m_i$, 
the baryon scalar density is $ n_s = \la\bar\psi \psi\ra$, the baryon density is 
$ n= \la \psi^\dagger \psi  \ra  = (k_{F_p}^3 + k_{F_n}^3)/(3\pi^2)$,  
and the baryon isovector density is
$  n_3 = \la \bar \psi \tau_3 \gamma^0 \psi\ra = (k_{F_p}^3 - k_{F_n}^3)/(3\pi^2)$.
The pressure and energy density for nuclear matter are obtained as
\begin{eqnarray}\label{eq:rmf}
P &=&  \frac{1}{3\pi^2} \sum_{n,p}\int_{0}^{k_F} 
\frac{k^4}{ \sqrt{k^2 + m_N^{*2} }}\,dk
- \frac{\Phi^2}{2 c_\sigma^2} 
+ \frac{W^2}{2 c_\omega^2 } 
+ \frac{R^2}{2 c_\rho^2} 
- \frac{D^2}{2 c_\delta^2}
- V_{\rm eff} (\Phi,W,R),  \nonumber\\
 \mc{E}  &= & \frac{1}{\pi^2} \sum_{n,p}\int_{0}^{k_F} k^2 \sqrt{k^2 + m_N^{*2}}\,dk
+ \frac{\Phi^2}{2 c_\sigma^2} 
+ \frac{W^2}{2 c_\omega^2 }
+ \frac{R^2}{2 c_\rho^2} 
+ \frac{D^2}{2 c_\delta^2}
+ V_{\rm eff} (\Phi,W,R)  \nonumber \\ && 
+  W n + \frac 12 R n_3,  
\end{eqnarray}
where $m_N^{*} = m_N - g_\sigma \langle \sigma \rangle 
\pm g_\delta \langle \delta_3 \rangle$  ($+$ : proton, $-$ : neutron).
Among various forms of meson effective potentials $V_{\rm eff}(\Phi, W, R)$ in the literature, 
we use a form given as
 \begin{equation}
V_{\rm eff}(\Phi, W, R)
= \frac{\kappa}{3!}\Phi^3
+ \frac{\lambda}{4!}\Phi^4
- \frac{\zeta}{4!}W^4
- \frac{\xi}{4!}R^4
- f(\sigma,\omega^\mu\omega_\mu)\,R^2\,,
\end{equation}
with
\begin{equation}
f(\sigma,\omega^{\mu}\omega_\mu) = \sum_{i=1}^{6}a_i \sigma^{i}
+ \sum_{j=1}^3 b_j (\omega^\mu\omega_\mu)^j\,.
\end{equation}
In the density dependent coupling constant model (DD-ME$\delta$), the coupling constant has the form of
\begin{equation}
g_\lambda (n) = g_\lambda (n_{0}) s_\lambda (x)\,,
\label{eq:gn}
\end{equation}  
where $x = n/n_{0}$, $\lambda = \sigma,\, \omega,\, \rho,\, \delta$ and
\begin{equation}
s_\lambda (x) = a_\lambda  \frac{1 + b_\lambda (x+d_\lambda )^2}{1 + c_\lambda (x+e_\lambda )^2}\,.
\end{equation} 
Numerical values of parameters of each model can be found in the references in Table~\ref{tab1}.
Even with density dependent couplings, the pressure and energy density in Eq.~\eqref{eq:rmf} do not change.
In the lower part of Table~\ref{tab1}, we summarize the 
basic nuclear matter properties of RMF models selected in this work. 
In the right panel of Figure~\ref{fig:nsmr}, masses and radii of 
neutron stars are summarized for the RMF models. 
 \subsection{Neutron star crust}

In the crust of neutron stars, heavy nuclei are expected to exist 
together with free gas of neutrons and electrons. A simple but appropriate 
description of this state is feasible by using liquid droplet 
formalism \cite{kaon2014,ls1991}. 
The total energy density (without electron contribution) is given by
\begin{equation}\label{eq:heavy}
F = \unifi + \frac{3\su}{\rn} \left [ \sigma(x_p) +\mus\nun \right ] 
+ \frac{4\pi}{5}(\rn\nii\xii e)^2 \cu + (1-u)\nno\foo\,\,,
\end{equation}
where $u$ is the volume fraction of heavy nuclei to Wigner-Seitz cell,
$n_i$ is the density inside of heavy nuclei, $f_i$ is 
the energy per baryon
of the heavy nuclei, $s(u)$ is the surface shape factor, 
$r_N$ is the radius of heavy nuclei,
$\sigma(x_p)$ 
is a surface tension as a function of proton fraction $x_p$,
$\mu_s$ is the neutron chemical potential on the surface,
$\nu_n$ is the neutron skin density on the surface, 
$x_i$ is the proton fraction of heavy nuclei,
$c(u)$ is the Coulomb shape function, 
$n_{no}$ is the neutron density outside of heavy nuclei,
and $f_o$ is the energy per baryon outside of heavy nuclei. 
Minimizing the energy density, we have four equations to solve 
\begin{equation}
\begin{aligned}
&P_i  -P_o - \beta\left(\mathcal{D}^{\prime} -\frac{2\mathcal D}{3u} \right) =0, \\
&\unixi - n Y_{p} = 0, \\
&\uni + \frac{2\beta}{3\sigma}\cald\nun + (1-u)\nno - n =0 ,\\
&\muni -\muno = 0,
\end{aligned}
\end{equation}
with four unknowns, $u$, $\nii$, $\nno$, and $\xii$. 
Here $\beta=9(\pi e^2x_i^2 n_i^2 \sigma^2/15)^{1/3}$,  
$\mathcal{D}^{\prime}=\pt \mathcal{D}/\pt u$, and $\mathcal{D}=[c(u)s^{2}(u)]^{1/3}$ 
is a geometric shape function which corresponds to nuclear pasta phase in liquid droplet model
 \cite{ls1991, rpw1983}.
$P_i$ ($P_o$) is pressure inside (outside) of the heavy nuclei, and
the total pressure is given by
\begin{equation}
P  = P_o -\beta(\cald -u\caldp)\,.
\end{equation}
The boundary between the crust and the core can be found by comparing
the energy density or energy per baryon of uniform nuclear matter and
heavy nuclei with free neutron and electron gas. In general case, the energy 
difference between two phases near the boundary is so small that the pressure
difference is negligible \cite{oya1993}.

\section{Neutron Star Cooling Mechanisms}\label{sec:cool}

Thermal evolution of a neutron star can be obtained by solving the coupled diffusion equations
\begin{align}
& \frac{L_r}{4\pi \kappa r^2}
= -\sqrt{1 -\frac{2Gm}{rc^2}}e^{-\Phi_g}\frac{\pt}{\pt r}
\left(T e^{\Phi_g}\right)\,,\\
& \frac{1}{4\pi r^2 e^{2\Phi_g}}
\sqrt{1-\frac{2Gm}{rc^2}}
\frac{\pt}{\pt r}\left(e^{2\Phi_g} L_{r}\right)
= - Q_{\nu} -\frac{C_{v}}{e^{\Phi_g}}\frac{\pt T}{\pt t}\,,
\end{align}
where $L_r$ is the local luminosity due to the non-neutrino heat flux \cite{thorne1977}, 
$T$ is the local temperature, 
$m=m(r)$ is the enclosed mass,
and $e^{\Phi_g}$ is the general relativistic metric function.
$\kappa$ is the total thermal conductivity, 
$Q_\nu$ is the total neutrino emissivity, $C_v$ is the total specific heat.
The first equation is the general relativistic definition of
photon luminosity and the second equation tells how the 
photon luminosity varies with neutrino emission. 
Below we briefly discuss about the neutrino emission, heat capacity, and 
thermal conductivity.

In the standard cooling mechanism,
neutrinos are emitted via various process, such
as the modified Urca process \cite{maxwell1979, yak1995} and
the Nucleon-Nucleon bresmmstrahlung \cite{maxwell1979, yak1995}.
The reduction factors from nuclear superfluidity were calculated by
Yakovlev and Levenfish and tabulated in Ref. \cite{yak1995}
There are also neutrino emission contribution from
electron-nucleus collision \cite{kaminker1999}, neutrino emission
from medium electrons \cite{ykgh2000}, and electron positron
annihilation \cite{ykgh2000}.
As an enhanced neutrino process, the direct Urca process \cite{lpph1991, ykgh2000}
is allowed if the proton fraction is high enough to satisfy 
momentum conservation.

The specific heat in the neutron star is given by the sum of its constituents,
neutrons, protons, electrons, and muons.
In the crust of neutron stars, the existence of heavy nuclei can contribute
the specific heat as an ion contribution \cite{yls1999}.
The heat capacity is also effected by superfluidity. Once the temperature 
decreases below the critical temperature, the most portion of heat capacity
comes from electrons since the critical temperature for electron superfluidity
is much lower than nucleon cases \cite{kaminker1999}.

The thermal conductivity arises from the collision phenomenon 
between particles for a given density and temperature. 
In the core,  the thermal conductivity consists of neutron, 
proton, electron and muon contributions \cite{bhy2001, gy1995}. On the other hand,
electrons are the main thermal conductivity factor and they 
collide with other electrons or heavy nuclei in the crust of neutron stars \cite{sy2006}.
As in the case of heat capacity, once the temperature drops below the critical 
temperature of nuclear superfluidity, the thermal conductivity 
caused by collisions between superfluid baryons or between electrons,
muons and superfluid baryons experience reductions. 
Thus the electron thermal conductivity dominates both in the core and crust.

\section{Results from the Standard Cooling}\label{sec:stancool} 

\subsection{Standard cooling and direct Urca}

\begin{figure}
\centerline{
\includegraphics[scale=0.4]{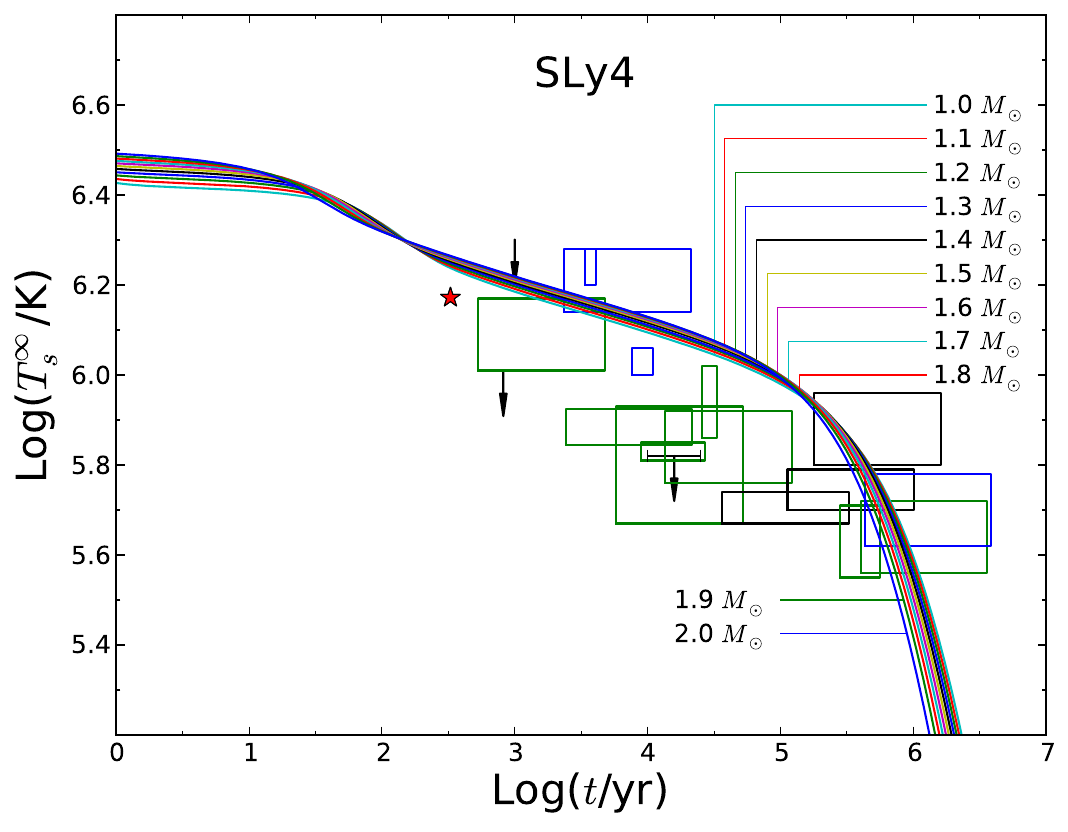} 
\includegraphics[scale=0.4]{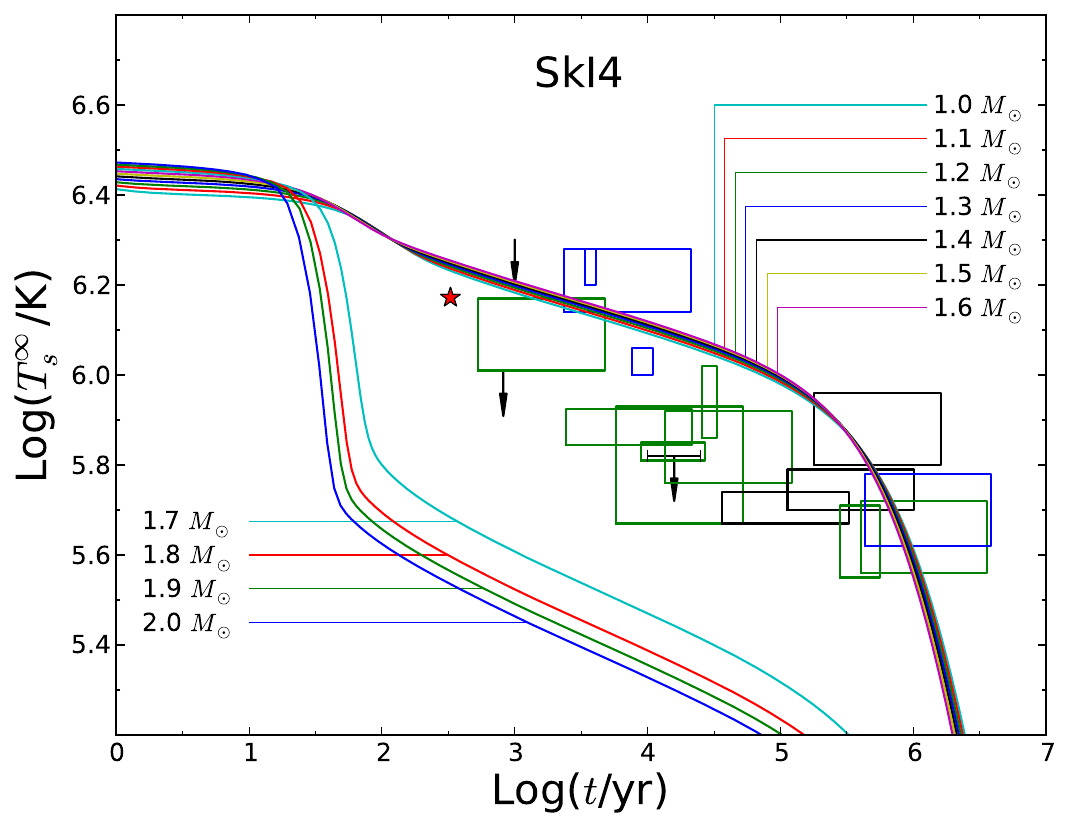} 
}
\centerline{
\includegraphics[scale=0.4]{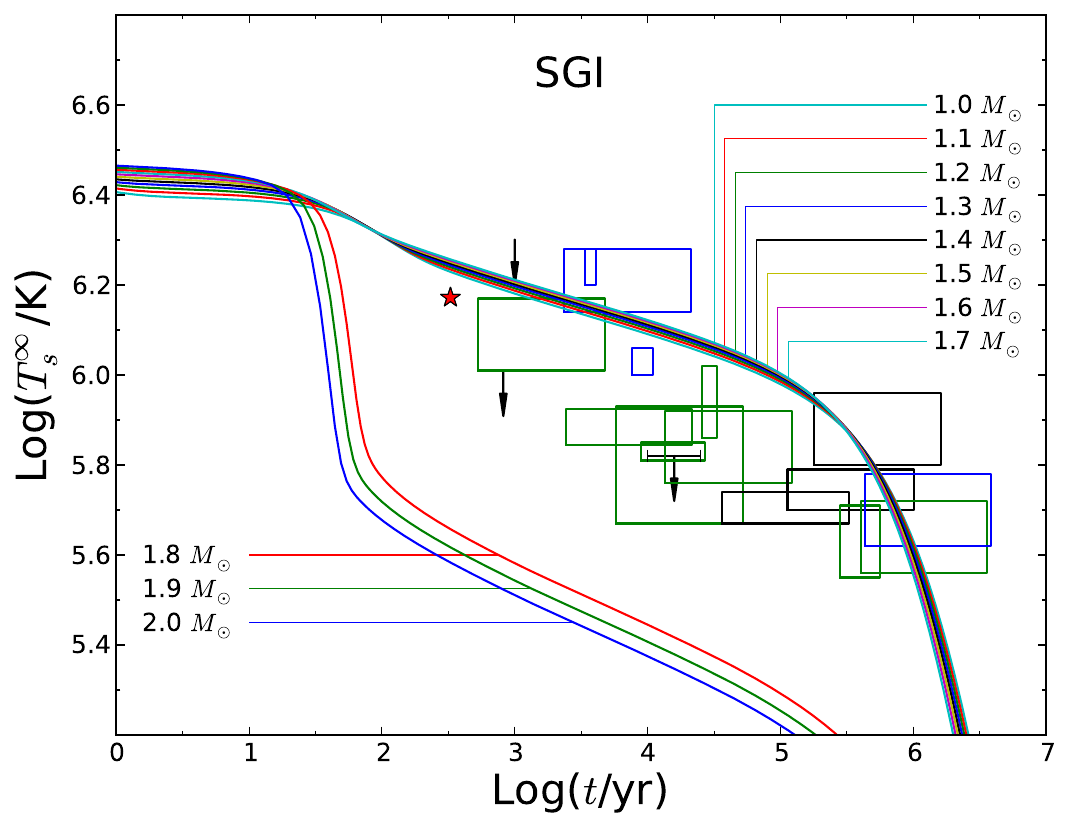} 
\includegraphics[scale=0.4]{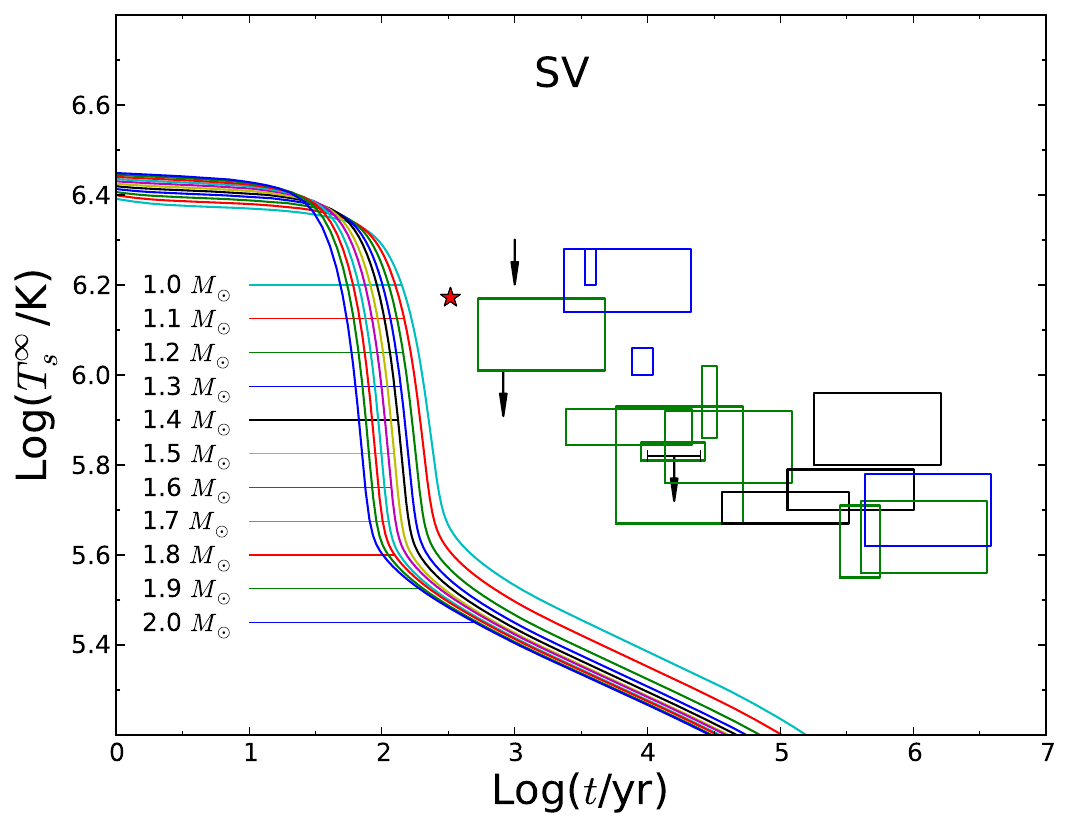} 
}
\centerline{
\includegraphics[scale=0.4]{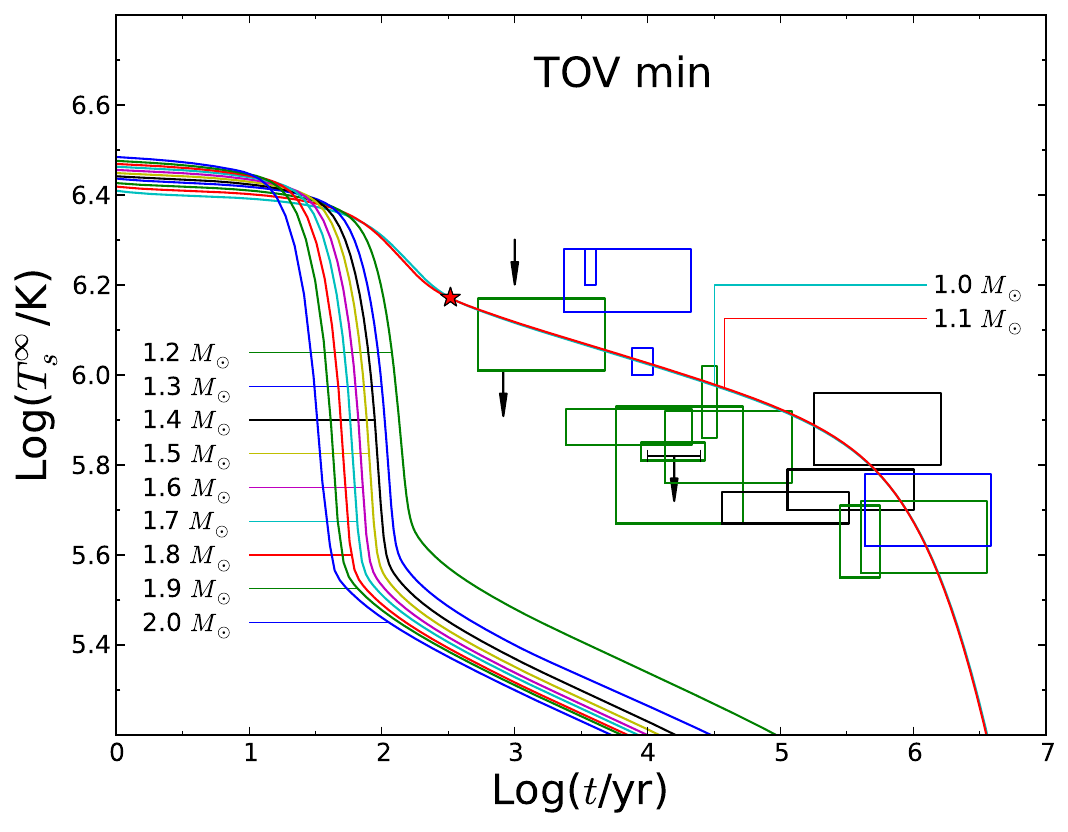} 
\includegraphics[scale=0.4]{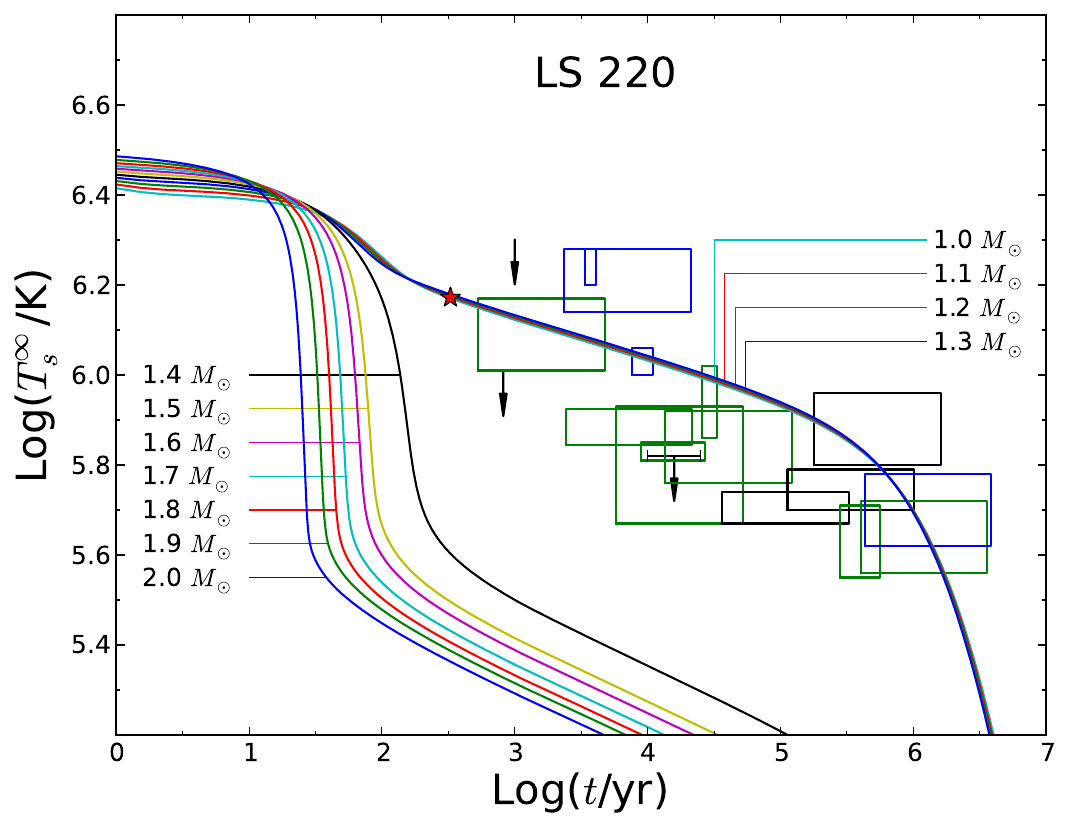}
}
\caption{(Color online) Surface temperature vs age without superfluidity effects
in the non-relativistic models. 
The symbol `$\star$' indicates the effective temperature of Cas A neutron star.
Each curve in the plot corresponds to different neutron star mass in the range of 
$1.0 M_{\odot}$ to $2.0 M_{\odot}$.
In case of SLy4, the direct Urca is not turned for any mass of neutron stars.
We use the (GPE) $T_s-T_{\rm b}$ relation in \cite{gpe1983}.
}\label{fig:sTL1}
\end{figure}

Figure~\ref{fig:sTL1} shows the cooling curves for the Skyrme force models.
SLy4 model shows similar behavior regardless of the masses of neutron stars, which is 
mainly due to the absence of the direct Urca process.
Were it not for the direct Urca, modified Urca is driving cooling mechanism in the
standard cooling scenario.
We can see that the modified Urca is good at reproducing the data for young (below $10^4$ yrs)
and old (above $10^5$ yrs) stars, but completely misses the middle-age ($10^4 \sim 10^5$ yrs) data.
This may imply that actual cooling will go through slow-quick-slow stages of the neutrino emission process.
For the SkI4 and SGI models, temperature drops abruptly if the masses of neutron stars are greater than 
$1.7 M_\odot$ and $1.8 M_\odot$, respectively.
This abrupt decrease of temperature is the signal for the ignition of the direct Urca.
Direct Urca is the fastest neutrino emission process ever known,
so once it is turned on, regardless of the existence of PBF or exotic states,
shape of the cooling curve is predominantly controlled by the direct Urca.
In Table~\ref{tb:crit_skyrme}, we summarize the critical 
densities for the electron and muon direct Urca process.
The result shows that the direct Urca is too fast that it fails to pass through any observation data.
On the other hand, if cooling is driven by the modified Urca, 
SLy4, SkI4 and SGI models show very similar thermal evolution trajectories.
In the SV model, the direct Urca occurs regardless of the mass of neutron stars, 
so the model cannot explain the temperature data at all.
For TOV min and LS220 models, the modified Urca is the main mechanism for low mass stars,
but the direct Urca starts to occur also in the low mass stars.
Assuming that most of the mass of the measured star in Figure~\ref{fig:obsns}
is in the range of $1.0 M_\odot \sim 1.6 M_\odot$,
TOV min can hardly explain the observed temperature profile.
It is striking that though TOV min model shows similar quality of mass-radius relation to
SLy4, SkI4 and SGI (Figure~\ref{fig:nsmr}), 
they predict quite different thermal evolution scenario.
Combining the empirical data from both mass-radius relation and temperature,
we can reduce the space for nuclear models which are suitable for the investigation of
superdense nuclear matter.
For this reason, we remove SV and TOV min models from the consideration hereafter.

\begin{figure}
\centerline{
\includegraphics[scale=0.4]{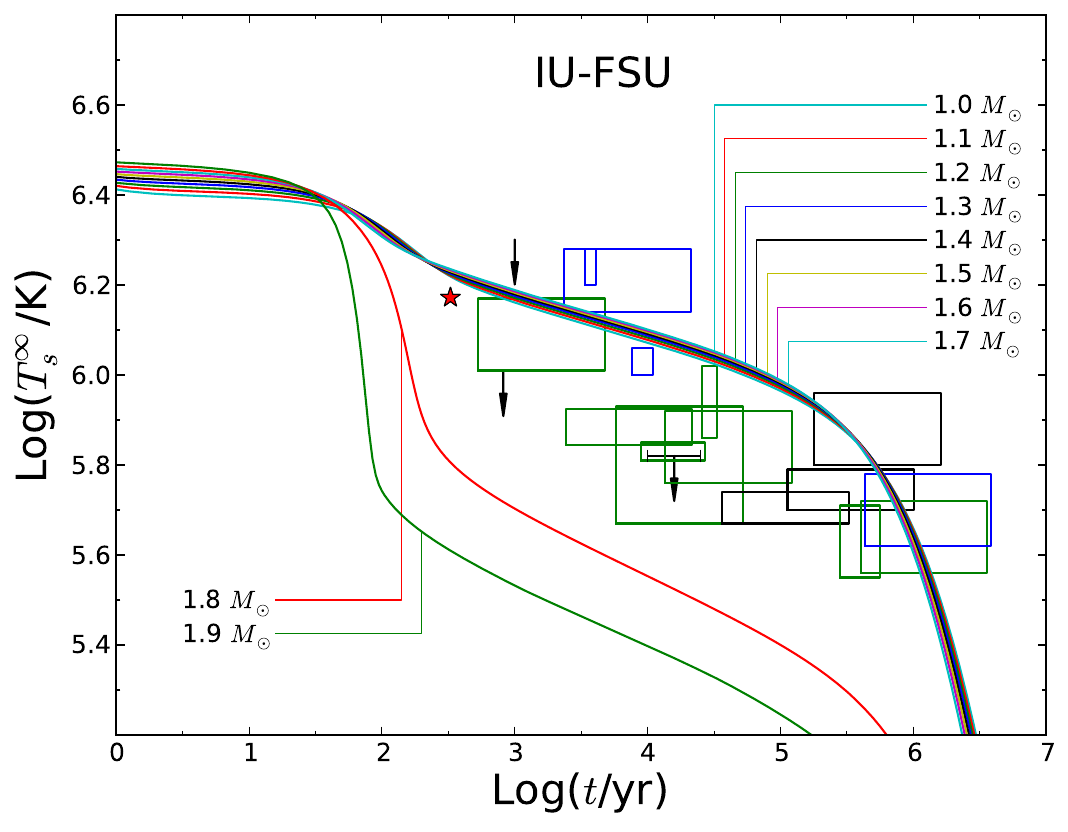} 
\includegraphics[scale=0.4]{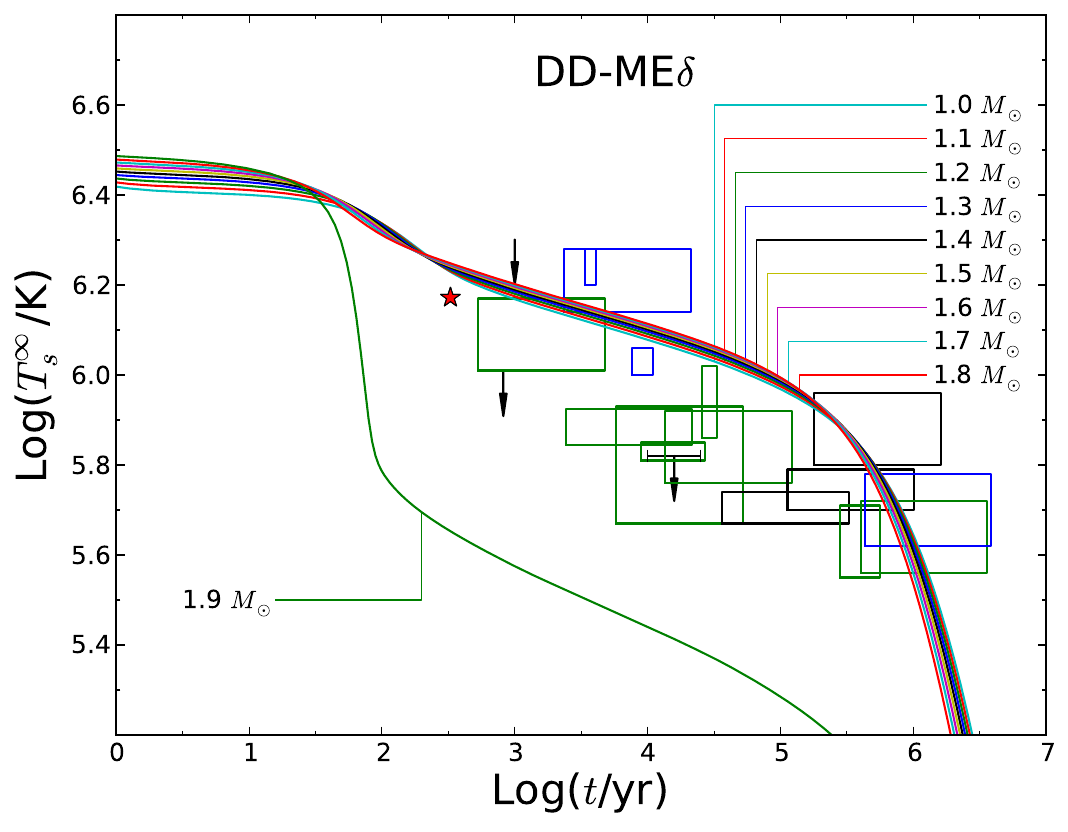}
}
\centerline{
\includegraphics[scale=0.4]{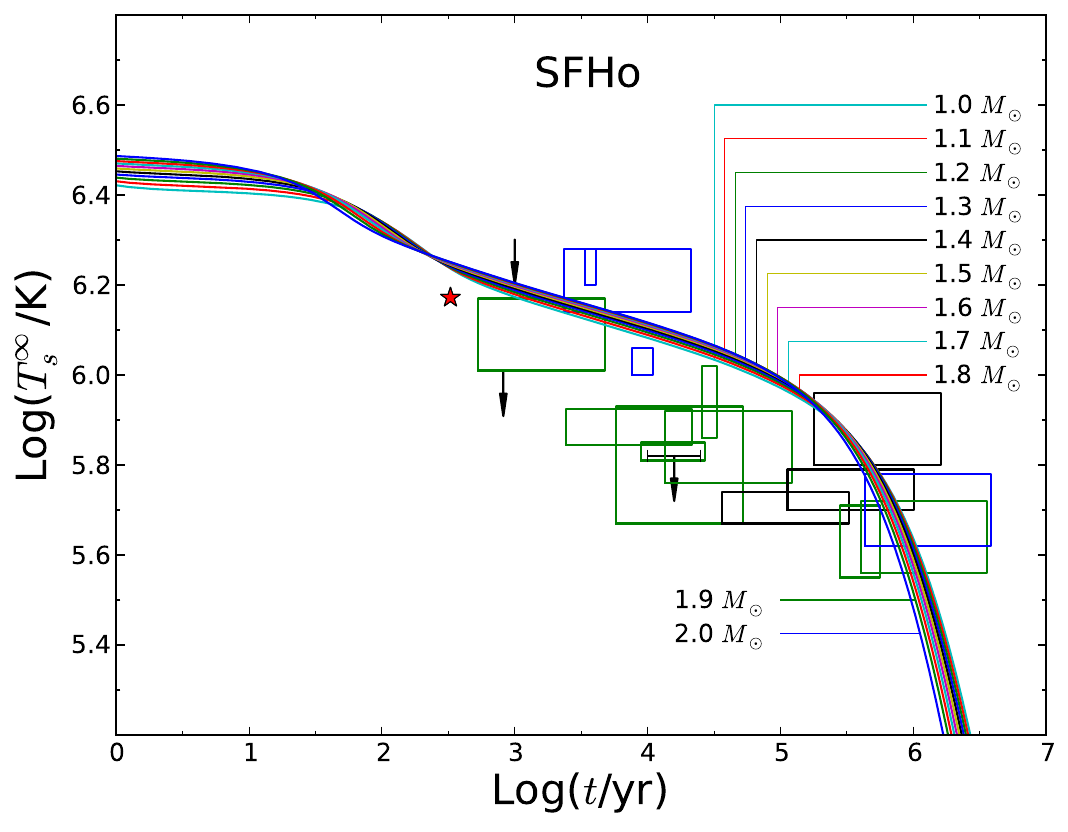} 
\includegraphics[scale=0.4]{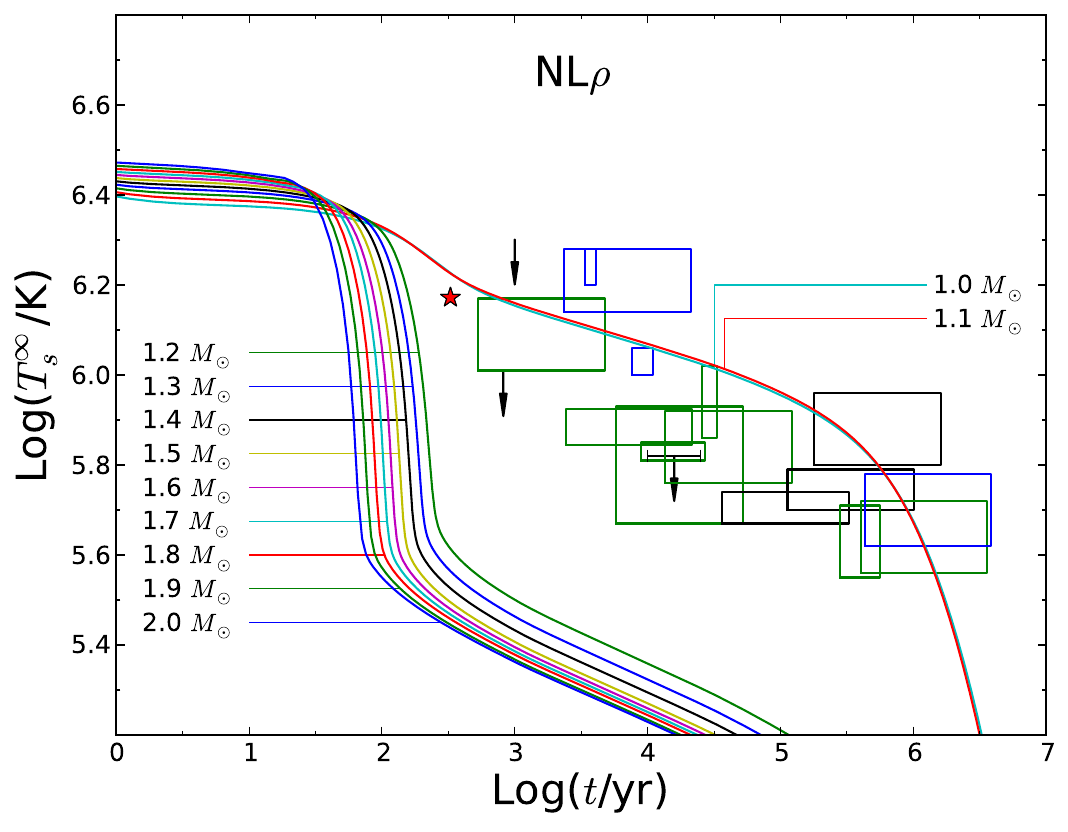} 
}
\centerline{
\includegraphics[scale=0.4]{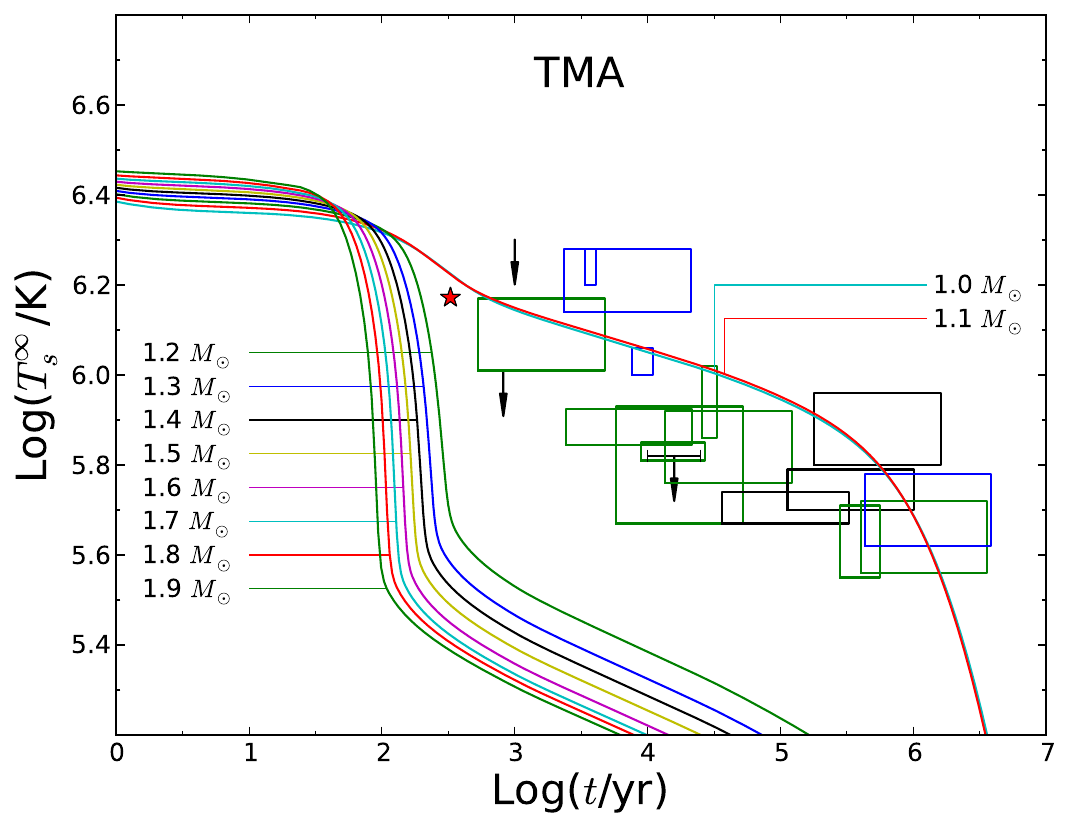} 
\includegraphics[scale=0.4]{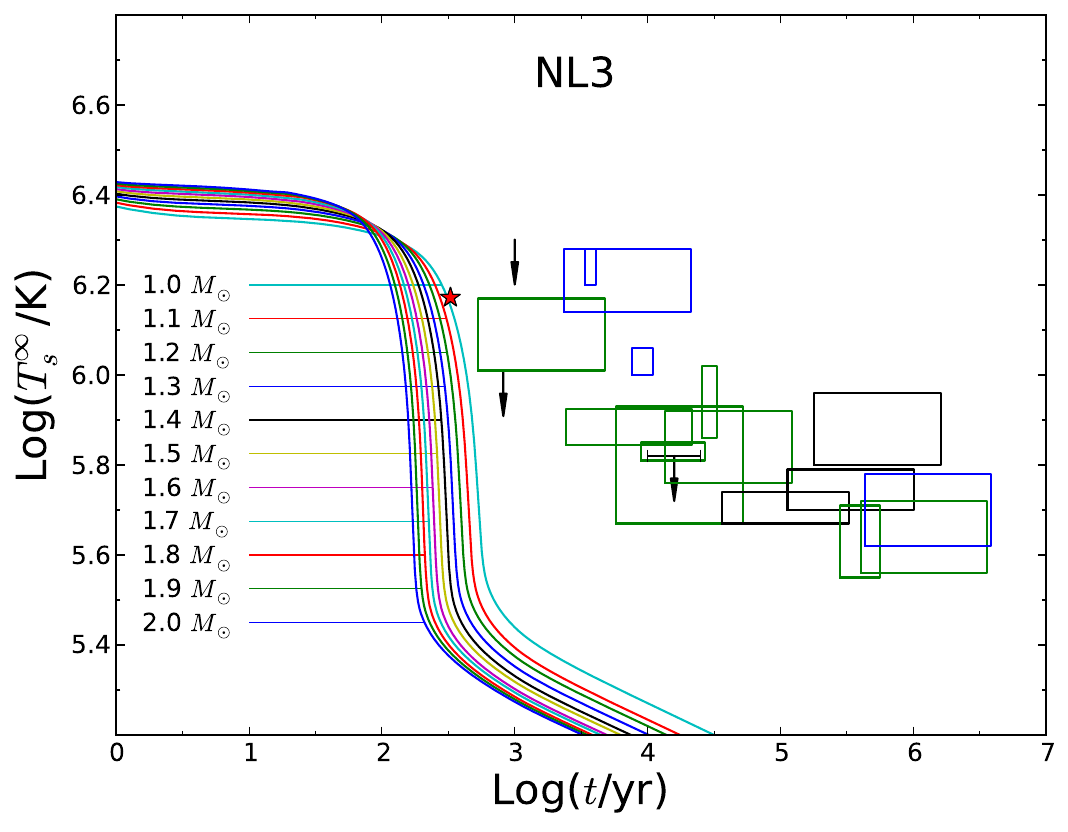}
}
\caption{ (Color online)
Same plot as in Figure~\ref{fig:sTL1} for RMF models.
For RMF models which have maximum mass less than
$2.0 M_{\odot}$, the cooling curve starts from $1.0 M_{\odot}$
and end up with $1.9 M_{\odot}$. 
For others, we draw up to $2.0 M_{\odot}$.
Critical neutron star mass for the direct Urca process in IU-FUS model is 
$1.77 M_{\odot}$ and significant effects can be seen for 
neutron stars, $M \ge 1.8 M_{\odot}$. GPE $T_s-T_b$ relation was used.}
\label{fig:sTL3}
\end{figure}

Figure \ref{fig:sTL3} presents the cooling curves for the RMF models.
We can see that the direct Urca is not working in the SFHo model,
and it is activated only in large mass stars in IU-FSU and DD-ME$\delta$ models.
We note that mass-radius relations with SFHo, IU-FSU and DD-ME$\delta$ models are
similar to those with SLy4, SkI4 and SGI models.
The similarity is kept for the cooling curves, which may support a strong correlation
between bulk properties of neutron stars and their thermal evolution.
Three stiff EoS models, NL$\rho$, TMA and NL3 cannot reproduce the observation data at all,
so we exclude them in the coming analyses.

\begin{table}
\begin{center}
\begin{tabular}{cccc}
\hline
\multirow{2}{*}{Model}    & \multicolumn{2}{c}{ $n_c$ ($M_{\rm crit}$) } & \multirow{2}{*}{$M_{\rm max}$ ($M_\odot$)} \\
\cline{2-3}
                                       &            $e$-dUrca  &  $\mu$-dUrca               &            \\
\hline
SLy4                 &  -            &    -         &  2.07 \\
SkI4                 &  0.502 (1.63) & 0.582 (1.83) &  2.19 \\
SGI                  &  0.492 (1.72) & 0.616 (2.00) &  2.25 \\
SV                   &  0.253 (0.97) &  0.315 (1.30) &  2.44 \\
TOV min          &  0.385 (1.12) & 0.458 (1.37) &  2.05 \\
LS220              &  0.433 (1.31) & 0.527 (1.55) &  2.04 \\
\hline
IU-FSU              & 0.611 (1.77) & 0.900 (1.94) & 1.94 \\
DD-ME$\delta$ & 0.764 (1.79) & 0.894 (1.89) & 1.96 \\
SFHo                 &-                    &  -                  & 2.06 \\
NL$\rho$           & 0.340 (1.11) & 0.414 (1.37) &  2.09 \\
TMA                   & 0.286 (1.14) & 0.358 (1.43) & 1.99 \\
 NL3                  & 0.205 (0.85)  & 0.255 (1.36) & 2.78 \\
\hline
\end{tabular}
\end{center}
\caption{Critical densities ($n_c$ in fm$^{-3}$) for the 
electron and muon direct Urca 
processes and the maximum mass of neutron stars ($M_{\rm max}$) for each model. 
Numbers in the parentheses ($M_{\rm crit}$ in unit of $M_\odot$) correspond to 
the neutron star masses at which the direct Urca processes start to occur.}
\label{tb:crit_skyrme}
\end{table}

%
%
%
%

%
%
%
%

\subsection{Radius and symmetry energy properties}

\begin{figure}[h]
\begin{center}
\includegraphics[scale=0.4]{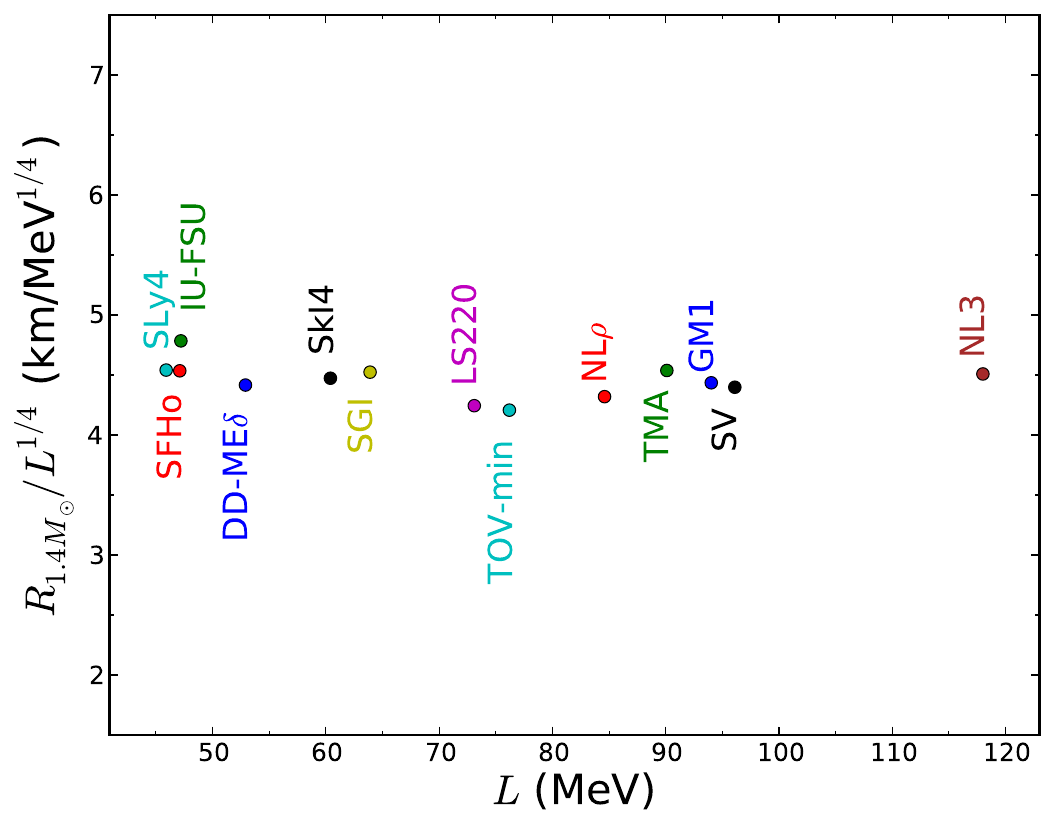}
\end{center}
\label{fig:slykaon}
\caption{(Color online) $R_{1.4\, M_{\odot}}/L^{1/4}$ for various nuclear models. 
$R_{1.4\, M_{\odot}}/L^{1/4}$ is nearly independent of models.}
\label{fig:rscale}
\end{figure}

The radii of neutron stars have a close relation with the pressure around 
nuclear saturation density, $R \propto P^{1/4}$ \cite{lp2007}. 
The energy per baryon and the pressure around saturation density can be expanded as
\begin{eqnarray}
E(n,\delta) &=& - B + \left(S_v + \frac{L}{3}\frac{n-n_0}{n_0} + \cdots \right)\delta^2
+ \cdots ,\\
P & = & n^2 \frac{\pt E}{\pt n} \simeq \frac{L}{3}\frac{n^2}{n_0}\delta^2,
\end{eqnarray}
where $\delta = \frac{n_n - n_p}{n}$.
This leads to a rough relation between the radius and the density derivative of symmetry energy $(L)$,
\begin{equation}
R \propto L^{1/4}\,.
\end{equation}
Proton fraction is determined from the ground state energy of nuclear matter. 
Symmetry energy, which is roughly an estimate of the energy difference 
between symmetric and asymmetric nuclear matter,
has a relation with proton fraction.
The algebraic relation $S(n) \simeq S_v + \frac{L}{3}\frac{n - n_0}{n_0}$ 
indicates that greater $L$ leads to larger the proton fraction. 
This implies that the direct Urca process is related to the radius of
neutron stars, and thus $L$ can be a good indicator
of the direct Urca process in the core of neutron stars.
Figure \ref{fig:rscale} shows that $\frac{R}{L^{1/4}}$ is insensitive to the choice of nuclear models.
From this observation, one can conclude that the radius may be a good indicator of the symmetry energy. 
Since the proton fraction increases as $L$ increases in general and the turn-on of direct Urca process 
strongly depends on the proton fraction, NS radius may be a good indicator of the existence of direct Urca process.
For example, a large radius ($R_{1.4 M_{\odot}} > 14$ km, 
or $L > 90$ MeV) is not favored
because the direct Urca process occurs even for a small mass NS.
This is also consistent with Steiner \textit{et al}. \cite{slb2010} in which
they estimated the range of neutron stars' masses and radii using X-ray burst data.

Lattimer and Lim \cite{ll2013} summarized symmetry energy properties ($S_v$, $L$)
with both experimental results and theoretical calculations.
The analysis from the nuclear mass fits, neutron skins, heavy-ion collisions,
giant dipole resonances and dipole polarizabilities gives an overlapped region.
Considering the theoretical calculation of pure neutron matter and
astrophysical observations of neutron stars, the allowed ranges
of symmetry energy ($S_v$) and its density gradient ($L$) are
29.0 MeV$ < S_v < $ 32.7 MeV and 40.5 MeV$ < L < $ 61.9 MeV.
Our result for neutron star cooling indicates that $L < 85$ MeV 
so that the direct Urca process should not be
turned on in the low mass neutron stars ($M < 1.2M_\odot$).
This is consistent with Lattimer and Lim's conclusion. 

\subsection{Effect of envelope elements}

\begin{figure}
\centerline{
\includegraphics[scale=0.4]{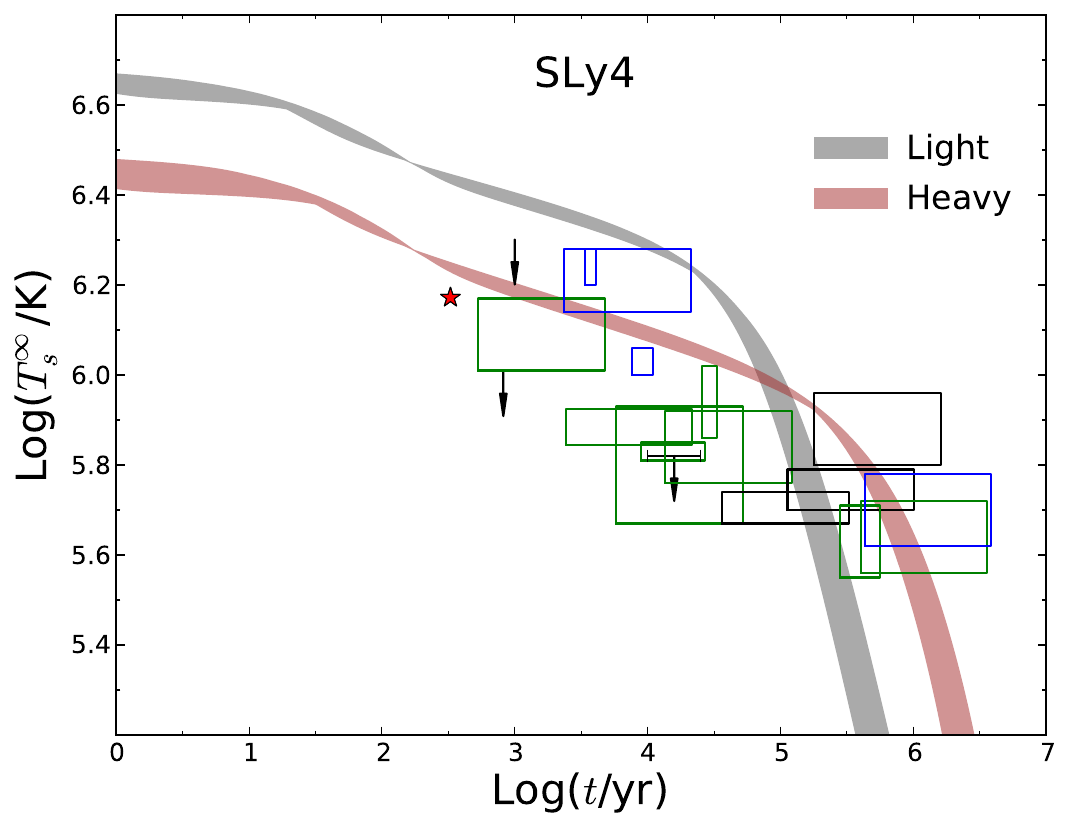} 
\includegraphics[scale=0.4]{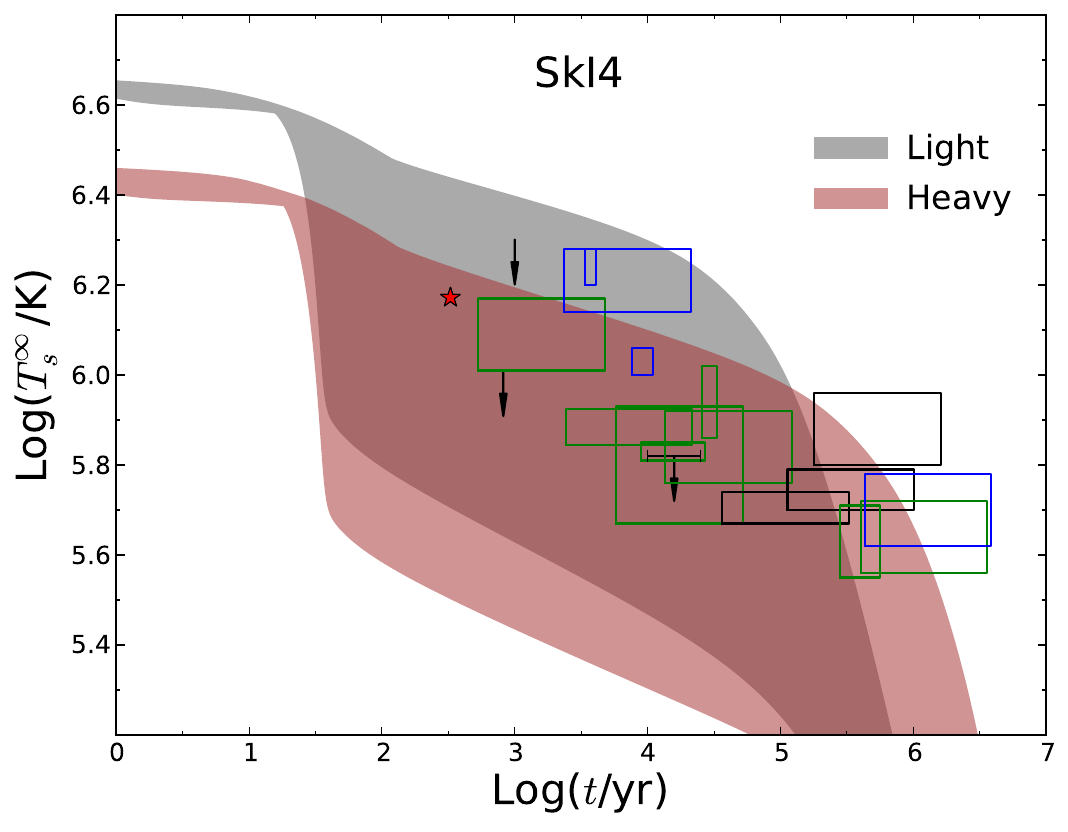}
}
\caption{(Color online) Band plot of $T_{s}^{\infty}$ with light elements envelope and
heavy elements envelope, respectively. 
Each band has the mass range between 1.2 $M_\odot$ and 2.0 $M_\odot$. 
SLy4 cannot explain some of data even 
if light and heavy elements are considered simultaneously
since the direct Urca process is not activated even 
in the maximum mass of a neutron star.
}
\label{fig:ts_band}
\end{figure}

It was shown that the surface temperature highly depends on the abundance
of light elements in the envelope region \cite{plps2004, pcy1997}. 
Figure~\ref{fig:ts_band} shows the band plot of $T_{s}^{\infty}$ both with the 
light and heavy elements. The bands in each plot indicate neutron star masses 
in the range of $1.2 M_{\odot} \sim  2.0 M_{\odot}$.
At early ages, the top curve represents the most massive star and at  later times, 
the curve from massive stars is the one with the lowest temperature. 

For a more realistic cooling process, one has to take into account 
the fraction of light elements in the envelope of neutron stars. 
The chemical evolution from the light elements
to the heavy elements or pulsar injection of light elements into the magneto sphere \cite{plps2004}
indicate that the real cooling curves may start from the band with light elements and move towards 
 the band with heavy elements as the neutron star evolves.
The mass of light elements is defined as
\begin{equation}
\Delta M (t) = \Delta M (t=t_i) e^{-(t-t_i)/\tau_{d}}\,,
\end{equation}
where $\tau_d$ is the reduction time scale of the mass fraction of light element. 
In the accreted envelope, the surface temperature can be fitted
as a function of the mass fraction of light elements
to the total mass of neutron stars \cite{pcy1997}.
If the surface is made of pure irons, 
\begin{equation}
T^4_{\mathrm{eff6,Fe}} = g_{14}[(7\zeta)^{2.25} + (\zeta/3)^{1.25}] \,,
\end{equation}
where $\zeta = T_{\mathrm{b9}} - (7 \, T_{\mathrm{b9}}\sqrt{g_{14}})^{1/2}/10^3$
and $g_{14} = \frac{1}{10^{14}}\frac{GM}{R}\left(1- \frac{2GM}{Rc^2}\right)^{-1/2}$.
We define $T_{\mathrm{b}}$  ($T_{\mathrm{b9}}=T_{\mathrm{b}}/10^9$K) 
is the temperature where the energy density is $10^{10}$ g cm$^{-3}$.
On the other hand, if there are only hydrogens, 
\begin{equation}
T^4_{\mathrm{eff6,a}} = g_{14}(18.1\, T_{\mathrm{b9}})^{2.42} \,.
\end{equation}
For partially accreted envelope, with the definition of $\eta \equiv g_{14}^2 \Delta M/ M$,
we have 
\begin{equation}
T_s = \left[ 
\frac{a T^{4}_{\mathrm{eff6, Fe}} + T^4_{\mathrm{eff,a}}}
{a + 1}
\right]^{1/4}\,,
\end{equation}
where
$a =[1.2 + (5.3\times 10^{-6}/\eta)^{0.38}]\,T_{\mathrm{b9}}^{5/3}$.
Using $T_s-T_{\rm b}$ relation in Ref. \cite{pcy1997}, 
we can find the surface temperature for given mass of light elements on the surface.  

\begin{figure}
\centerline{
\includegraphics[scale=0.4]{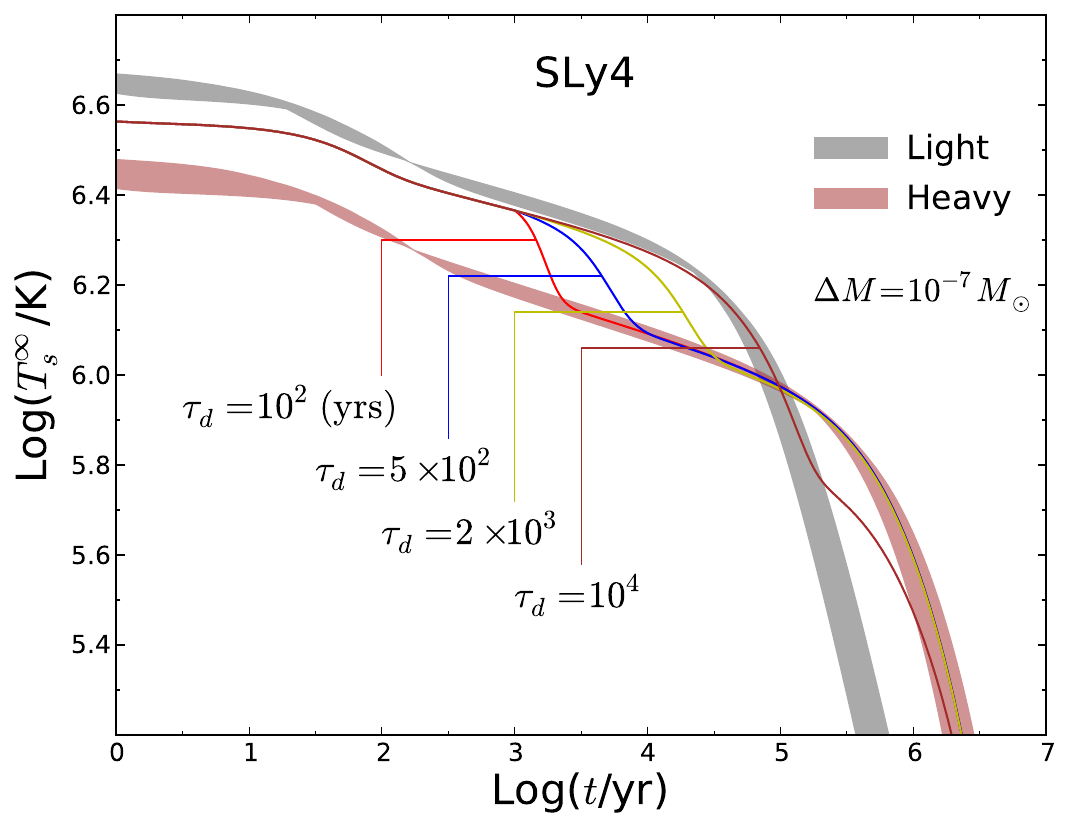} 
\includegraphics[scale=0.4]{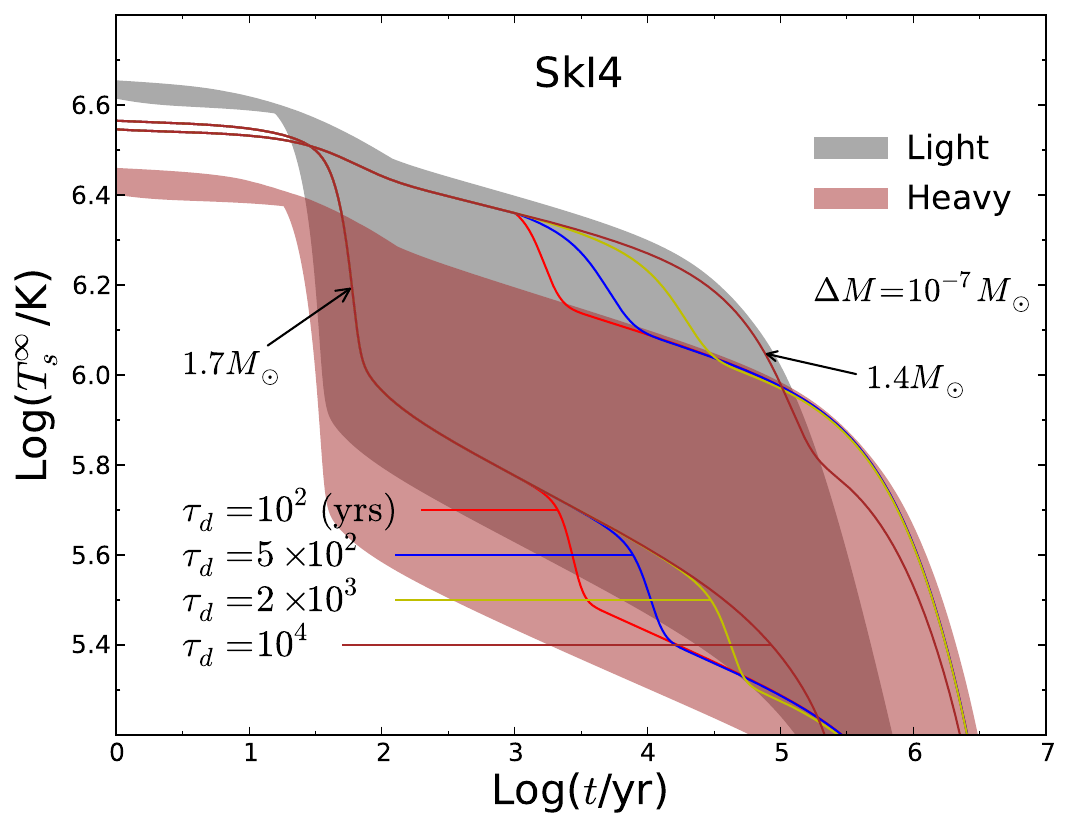} 
}
\caption{(Color online) Light element decay and neutron star cooling curve. Left panel shows
 $1.4 M_{\odot}$ neutron star cooling path with SLy4 model. The initial mass
 of light elements is assumed to be $10^{-7} M_{\odot}$ and the decay
 starts at $t = 10^3$ years. The right panel shows the same evolution path
 but with two different masses of neutron stars ($1.4 M_{\odot}$, $1.7 M_{\odot}$)
 using SkI4 model.}
\label{fig:ts_evolve}
\end{figure}

In Figure \ref{fig:ts_evolve}, we show the cooling paths obtained by taking 
into account the reduction in the mass fraction of light elements. Depending on the amount of
light elements and the reduction time scale ($\tau_d$), the actual neutron star cooling curve
will be located between the band of light and the band of heavy elements.
In this figure,  the initial mass of light elements is assumed to be $10^{-7} M_{\odot}$ and
the mass of light elements starts to decrease when $t =10^3$ years after the birth of neutron stars. 
In case of SLy4, the cooling curve is almost identical 
for all mass of neutron stars, thus a typical mass $1.4 M_{\odot}$ is chosen to see the evolution path.
For SkI4, $1.4 M_{\odot}$  and $1.7 M_{\odot}$ cooling paths are shown to compare the dependences 
on the reduction time scale $\tau_d$. 
The results show that the rapid drop of surface temperature can occur 
when the mass fraction of light elements decreases.
Note that the EoS which doesn't allow the direct Urca process (e.g. SLy4) cannot
explain middle-age low temperature neutron stars without other fast cooling mechanisms
such as cooper pair emission or Bose condensation. 
This implies that elements composition, abundances of light elements, and the direct Urca process can be 
used selectively to explain the observed data. 

\section{Results with Superfluidity}\label{sec:super}

As discussed by Page and Applegate \cite{pa1992}, and
Yakovlev \textit{et al}. \cite{yls1999}, the direct Urca process is active in the really
narrow mass range. 
As shown in Figures \ref{fig:sTL1}, \ref{fig:sTL3},
the direct Urca process imposes huge effects on the cooling curve. If the mass of
a neutron star is slightly greater than the critical mass for the direct Urca process
(e.g. $M > M_{D} + 0.01 M_{\odot}$), the effect of direct Urca is manifest
\footnote{Numerically, we should increase grid points in the core of neutron
star to treat the direct Urca process properly. In this work, we used 16 times
more grid points to see the split of the curves between $1.30 M_{\odot}$
and $1.40 M_{\odot}$.}.
\begin{figure}
\centerline{
\includegraphics[scale=0.4]{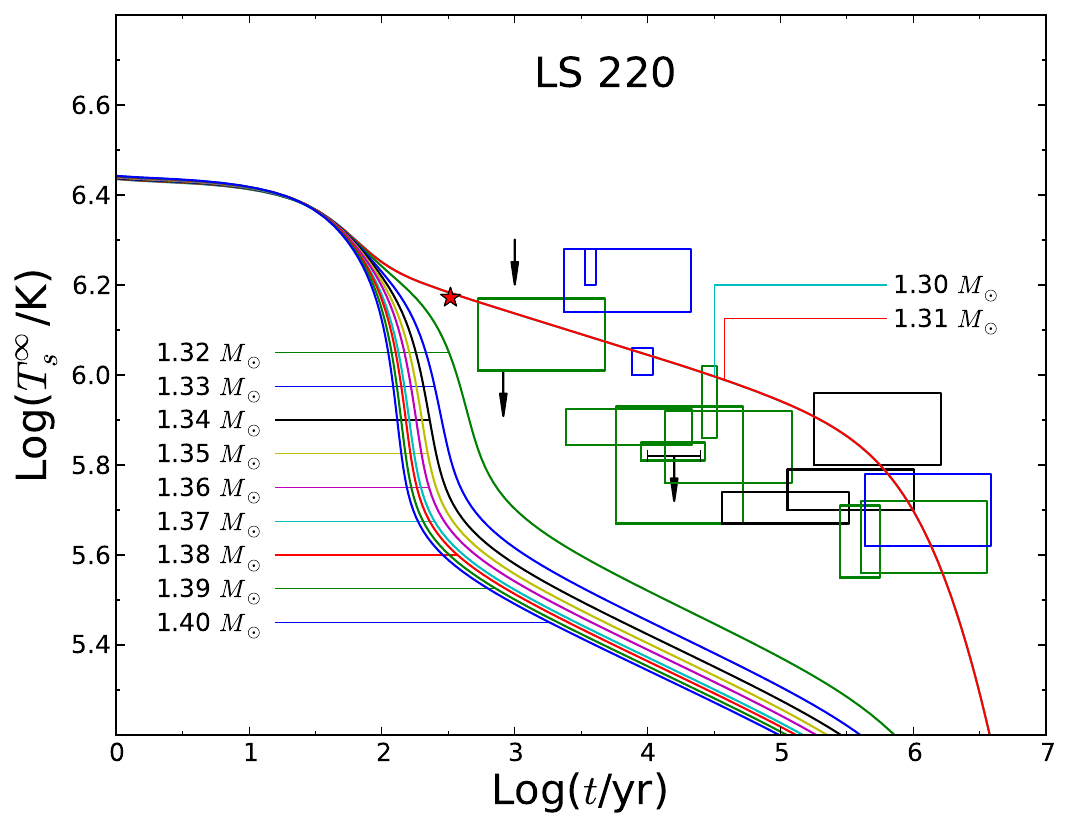} 
\includegraphics[scale=0.4]{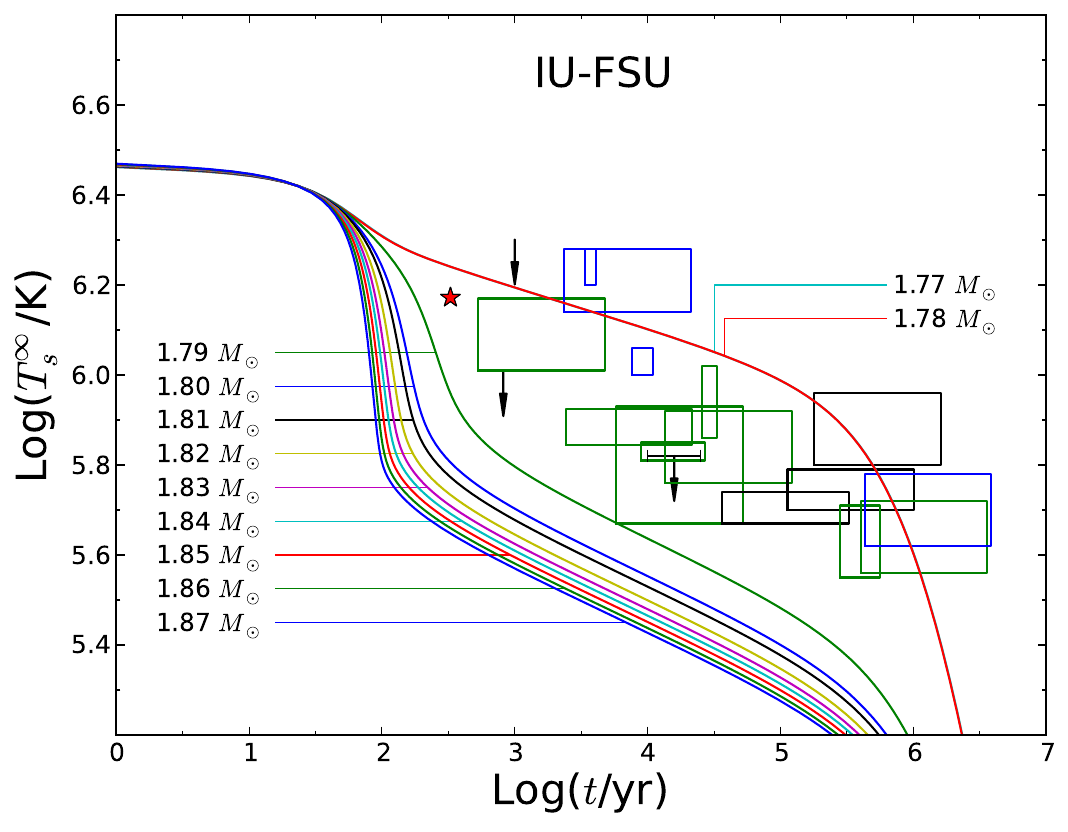} 
}
\caption{(Color online) 
The direct Urca Process effects on LS200 and IU-FSU. For both models,
most of the observations are in the very narrow mass range 
$(\Delta M = 0.01 \,M_{\odot})$. GPE $T_s-T_b$ relation was used.} 
\label{fig:ts_durca}
\end{figure}
In Figure \ref{fig:ts_durca} the effects of the direct Urca process  
on neutron star cooling are summarized with two models, LS220 and IU-FSU. 
The results show that most of the observed data are located between
the two curves of  $1.31 M_\odot$ and $1.32 M_\odot$ for LS220, and 
$1.78 M_\odot$ and $1.79 M_\odot$ for IU-FSU. This may imply that the
mass of observed middle-age stars is concentrated in a very narrow range
of $0.01 M_\odot$.
In the statistical point of view, it is very unlikely
that almost all the observations are in the range of
$0.01M_{\odot}/M_{\odot} \approx 1\%$, because the mass
distribution of the observed neutron stars has a broader range \cite{lattimer2012}.
This problem can be managed if we employ the pairing effects. 

It is believed that ${}^{1}S_0$ neutron superfluidity exists
in the inner crust, and ${}^{1}S_0$ proton and ${}^{3}P_2$ neutron superfluid states appear in the
core of neutron stars. 
Since there is no free proton in the crust, the superfluidity involving
proton can exist only in the core region. 
If the local temperature drops below the critical temperature
for the superfluidity, the proton superfluidity delays the surface temperature drop
due to significant reduction of neutrino emissivity, heat capacity, and thermal conductivity.
For instance, the reduction factor for the direct Urca
process behaves like $\exp(-\Delta/T)$
where $\Delta$ is a pairing gap energy.
\footnote{For the modified Urca process, the reduction factor behaves as $\exp(- 2 \Delta/T)$} 
In general, the reduction factors can be obtained through the 
numerical calculation \cite{yls1999,bhy2001}.

Along with reduction from the superfluidity state,
the superfluidity opens new neutrino emission process which is called
pair breaking and formation (PBF) process. 
The neutrino emissivity formula for the PBF process is given by \cite{yak1999, plps2009}
\begin{equation}
Q_{\rm PFB}  =  3.51\times 10^{21} \left(\frac{m_i^\star}{m_i}\right)
\left(\frac{p_{Fi}}{m_{i}C}\right)T_{9}^7 a_{i,j}F_{j}
\left[\frac{\Delta_i(T)}{T}\right] \,
\frac{\mathrm{erg}}{\mathrm{cm}\cdot\mathrm{s}}\,,
\end{equation}
where $i$ represents type of nucleons ($i=n, p$) and  $j$ stands for singlet ($j=s$) or triplet ($j=t, m_J=0$)
pairing. $F_{s}$ and $F_{t}$ are given in Ref. \cite{yak1999} as
\begin{equation}
F_{s} = y^2 \int_{0}^{\infty} \frac{z^4\, dx}{(1+e^z)^2}\,,
\quad 
F_{t} = \frac{1}{4\pi}\int d\Omega\,y^2\int_{0}^{\infty} \frac{z^4\, dx}{(1+e^z)^2}\,,
\end{equation}
where $y = \Delta_i(T)/T $, $z =\sqrt{x^2+y^2}$, and
$\int d\Omega$ is the solid angle integration. The fitting functions for $F_s$ and $F_t$
are also given in Ref. \cite{yak1999}. 
When vector current is conserved, $a_{i,j}$'s are given in Ref. \cite{plps2009},
and their forms without the vector current conservations are given in \cite{yak1999}.
Since the core of a neutron star is dominated by neutrons  
($p_{Fn} \gg p_{Fp}$) and the magnitude of $a_{n,t}$ and $a_{p,s}$ are
comparable, the triplet PBF is the main neutrino emission agent once the
superfluidity occurs. (Note that $^3P_2$ neutron and $^1S_0$ proton pairing
are expected to exist in the core of neutron stars.) When the temperature drops below
the critical temperature, the modified Urca and bremsstrahlung neutrino
emission processes are highly suppressed and PBF process overwhelmes the other neutrino
emission processes \cite{plps2009}.

In order to make the calculation or nuclear superfluidity simple and efficient, 
we introduce a phenomenological critical temperature formula to see the effect 
of gap size and the density range,
\begin{equation}\label{eq:tcform}
T_c(k_f) =
\begin{cases}
T_{c}^{\text{max}} \cdot \mathcal{N} \cdot (k_f - k_0)^{\alpha_c}(k_2 - k_f)^{\beta_c}
  & \text{if} \quad k_0 < k_f < k_2 ; \\
        0 & \text{if otherwise},
\end{cases}
\end{equation}
where $T_c^{\mathrm{max}}$ is the maximum critical
temperature for superfluidity for a given
$k_f$ (the Fermi wave number for a total bayron number density).
$k_0$ ($k_2$) is the starting (ending) wave number for a given type of
pairing. $\mathcal{N}$ is the normalization factor for the critical temperature.

\begin{figure}
\centerline{
\includegraphics[scale=0.4]{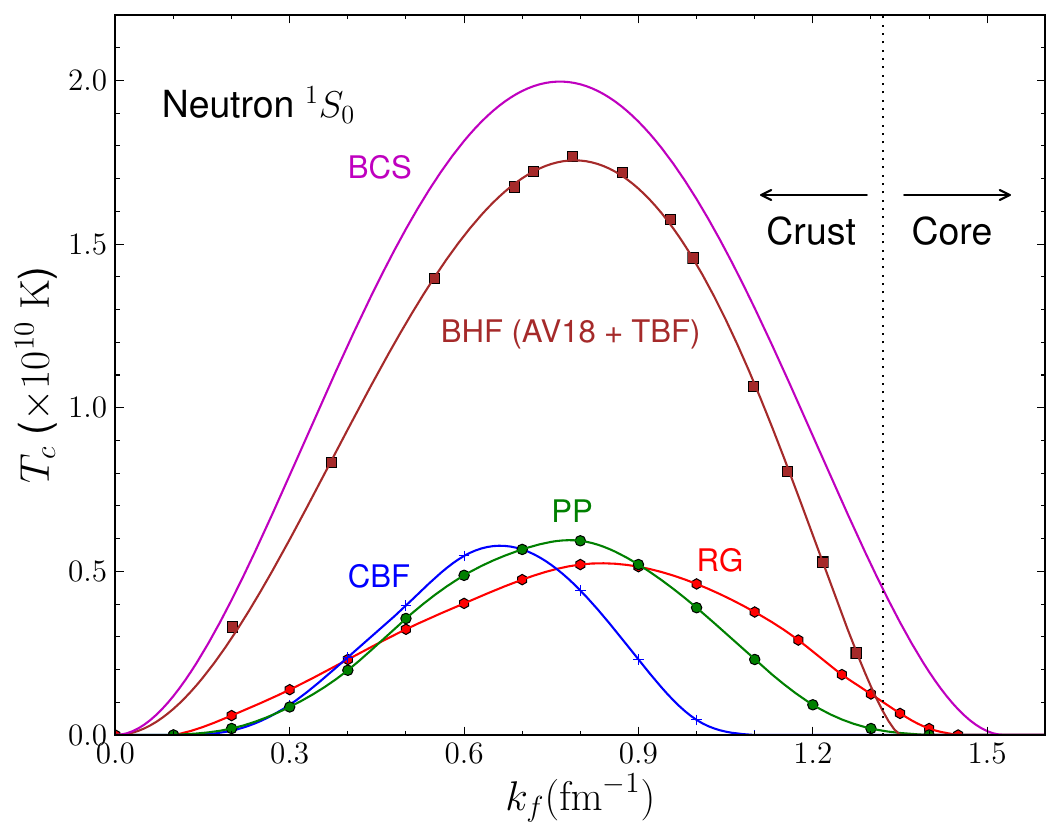} 
\includegraphics[scale=0.4]{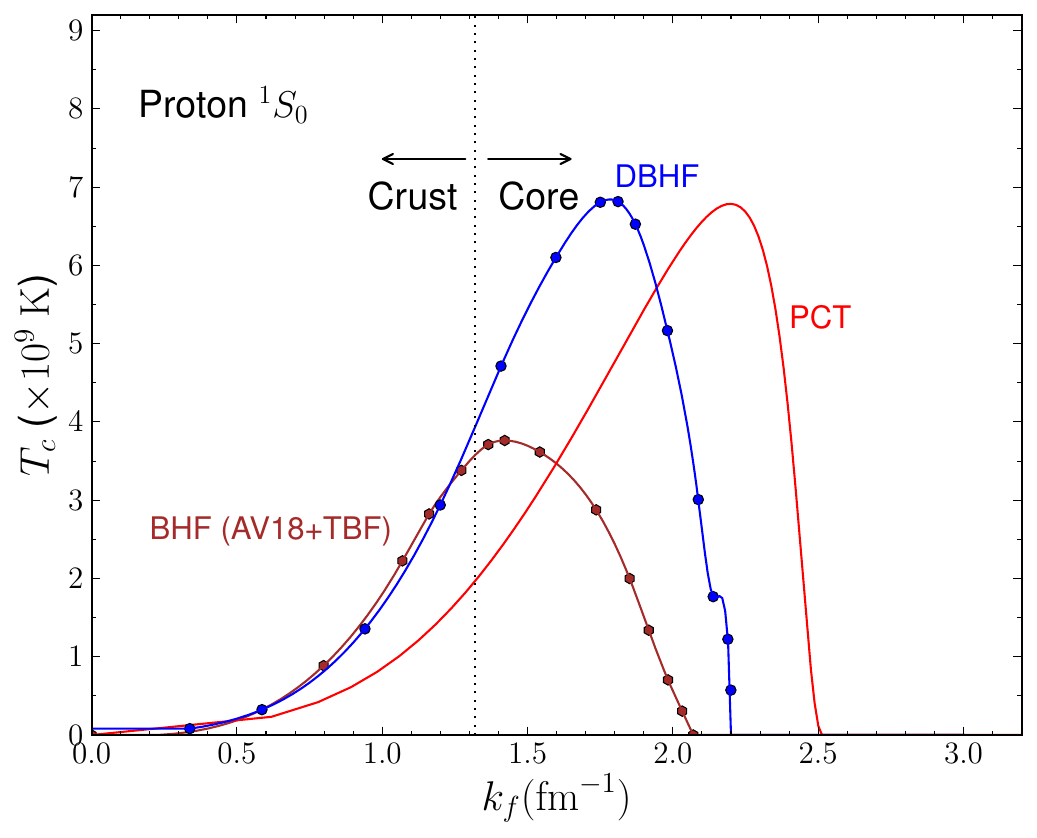} 
}
\centerline{
\includegraphics[scale=0.4]{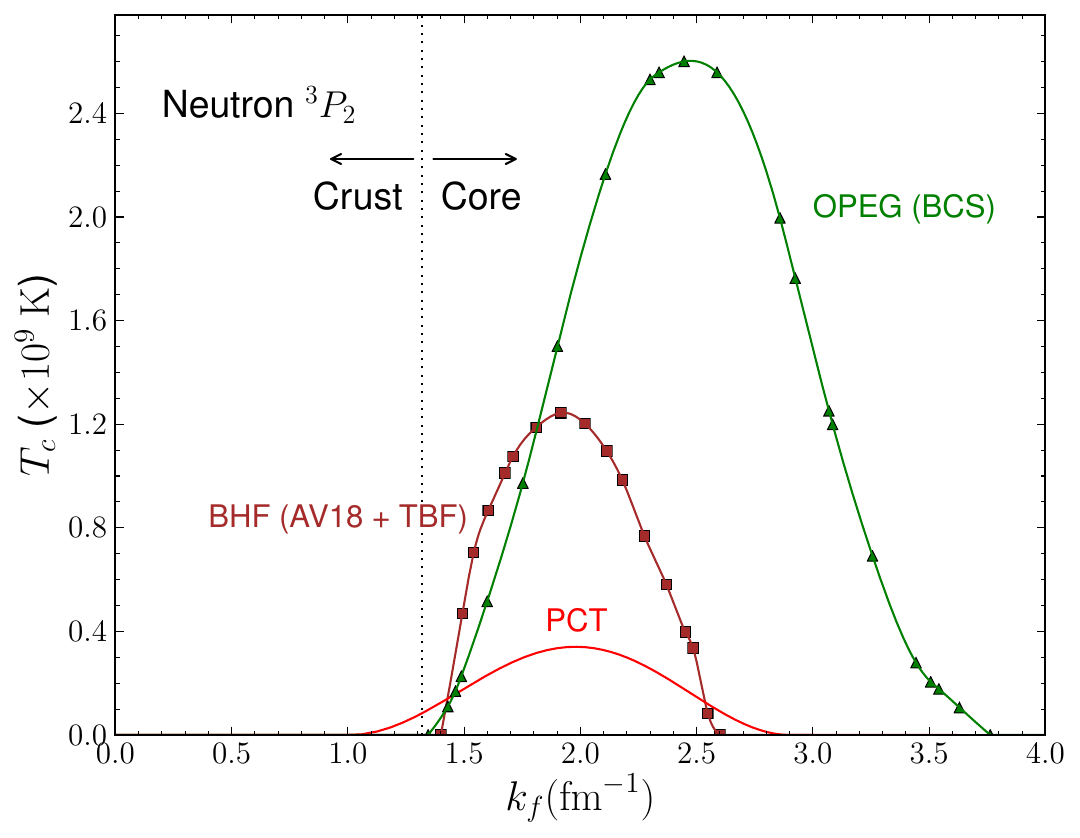} 
\includegraphics[scale=0.4]{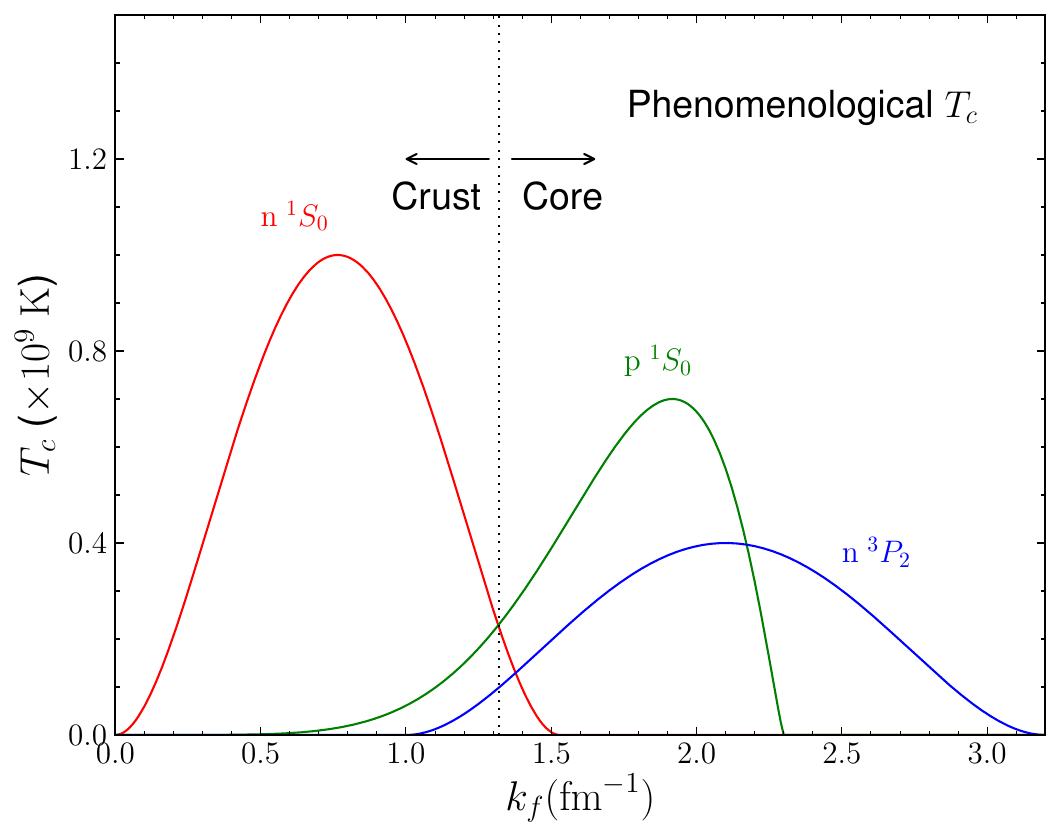} 
}
\caption{(Color online) Critical temperature for different type of pairing in beta-stable nuclear matter. 
$k_f$ is the wave number of Fermi-momentum (in unit of $\text{fm}^{-1}$) for the total baryon number density. 
The core-crust boundary is obtained assuming the phase transition happens at 
$\rho \simeq 0.5\rho_0 = 0.08 \,\,\text{fm}^{-3}$.
Each type of pairing
can be obtained using different methods. ${}^{1}S_0$ neutron : 
CBF - Correlated Basis Function \cite{ccdk1993}, 
PP - Polarization Potential \cite{wap1993},
BHF - Brueckner Hartree Fock \cite{zuo2004},
RG - Renormalization Group \cite{sfb2003}.
${}^{1}S_0$ proton : DBHF - Dirac Brueckner Hartree Fock \cite{eeho1996},
PCT - Parameterized Critical Temperature \cite{khy2001}, BHF \cite{zuo2004}.
${}^{3}P_2$ neutron : BHF \cite{zuo2008}, PCT \cite{khy2001},
OPEG (BCS) - One Pion Exchange Gaussian with generalized BCS \cite{aom1}.
The right bottom figure shows the critical temperatures for each type of
pairing. All curves were obtained from Eq.~\eqref{eq:tcform}
with $\alpha_c=2$ and $\beta_c=2$ for ${}^{1}S_{0}$ and ${}^{3}P_{2}$ neutron pairing
and with $\alpha_c=6$ and $\beta_c=1.2$ for ${}^{1}S_{0}$ proton pairing.
}\label{fig:tc_all}
\end{figure}

In Figure \ref{fig:tc_all}, the critical temperatures are summarized for a given $k_f$ in
beta-stable nuclear matter. 
For the $^{3}P_2$ neutron pairing (left bottom),
$k_f$ may be different from the original paper since we use $k_f$ for the total baryon
number density not for the pure neutron matter. If the gap calculation
is done in the pure neutron matter ($k_{Fn}$), the proton fraction is given
by APR EoS \cite{aprEoS} to recover $k_f$ of the total baryon number density for this figure. 
In each gap calculation, the critical temperature strongly depends on the methodology. 
Considering this fact, we use the phenomenological critical temperature formula and see the cooling
curve how it depends on it. 
For $^{1}S_0$ and $^{3}P_{2}$ neutron pairing,
$\alpha_c=2$ and $\beta_c=2$ are suitable to represent the critical temperatures.
For ${}^{1}S_0$ proton pairing, we adopt $\alpha_c=6$ and $\beta_c=1.2$ to
mimic the behavior of the critical temperature in this example.
\begin{figure}[tbp]
\centerline{
\includegraphics[scale=0.4]{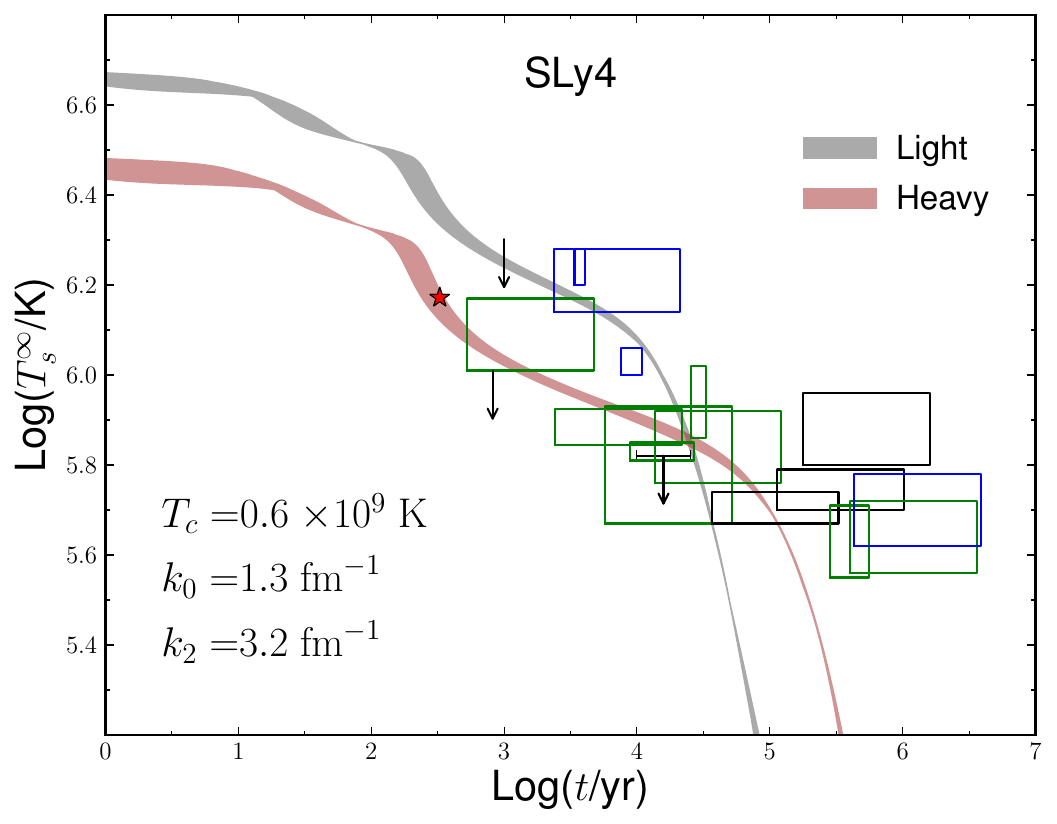} 
\includegraphics[scale=0.4]{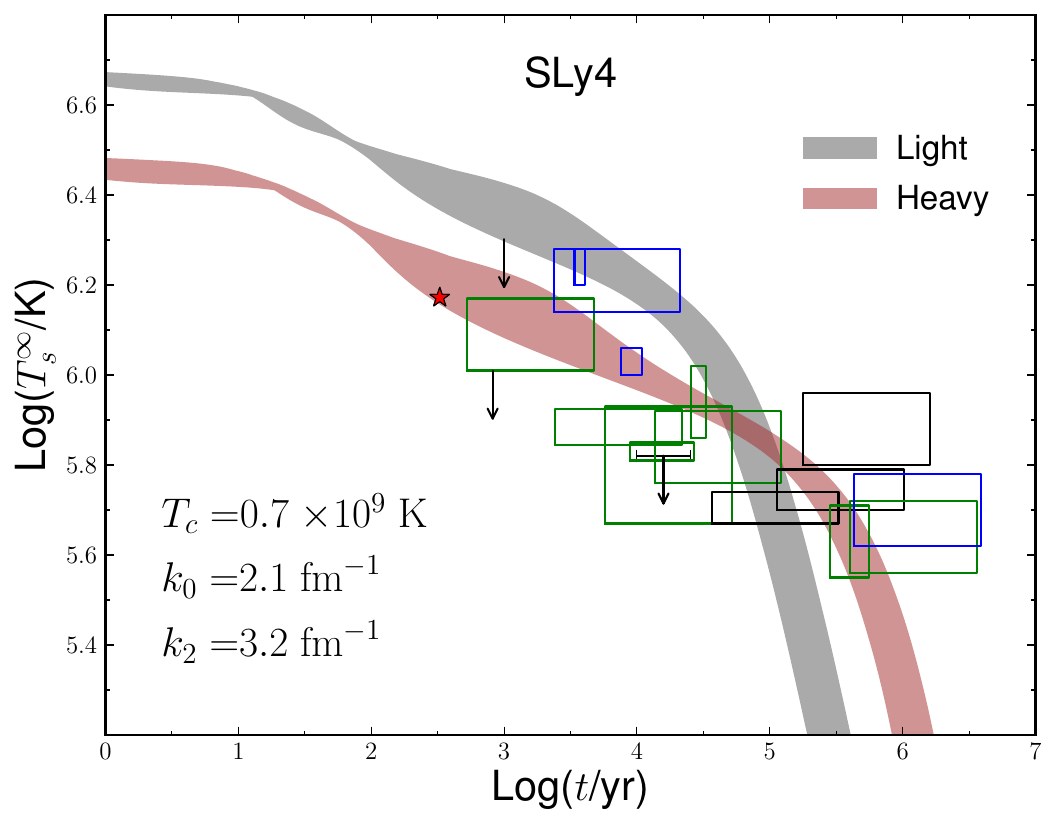}
}
\caption{(Color online) The effect of superfluidity for the SLy4 model.}\label{fig:sly4_sup}
\end{figure}

\begin{figure}[h]
\centerline{
\includegraphics[scale=0.4]{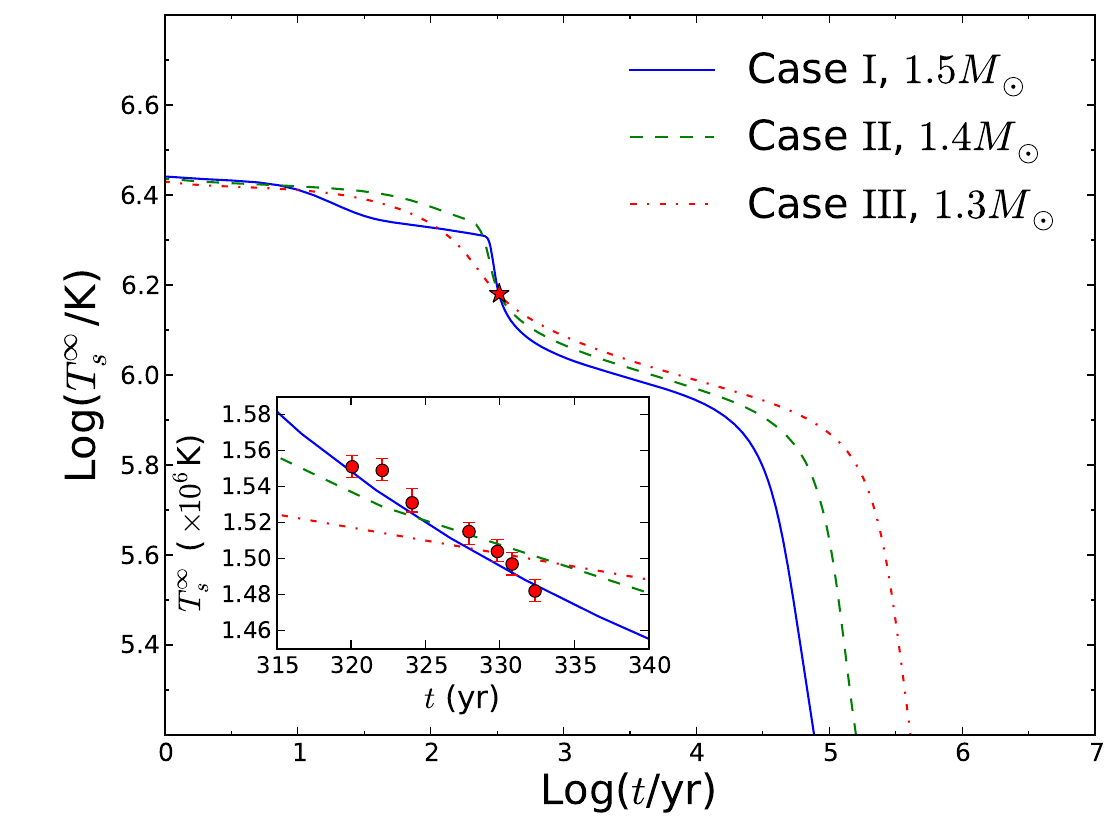}
\includegraphics[scale=0.4]{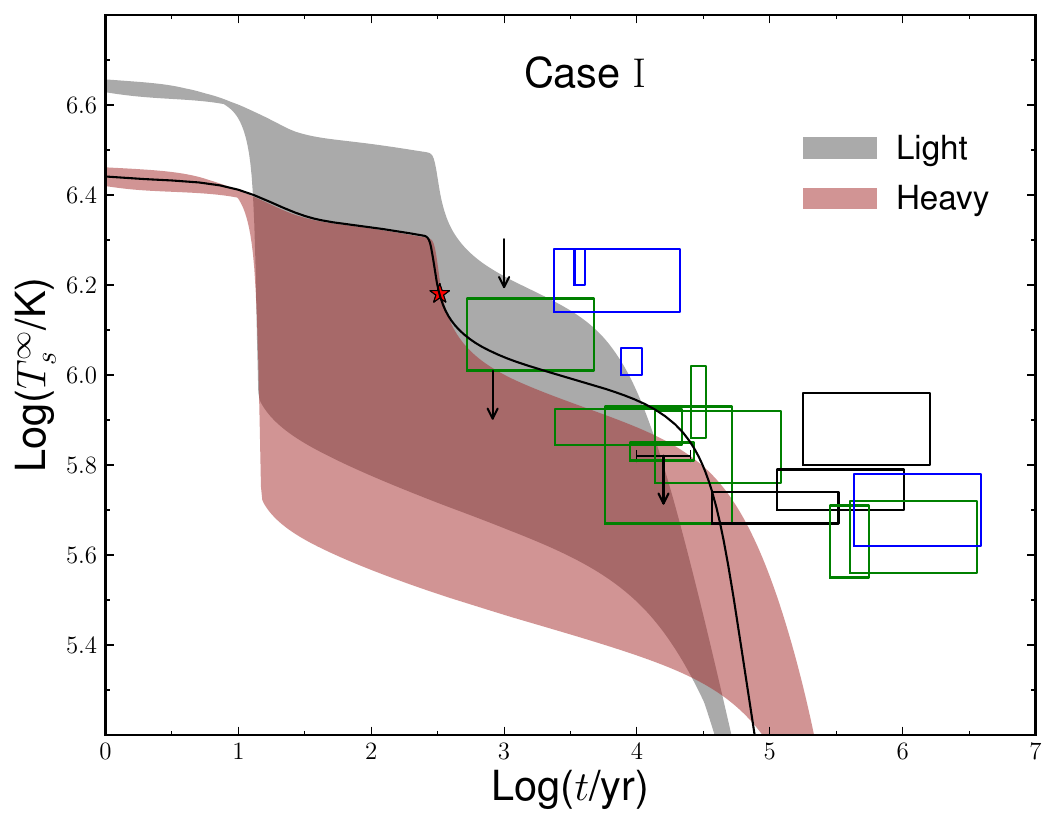}
}
\centerline{
\includegraphics[scale=0.4]{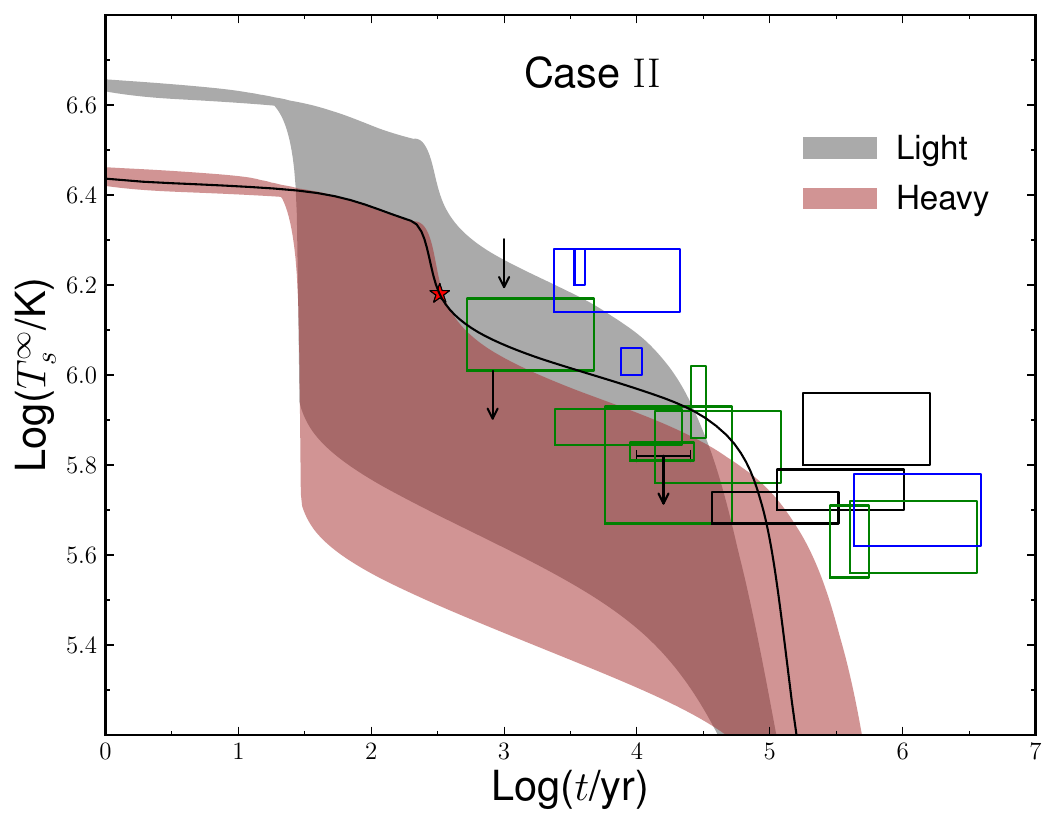} 
\includegraphics[scale=0.4]{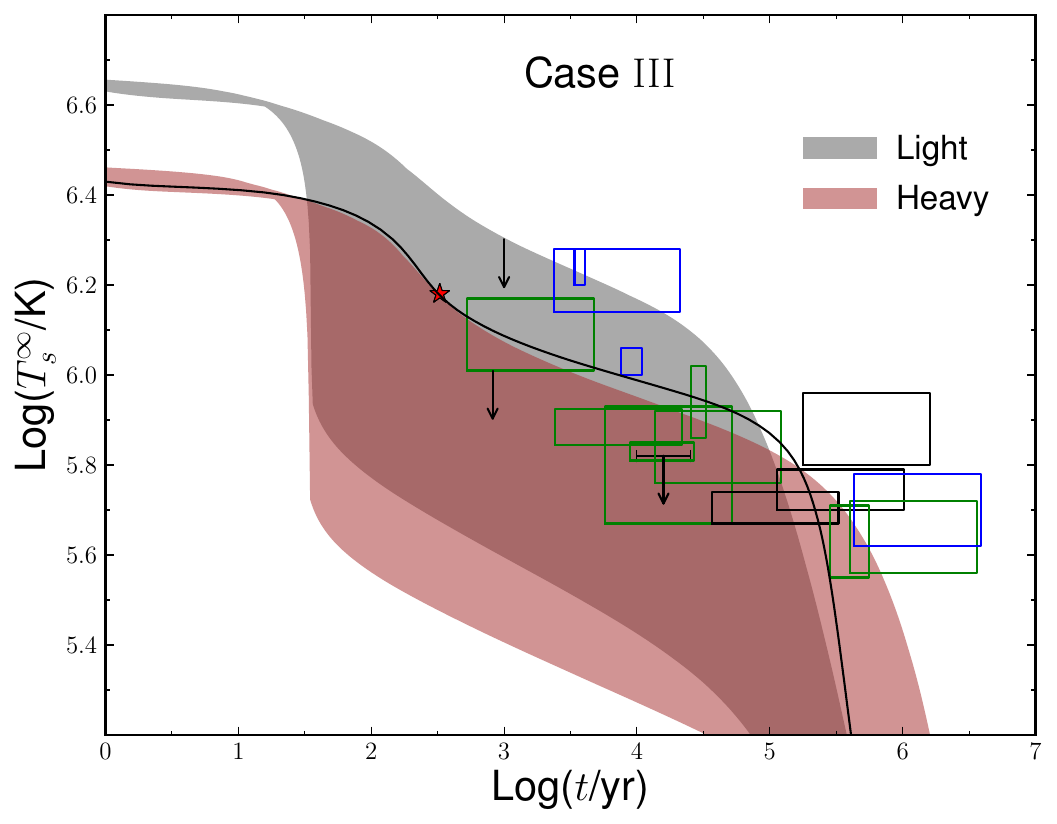} 
}
\caption{
(Color online) Superfluidity effects with the SkI4 model. 
Each case has different critical temperatures so different $d\ln T/d\ln t$. 
In the top-left panel, we compare the cooling curves of the case I, II, and III
for the mass $1.5 M_\odot$, $1.4 M_\odot$, and $1.3 M_\odot$ neutron stars, respectively. 
Mass of light elements in the envelope is assumed to be $\Delta M = 5 \times 10^{-13} M_\odot$. 
The error bars in the top-left plot denote the analysis of ACIS-S (Graded Mode) in Ref. \cite{cas2013}. 
The solid lines in the other panels correspond to the curves in the top-left panel. 
}\label{fig:ski4_sup}
\end{figure}

To see the effect of nuclear pairing, we choose SLy4 and SkI4 models to simulate
neutron star cooling. 
Note that they satisfy the mass-radius constraint zone \cite{slb2010}, and
SLy4 does not turn on the direct Urca process while SkI4 does 
if the mass of a neutron star is greater than $1.63$ $M_{\odot}$.
Figure \ref{fig:sly4_sup} shows the results for the SLy4 model.
The cooling curves show that
the early start of ${}^{3}P_{2}$ pairing gives the narrow band of cooling curves (left panel). 
The early start means that the pairing is happening close to the 
boundary between crust and core.
Thus, the core of neutron stars has superfluidity regardless
of its mass. Therefore, small $k_0$ gives sharp drop of temperature in young-age 
neutron stars. In this case, the observational data for old-age neutron 
star cannot be explained with cooling simulation.
On the other hand, if the $^{3}P_{2}$ neutron pairing appears 
at higher densities (right panel), the low mass neutron stars do not show sharp temperature
drop since $^{3}P_2$ pairing is not available in the core of neutron stars.
%

Figure \ref{fig:ski4_sup} shows the cooling curves with PBF for the SkI4 model. 
Three sets of input parameters given in Table~\ref{tb:ct} are considered.
In the top-left panel, we compare the cooling curves of three cases with
neutron star masses $1.5 M_\odot$, $1.4 M_\odot$, and $1.3 M_\odot$, respectively.
For all cases, the mass of light elements in the neutron star envelope is assumed to be 
$\Delta M = 5 \times 10^{-13} M_\odot$. 
We assume vector current conservation for the cases II and III and no conservation for the case I.
Three cases predict distinct thermal histories, showing a tendency that
more abrupt and rapid drop of temperature at young ages is connected to
lower temperatures in the old-age stars.
Inset in the top-left panel compares the three cases with the data from Cas A in details.
The remaining three panels show the results where the effect of envelope element is combined.
In the detailed comparison with the data of Cas A,
Case I is the best fit among the three cases. 
However, if the comparison is extended to the whole data in the figure, 
the best agreement can be obtained from Case III.
Though Case I and II can reproduce the Cas A data and cover substantial
portion of young- and middle-age stars, they can hardly explain the data of stars
aging more than 10$^5$ yrs.

\begin{table}[tbp]
\begin{center}
\begin{tabular}{ccccc}
\hline
& & Case $\mathrm{I}$ & Case $\mathrm{II}$ & Case $\mathrm{III}$ \\
\hline 
\multirow{3}{*}{~~${}^{3}P_{2}$ $n$~~} & ~~$T_{c}^{\text{max}}$ (K)~~ & ~~$6.65\times 10^{8}$~~  & ~~$6.82\times 10^{8} $~~ & ~~$5.95\times 10^8$~~ \\
                                   & $k_0$ (fm$^{-1}$)        & 0.99                & 1.45                  & 1.6 \\
                                   & $k_2$ (fm$^{-1}$)        & 2.8                 & 2.7                 & 2.4 \\            
\hline
\multirow{3}{*}{${}^{1}S_{0}$ $p$} & $T_{c}^{\text{max}}$ (K) & $6.48\times 10^{9}$  & $7.50\times 10^{9} $ & $0.7 \times 10^9$ \\
                                   & $k_0$ (fm$^{-1}$)        & 0.1                 & 0.1                  & 1.5 \\
                                   & $k_2$ (fm$^{-1}$)        & 2.5                 & 2.5                  & 2.6 \\            
\hline
\multirow{3}{*}{${}^{1}S_{0}$ $n$} & $T_{c}^{\text{max}}$ (K) & $3.2\times 10^{9}$  & $2.0\times 10^{9} $ & $1.0\times 10^9$ \\
                                   & $k_0$ (fm$^{-1}$)        & 0.0                 & 0.0                  & 0.0 \\
                                   & $k_2$ (fm$^{-1}$)        & 1.3                 & 1.3                  & 1.3 \\
\hline
\multicolumn{2}{c}{$d \ln T/ d \ln t$ }                       & $-1.087$            & $-0.673$             & $-0.313$ \\
\hline
\end{tabular}
\end{center}
\caption{The critical temperatures and  parameters of superfluidity 
for cooling simulations with the SkI4 model.} \label{tb:ct}
\end{table}

\section{Conclusion}\label{sec:con}
We investigated the consistency of nuclear models with 
observation of neutron star temperatures.
First, model selection was performed by constraining the predicted maximum mass of 
neutron stars to be at least  $2 M_\odot$.
We picked up 6 models among the non-relativistic Skyrme force models,
and 6 models among the RMF models.
In the second step, with the selected 12 models, we calculated the cooling curves with
only standard cooling mechanisms and the direct Urca process.
The result manifestly shows dependence on the EoS.
We found that the standard cooling processes reproduce the observation data
for the ages less than $10^4$ years or more than $10^5$ years.
However, no model could explain the data in the age of $10^4 \sim 10^5$ years,
and the standard cooling processes
always give temeperature higher than the observed ones.
On the other hand, if the direct Urca is operating inside of the neutron star, 
the star cools down so fast that 
the calculated cooling curves are located at temperatures much below the observation data.
In other words, real fine tuning is required to be consistent with the observation.
We showed that nuclear models with large symmetry energy gradient ($L > 85$ MeV) are not 
consistent with temperature observation mainly because the direct Urca process is turned on 
even for low mass neutron stars. 
As a result, we could sort out the nuclear models 
compatible with both mass-radius relations and temperature data.

Surface temperature heavily depends on the
composition of elements in the envelope.
We showed that if the direct Urca is absent  (e.g. SLy4 model), the effect of 
elements in the envelope is limited, and exhibits negligible dependence on the 
mass of neutron stars.
On the other hand, in a model which allows the direct Urca process for heavy mass
neutron stars (e.g. SkI4 model), we have wide band of cooling curves for both
light and heavy elements in the envelope.
We investigated the effect of evolution of elements in the envelope from light to heavy ones.
Important parameters are the mass of the envelope, the mass of neutron star, and the
reduction time scales of the mass of light elements.
We showed that the depletion of light elements leads to sudden and fast decrease of
surface temperature regardless of the existence of exotic states or superfluidity.

We have explored the effect of nuclear superfliudity by combining the PBF to models that 
are qualified with the standard cooling mechanisms.
Cooling curves with PBF are sensitive to the 
physical inputs such as pairing gaps, critical temperature for PBF, neutron star mass,
and vector current conservation.
Moreover, the temperatures predicted from the curves 
that are consistent with the middle age and Cas A data are much lower than the data
for the stars with ages more than $10^5$ years.
Measurement of temperature change of Cas A in the next decade will shed some light on resolving
these problems and we expect to reduce the uncertainties in the underlying physics. 
In conclusion, we could confirm that the existing mass-radius relation and thermal evolution
history of neutron stars provide critical test grounds with which one can find more realistic nuclear 
models that are suitable for dense nuclear matter.

\section*{Acknowledgements}
YL would like to thank Prof. D. Page (UNAM) for many helpful discussions
and advices when he began to study the neutron star cooling. He is also 
grateful to Prof. J.~M. Lattimer (Stony Brook) for providing 
the summary of neutron star's mass distribution.
YL was supported by the Rare Isotope Science Project of Institute 
for Basic Science funded by Ministry of Science, ICT 
and Future Planning and National Research Foundation of Korea (2013M7A1A1075764).
%
CHH expresses sincere gratitude to IBS, where part of the work was performed.
Work of CHH was supported by Basic Science Research Program through the
National Research Foundation of Korea (NRF) funded by the Ministry of Education (2014R1A1A2054096).
CHL was supported by the National Research Foundation of Korea (NRF) grant funded by the
Korea government (MSIP)  (No. 2015R1A2A2A01004238 and No. 2016R1A5A1013277).

\appendix

\section{Numerical Solution for Neutron Star Cooling}
In this section, we explain the numerical methods implemented in our code 
for solving thermal evolution of isolated 
neutron stars. Several numerical methods were attempted, such as
tri-diagonal scheme, penta-diagonal scheme, and tri-digonal scheme with every grid point. 
We compare the convergence between them.\\
The analysis for numerical solution of the diffusion equations
can be found in the appendix of D. Page's thesis \cite{page_thesis}. Here
we adopt the same notation with Page's thesis
in which tri-diagonal scheme is used to solve the coupled diffusion
equations. To solve the diffusion equations, $L_r\, (T)$ is defined on the even (odd) grid.
\footnote{In penta-diagonal scheme $L_r$ ($T$) is defined in even (odd) grid and intermediate time step, 
and in tri-diagonal scheme with all grid points, we mean $L_r$ and $T$ are defined
in all grid points.}
The penta-diagonal scheme for a neutron star thermal evolution
was first tried by Van Riper~\cite{vanriper1991}. The penta diagonal scheme is expected
more stable as time increases in his work.


\subsection{Penta-diagonal scheme}
Numerical method using the penta-diagonal scheme is described in ref. \cite{vanriper1991}
in detail. 
The first diffusion equation is
\begin{align}
\mc{L} = - \kappa S \frac{d\mcT}{da}
\end{align}
where
\begin{equation}
\mcL = e^{2\Phi_g}L_{r}\,, \quad
\mcT = e^{\Phi_g}T\,, \quad
S = (4\pi r^2)^2 e^{\Phi_g}n\,.
\end{equation}
Here $a(r)$ is the enclosed baryon number in radius $r$ and
it is related by
\begin{equation}
\frac{da}{dr} = 4\pi r^2 n \sqrt{1 -\frac{2Gm}{rc^2}}\,,
\end{equation}
where $n$ is a baryon number density.
The second diffusion equation becomes
\begin{equation}
\frac{d\mcT}{dt} = -Q \frac{d\mcL}{d a} - R
\end{equation}
where
\begin{equation}
Q = \frac{n}{C_v}\,, \quad
R = e^{2\Phi}\frac{Q_{\nu}}{C_{v}}\,.
\end{equation}
As usual, we define $\mcL$ on the even grids
$(\mcL_0, \mcL_2, \dots, \mcL_{2I+2})$ and $\mcT$ on 
the odd grids $(\mcT_1, \mcT_3, \dots, \mcT_{2I+1})$ since 
we should use the boundary condition at the center ($\mcL_{0} =0$).
The main feature of the penta-diagonal scheme is that the 
luminosity and temperature profile can be obtained even in the mid time interval.
Thus the unknowns for numerical equations are
\begin{equation*}
(\mcL_{0}^{n + 1/2}, \mcL_{0}^{n+1}, 
\mcT_{1}^{n+1/2}, \mcT_{1}^{n+1},
 \mcL_{2}^{n+1/2}, \mcL_{2}^{n+1},
\dots, \mcT_{2I+1}^{n+1/2}, \mcT_{2I+1}^{n+1}, 
\mcL_{2I+2}^{n+1/2}, \mcT_{2I+2}^{n+1})\,,
\end{equation*}
where the subscript means spatial dimension and superscript
represents time step. 
The total number of unknown is $4I + 6$\,.
Using these mid time interval variables and Henyey method, we
write the thermal evolution of a neutron star time index from
$n$ to $n+1$
as
\begin{equation}
\mcT^{n+1}_{2i+1}
=\mcT^{n}_{2i+1}
- \Delta t
\left[
Q_{2i+1}^{n+1/2}\frac{d\mcL}{da}\bigg\vert_{2i+1}^{n+1/2}
+ R_{2i+1}^{n+1/2}
\right]\,,
\end{equation}
where
\begin{equation}
\mcL_{2i}^{n+1/2} = -\kappa_{2i}^{n+1/2}S_{2i}^{n+1/2}
\frac{d\mcT}{da}\bigg\vert_{2i}^{n+1/2}\,.
\end{equation} 
The thermal evolution of time index from $n+1/2$ to $n+1$ is given by
\begin{equation}
\mcT^{n+1}_{2i+1}
=\mcT^{n+1/2}_{2i+1}
- \frac{1}{2}\Delta t
\left[
Q_{2i+1}^{n+1}\frac{d\mcL}{da}\bigg\vert_{2i+1}^{n+1}
+ R_{2i+1}^{n+1}
\right]\,,
\end{equation}
where
\begin{equation}
\mcL_{2i}^{n+1} = -\kappa_{2i}^{n+1}S_{2i}^{n+1}
\frac{d\mcT}{da}\bigg\vert_{2i}^{n+1}\,.
\end{equation} 
In general the thermal conductivity $\kappa$ is a function of temperature $T$, 
and the temperature is 
defined only on the odd grid in the numerical scheme, thus
we use the average $\kappa_{2i} = \frac{1}{2}(\kappa_{2i-1} + \kappa_{2i+1})$.
Unlike the iterative scheme in Van-Riper's work~\cite{vanriper1991}, we
use numerical Newton-Raphson scheme to solve non-linear equations.
 From the
above equations and $\kappa_{2i}$, we have equations to solve
\begin{align}
F_{4i+1} & =
F_{4i+1}(\mcL_{2i}^{n+1/2}, \mcT_{2i+1}^{n+1/2}, \mcT_{2i+1}^{n+1},
\mcL_{2i+2}^{n+1/2}) \notag \\ 
& =  R^{n+1/2}_{2i+1}
+ Q^{n+1/2}_{2i+1}\,\cdot
\frac{\mcL^{n+1/2}_{2i+2} -\mcL_{2i}^{n+1/2}}{da_{2i}+da_{2i+1}}
+\frac{\mcT_{2i+1}^{n+1} - \mcT_{2i+1}^{n}}{\Delta t} = 0\,, \\
F_{4i+2} & =
F_{4i+2}(\mcL_{2i}^{n+1}, \mcT_{2i+1}^{n+1/2}, \mcT_{2i+1}^{n+1},
\mcL_{2i+2}^{n+1}) \notag \\ 
& =  R^{n+1}_{2i+1}
+ Q^{n+1}_{2i+1}\,\cdot
\frac{\mcL^{n+1}_{2i+2} -\mcL_{2i}^{n+1}}{da_{2i}+da_{2i+1}}
+\frac{\mcT_{2i+1}^{n+1} - \mcT_{2i+1}^{n+1/2}}{\Delta t/2} = 0\,, \\
F_{4i-1} & =
F_{4i-1}(\mcT_{2i-1}^{n+1/2},\mcL_{2i}^{n+1/2}, \mcT_{2i+1}^{n+1/2}) \notag \\ 
& =  \mcL^{n+1/2}_{2i}
+ \frac{\kappa_{2i-1}^{n+1/2} + \kappa_{2i+1}^{n+1/2}}{2}\cdot S_{2i}
\cdot
\frac{\mcT^{n+1/2}_{2i+1} -\mcT_{2i-1}^{n+1/2}}{da_{2i}+da_{2i+1}} =0 \,,\\
F_{4i} & =
F_{4i}(\mcT_{2i-1}^{n+1},\mcL_{2i}^{n+1}, \mcT_{2i+1}^{n+1}) \notag \\ 
& =  \mcL^{n+1}_{2i}
+ \frac{\kappa_{2i-1}^{n+1} + \kappa_{2i+1}^{n+1}}{2}\cdot S_{2i}
\cdot
\frac{\mcT^{n+1}_{2i+1} -\mcT_{2i-1}^{n+1}}{da_{2i}+da_{2i+1}} =0\,.
\end{align}
These four types of equations can be solved by the multi-dimensional 
Newton-Raphson method. The variations of $\dmcL$'s and $\dmcT$'s
can be obtained by solving linearized Newton-Raphson method,
\begin{align}
& F_{4i+1} + A_{4i+1}\dmcL_{2i}^{n+1/2}
+ C_{4i+1}\dmcT_{2i+1}^{n+1/2} + D_{4i+1}\dmcT_{2i+1}^{n+1}
+ E_{4i+1}\dmcL_{2i+2}^{n+1/2} = 0 \,, \\
& F_{4i+2} + A_{4i+2}\dmcL_{2i}^{n+1}
+ B_{4i+2}\dmcT_{2i+1}^{n+1/2} + C_{4i+2}\dmcT_{2i+1}^{n+1}
+ E_{4i+2}\dmcL_{2i+2}^{n+1} = 0 \,, \\
& F_{4i-1} + A_{4i-1}\dmcT_{2i-1}^{n+1/2} 
+ C_{4i-1}\dmcL_{2i}^{n+1/2}
+ E_{4i-1}\dmcT_{2i+1}^{n+1/2} = 0 \,, \\
& F_{4i} + A_{4i}\dmcT_{2i-1}^{n+1}
+ C_{4i}\dmcL_{2i}^{n+1}
+ E_{4i}\dmcT_{2i+1}^{n+1} = 0\,,
\end{align}
where
\begin{align}
A_{4i+1} & = \frac{\pt F_{4i+1}}{\pt \mcL_{2i}^{n+1/2}}
= - \frac{Q_{2i+1}^{n+1/2}}{da_{2i}+da_{2i+1}} \,, \\
C_{4i+1} & = \frac{\pt F_{4i+1}}{\pt \mcT_{2i+1}^{n+1/2}}
= \frac{\pt R}{\pt \mcT}\bigg\vert_{2i+1}^{n+1/2}
+ \frac{\pt Q}{\pt \mcT}\bigg\vert_{2i+1}^{n+1/2}\cdot
\frac{\mcL^{n+1/2}_{2i+2} -\mcL_{2i}^{n+1/2}}{da_{2i}+da_{2i+1}}\,,
\\
D_{4i+1} & =
\frac{\pt F_{4i+1}}{\pt \mcT_{2i+1}^{n+1}}
= \frac{1}{\Delta t} \,, \\
E_{4i+1} &=\frac{\pt F_{4i+1}}{\pt \mcL_{2i+2}^{n+1/2}}
= \frac{Q_{2i+1}^{n+1/2}}{da_{2i}+da_{2i+1}} \,, \\
A_{4i+2} & = \frac{\pt F_{4i+2}}{\pt \mcL_{2i}^{n+1}}
= - \frac{Q_{2i+1}^{n+1}}{da_{2i}+da_{2i+1}} \,, \\
B_{4i+2} & = \frac{\pt F_{4i+2}}{\pt \mcT_{2i+1}^{n+1/2}}
 = -\frac{2}{\Delta t} \,, \\
C_{4i+2} & = \frac{\pt F_{4i+2}}{\pt \mcT_{2i+1}^{n+1}}
= \frac{\pt R}{\pt \mcT}\bigg\vert_{2i+1}^{n+1}
+ \frac{\pt Q}{\pt \mcT}\bigg\vert_{2i+1}^{n+1}\cdot
\frac{\mcL^{n+1}_{2i+2} -\mcL_{2i}^{n+1}}{da_{2i}+da_{2i+1}}
 + \frac{2}{\Delta t} \,,
\\
E_{4i+2} &=\frac{\pt F_{4i+2}}{\pt \mcL_{2i+2}^{n+1}}
= \frac{Q_{2i+1}^{n+1}}{da_{2i}+da_{2i+1}} \,, \\
A_{4i-1} & = \frac{\pt F_{4i-1}}{\pt \mcT_{2i-1}^{n+1/2}} \notag \\
& = \frac{1}{2}
 \frac{S_{2i}}{da_{2i}+da_{2i+1}} 
 \left[ \frac{\pt \kappa}{\pt \mcT}\bigg\vert_{2i-1}^{n+1/2}
 \left( \mcT_{2i+1}^{n+1/2}-\mcT_{2i-1}^{n+1/2}\right)
- \kappa_{2i+1}^{n+1/2} + \kappa_{2i-1}^{n+1/2}\right], \\
C_{4i-1} & = \frac{\pt F_{4i-1}}{\pt \mcL_{2i+1}^{n+1/2}}
= 1 \,, \\
E_{4i-1} &=
\frac{\pt F_{4i-1}}{\pt \mcT_{2i+1}^{n+1/2}} \notag \\
& = \frac{1}{2}
 \frac{S_{2i}}{da_{2i}+da_{2i+1}} 
 \left[ \frac{\pt \kappa}{\pt \mcT}\bigg\vert_{2i+1}^{n+1/2}
 \left( \mcT_{2i+1}^{n+1/2}-\mcT_{2i-1}^{n+1/2}\right)
+ \kappa_{2i+1}^{n+1/2}-\kappa_{2i-1}^{n+1/2}\right] \,, \\
A_{4i} & = \frac{\pt F_{4i}}{\pt \mcT_{2i-1}^{n+1}} \notag \\
& = \frac{1}{2}
 \frac{S_{2i}}{da_{2i}+da_{2i+1}} 
 \left[ \frac{\pt \kappa}{\pt \mcT}\bigg\vert_{2i-1}^{n+1}
 \left( \mcT_{2i+1}^{n+1}-\mcT_{2i-1}^{n+1}\right)
- \kappa_{2i+1}^{n+1} + \kappa_{2i-1}^{n+1}\right] \,, \\
C_{4i} & = \frac{\pt F_{4i}}{\pt \mcL_{2i+1}^{n+1}}
= 1 \,, \\
E_{4i} &=
\frac{\pt F_{4i}}{\pt \mcT_{2i+1}^{n+1}} \notag \\
& = \frac{1}{2}
 \frac{S_{2i}}{da_{2i}+da_{2i+1}} 
 \left[ \frac{\pt \kappa}{\pt \mcT}\bigg\vert_{2i+1}^{n+1}
 \left( \mcT_{2i+1}^{n+1}-\mcT_{2i-1}^{n+1}\right)
+ \kappa_{2i+1}^{n+1}-\kappa_{2i-1}^{n+1}\right]\,.
\end{align}
In the above equations to solve $F_{1}, \dots, F_{4I+2}$,
we have unknowns $\mcL_{0}^{n+1/2}$, $\mcL_{0}^{n+1}$,
$\mcT_{1}^{n+1/2}$, $\mcT_{1}^{n+1}$, $\dots$, 
$\mcL_{2I+2}^{n+1/2}$, and $\mcL_{2I+2}^{n+1}$\,.
There are $4I+6$ unknowns quantities  with only $4I+2$ equations,
but one can obtain the solutions with four additional boundary conditions. 
\footnote{$L_r(r=0) = 0$ reduces two unknowns ($\mcL_0^{n+1/2}=\mcL_{0}^{n+1}=0$),
and the uniform luminosity approximation ($L_{2I+2}=L_{2I+1}$) reduces
$\mcL_{2I+2}^{n+1/2}$ and $\mcL_{2I+2}^{n+1}$ as a function of
$\mcT_{2I+1}^{n+1/2}$ and $\mcT_{2I+1}^{n+1}$ respectively. }
With the $T_s - T_b$ relation \cite{gpe1983, pcy1997}, we can have
\begin{equation}
\mcL_{2I+2} 
= e^{2\Phi_{2I+2}}(4\pi R^2 \sigma_{B})T_{s}^4
= e^{2\Phi_{2I+2}}(4\pi R^2 \sigma_{B})f(\mcT_{2I+1})\,.
\end{equation}

We have matrix equations to solve
\begin{equation}
\left( 
\begin{matrix}
C_1 & D_1 & E_1 & 0   &\cds &     &      &     &     &      
0_{\phantom{1}}^{\phantom{/}} \\
B_2 & C_2 & 0   & E_2 & 0   &\cds &      &     &     &      
0_{\phantom{2}}^{\phantom{/}} \\
A_3 &  0  & C_3 & 0   & E_3 & 0   &\cds  &     &     &      
0_{\phantom{3}}^{\phantom{/}}\\
0   & A_4 & 0   & C_4 & 0   & E_4 & 0    &\cds &     &      
0_{\phantom{4}}^{\phantom{/}} \\
\\
    &     &     &     &     &     & \vds &     &     &        \\
\\
0   &\cds &          &   0  & A_{4I-1}& 0 & C_{4I-1}& 0   &E_{4I-1}&
0_{\phantom{I}}^{\phantom{/}} \\
0   &\cds &          &      & 0   & A_{4I}& 0 & C_{4I}& 0   &
E_{4I}^{\phantom{+}} \\
0   &\cds &          &      &     &  0  & A_{4I+1}& 0 & C_{4I+1}&
D_{4I+1}^{\phantom{/}} \\
0   &\cds &          &      &     &     &  0  & A_{4I+2}& B_{4I+2} & 
C_{4I+2}^{\phantom{+}}
\end{matrix}
\right)
\left(
\begin{array}{l}
\dmcT_{1}^{n+1/2} \\
\dmcT_{1}^{n+1} \\
\dmcL_{2}^{n+1/2} \\
\dmcL_{2}^{n+1} \\
\\
\phantom{aa}\vdots \\
\\
\dmcL_{2I}^{n+1/2} \\
\dmcL_{2I}^{n+1} \\
\dmcT_{2I+1}^{n+1/2} \\
\dmcT_{2I+1}^{n+1}
\end{array}
\right)
= -\left(
\begin{array}{l}
F_{1}^{\phantom{/}} \\
F_{2}^{\phantom{+}} \\
F_{3}^{\phantom{/}} \\
F_{4}^{\phantom{+}} \\
\\
\phantom{aa}\vdots \\
\\
F_{4I-1}^{\phantom{/}} \\
F_{4I}^{\phantom{+}} \\
F_{4I+1}^{\phantom{/}} \\
F_{4I+2}^{\phantom{+}}
\end{array}
\right).
\end{equation}

This penta-diagonal
linear equation can be solved by $L-U$ decomposition
or Gaussian elimination method.

\subsection{Tri-diagonal scheme with every grid point }
Another method to solve the diffusion equation is to use
every grid point $({\cal L}_0,$ ${\cal L}_1$, $\cdots$, ${\cal L}_N,$ ${\cal T}_1,$ ${\cal T}_2,$ $\cdots$,
${\cal T}_N)$.
In this case, we mix forward and backward numerical differentiation to make
tri-diagonal matrix. 
For the luminosity equation,
\begin{equation}
\mcL = -\kappa S \frac{d\mcT}{da} \quad \rightarrow \quad
\mcL_i + \kappa_i S_i \frac{\mcT_{i+1} - \mcT_i}{da_{i+1}} = 0 \,,
\end{equation}
and the temperature evolution equation becomes,
\begin{equation}
\frac{d\mcT}{dt} = - Q\frac{d\mcL}{da} - R \quad \rightarrow \quad
R_{i} + Q_i \frac{\mcL_{i}-\mcL_{i-1}}{da_i} + \frac{T_i - T_i^{old}}{\Delta t} =0 \,.
\end{equation}
Thus, the numerical equations to solve are
\begin{eqnarray}
F_{2i - 1}(\mcL_{i-1}^{n+1}, \mcT_i^{n+1}, \mcL_i^{n+1}) &= & R_i^{n+1} 
+ Q_{i}^{n+1} \cdot \frac{\mcL_i^{n+1} -\mcL_{i-1}^{n+1}}{d a_i } 
+ \frac{\mcT_i^{n+1} -\mcT_i^n}{\Delta t}\,,
\\
F_{2i}(\mcT_i^{n+1},\mcL_i^{n+1},\mcT_{i+1}^{n+1}) &=& 
\mcL_{i}^{n+1} + \kappa_i^{n+1}\cdot S_i^{n+1} \cdot
 \frac{\mcT_{i+1}^{n+1} - \mcT_i^{n+1}}{da_{i+1}}\,.
\end{eqnarray}
In this scheme, the unknowns are $(\mcT_1, \mcL_1, \mcT_2, \cdots, \mcL_{N-1}, \mcT_N)$
and the final equations to solve are $F_{2N-1}$ instead of $F_{2N}$ since we don't have
$T_{N+1}$ as unknown. We also use the same boundary condition as  in Penta-diagonal
scheme, $\mcL_0 = 0$ and $\mcL_{N} = e^{2\Phi}4\pi R^2 \sigma_B T_{s}^4 = 
e^{2\Phi}4\pi R^2 \sigma_B f(\mcT_{N}) $ . Therefore,
\begin{eqnarray}
F_{1} &=& R_1^{n+1} + Q_1^{n+1}\frac{\mcL_1}{a_1} + \frac{T_1^{n+1} - T_1^n}{\Delta t} \,,\\
F_{2N-1} &=& R_N^{n+1} + Q_N^{n+1}\frac{\mcL_{N}(\mcT_N)-\mcL_{N-1}}{da_N} 
+ \frac{T_N^{n+1} - T_N^n}{\Delta t} \,.
\end{eqnarray}
Newton-Raphson iteration method gives the equations,
\begin{eqnarray}
&& F_{2i-1} + A_{2i-1} \delta \mcL_{i-1}^{n+1} 
+ B_{2i-1}\delta \mcT_{i}^{n+1}
+ C_{2i-1}\delta \mcL_{i}^{n+1} = 0\,, \\
&& F_{2i} + A_{2i} \delta \mcT_{i}^{n+1}
    + B_{2i}\delta \mcL_{i}^{n+1}
    + C_{2i}\delta \mcT_{i+1}^{n+1} = 0\,,
\end{eqnarray}
where
\begin{eqnarray}
A_{2i-1} &=& \frac{\pt F_{2i-1}}{\pt \mcL_{i-1}^{n+1}} = - \frac{Q_i^{n+1}}{da_i}\,, \\
B_{2i-1} &=& \frac{\pt F_{2i-1}}{\pt \mcT_{i}^{n+1}} 
= \frac{\pt R}{\pt \mcT}\bigg\vert_{i}^{n+1}
+ \frac{\pt Q}{\pt \mcT}\bigg\vert_{i}^{n+1} \cdot 
\frac{\mcL_i^{n+1}-\mcL_{i-1}^{n+1}}{da_i} + \frac{1}{\Delta t}\,,\\
C_{2i-1} &=& \frac{\pt F_{2i-1}}{\pt \mcL_{i}^{n+1}} = \frac{Q_i^{n+1}}{da_i}\,, \\
A_{2i} &=& \frac{\pt F_{2i}}{\pt \mcT_{i}^{n+1}} 
= \frac{\pt \kappa}{\pt \mcT}\bigg\vert_{i}^{n+1}\cdot S_{i}\cdot
\frac{\mcT_{i+1}^{n+1}-\mcT_{i}^{n+1}}{da_i}
- \frac{\kappa_{i}^{n+1}\cdot S_{i}^{n+1}}{da_i}\,,\\
B_{2i} &=& \frac{\pt F_{2i}}{\pt \mcL_{i}^{n+1}} =1\,,\\
C_{2i} &=& \frac{\pt F_{2i}}{\pt \mcT_{i+1}^{n+1}} 
= \frac{\kappa_{i}^{n+1}\cdot S_{i}^{n+1}}{da_i}\,.
\end{eqnarray}
Special case is needed for the boundary grid points.
\begin{eqnarray}
A_1 &=& 0, \\
 C_{2N-1} &=& 0, \\
 B_{2N-1} &=& \frac{\pt R}{\pt \mcT}\bigg\vert_{N}^{n+1}
+ \frac{\pt Q}{\pt \mcT}\bigg\vert_{N}^{n+1} \cdot 
\frac{\mcL_N^{n+1}-\mcL_{N-1}^{n+1}}{da_i} + \frac{1}{\Delta t}
+ \frac{Q_N^{n+1}}{da_N}\frac{\pt \mcL_N^{n+1}}{\pt \mcT_N^{n+1} }\,.
\end{eqnarray}
The tri-diagonal matrix becomes
\begin{equation}
\left( 
\begin{matrix}
B_1 & C_1 & 0 &\cds &     &      &     &     &      
0_{\phantom{1}}^{\phantom{/}} \\
A_2 & B_2 & C_2   & 0   &\cds &      &     &     &      
0_{\phantom{2}}^{\phantom{/}} \\
0 &  A_3  & B_3 & C_3   & 0  & \cds    &     &   &   
0_{\phantom{3}}^{\phantom{/}}\\
\\
    &     &     &     &     &     & \vds &     &     &        \\
\\
0   &\cds &                &  & 0 & A_{2N-3} & B_{2N-3} & C_{2N-3}   &
0_{\phantom{I}}^{\phantom{/}} \\
0   &\cds &          &     &  & 0 & A_{2N-2} & B_{2N-2} & C_{2N-2}    \\
0   &\cds &          &      &     &   & 0  & A_{2N-1}& B_{2N-1} 
\end{matrix}
\right)
\left(
\begin{array}{l}
\dmcT_{1}^{n+1} \\
\dmcL_{1}^{n+1} \\
\dmcT_{2}^{n+1} \\
\\
\phantom{aa}\vdots \\
\\
\dmcT_{2N-1}^{n+1} \\
\dmcL_{2N-1}^{n+1} \\
\dmcT_{2N}^{n+1}
\end{array}
\right)
= -\left(
\begin{array}{l}
F_{1}^{\phantom{/}} \\
F_{2}^{\phantom{+}} \\
F_{3}^{\phantom{/}} \\
\\
\phantom{aa}\vdots \\
\\
F_{2N-3}^{\phantom{/}} \\
F_{2N-2}^{\phantom{+}} \\
F_{2N-1}^{\phantom{/}}
\end{array}
\right)\,.
\end{equation}

\subsection{Comparison}
\begin{figure}
\centerline{
\includegraphics[scale=0.4]{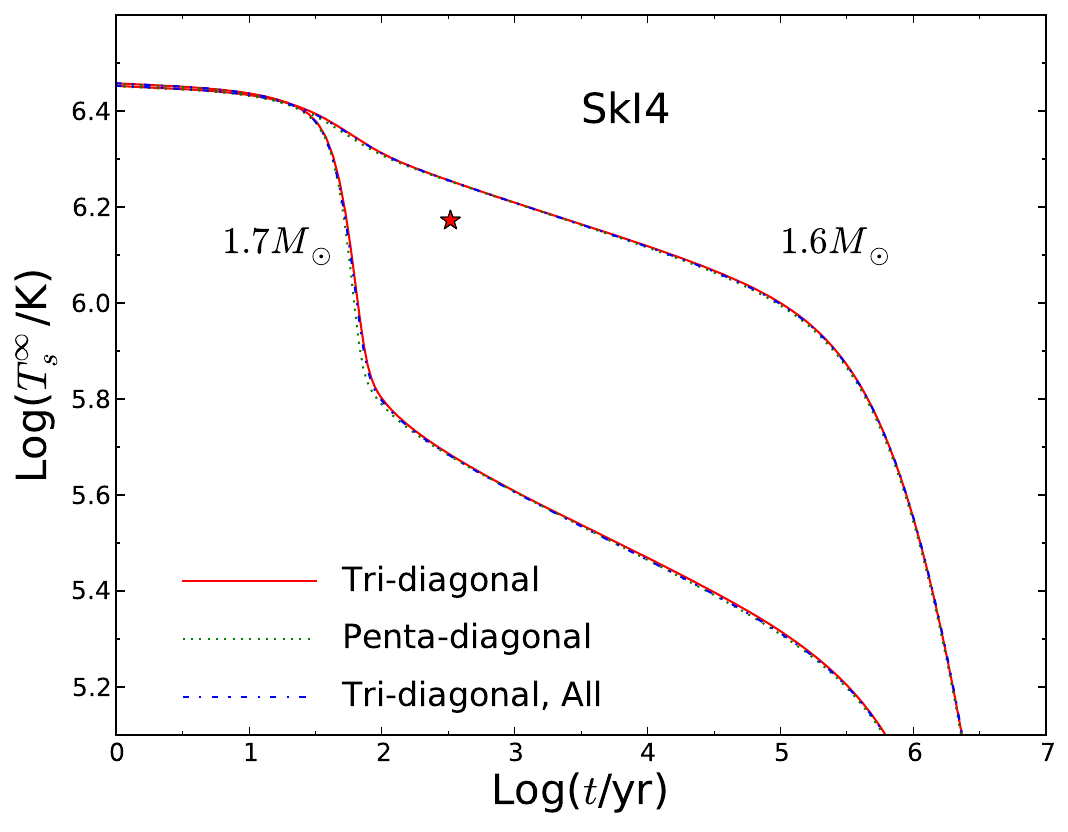} 
\includegraphics[scale=0.4]{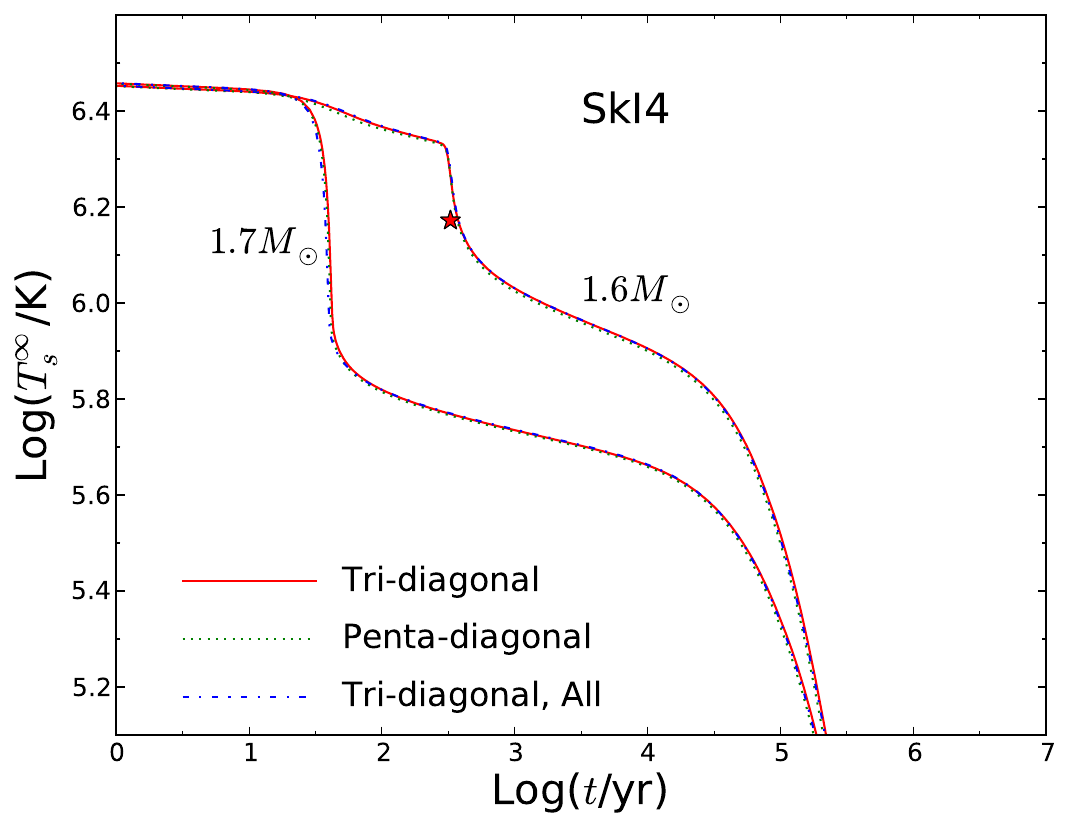}
}
\caption{(Color online) Curves for each numerical method. SkI4 is used to simulate neutron star
cooling. All three methods shows the identical results. The left figure shows the
normal non-superfluid phase and the right figure shows the superconducting phase.}
\label{fig:num_comp}
\end{figure}

Each numerical solution (tri-diagonal, penta-diagonal, and tri-diagoal all grids)
gives the similar solution if the initial condition is identical for each 
simulation. In the point of view of numerical cost, tri-diagonal even ($L_r$)-odd ($T$) method
is superior to penta-diagonal even ($L_r$)-odd ($T$) and tri-diagoal all grids method.
Figure \ref{fig:num_comp} shows neutron star cooling curves with SkI4 model.
Three different numerical methods show almost identical results.
The difference in early stage is caused by the difference in the time step $\Delta t$ in each simulation.
That is, penta-diagonal scheme,
for example, for some case, $t_{n+1/2}$ is normal state and $t_{n+1}$ can be 
superfluidic phase because of temperature difference in each step. Thus the
time step should be adjusted to solve the diffusion equations. 
For normal phase, all three methods give no difficulty in the simulation. 
However, in superconducting phase, 
the most stable numerical method is tri-diagonal with even ($L_r$) and odd ($T$)
scheme since it is free from the intermediate time step for sudden decrease of 
temperature.
\begin{figure}
\centerline{
\includegraphics[scale=0.4]{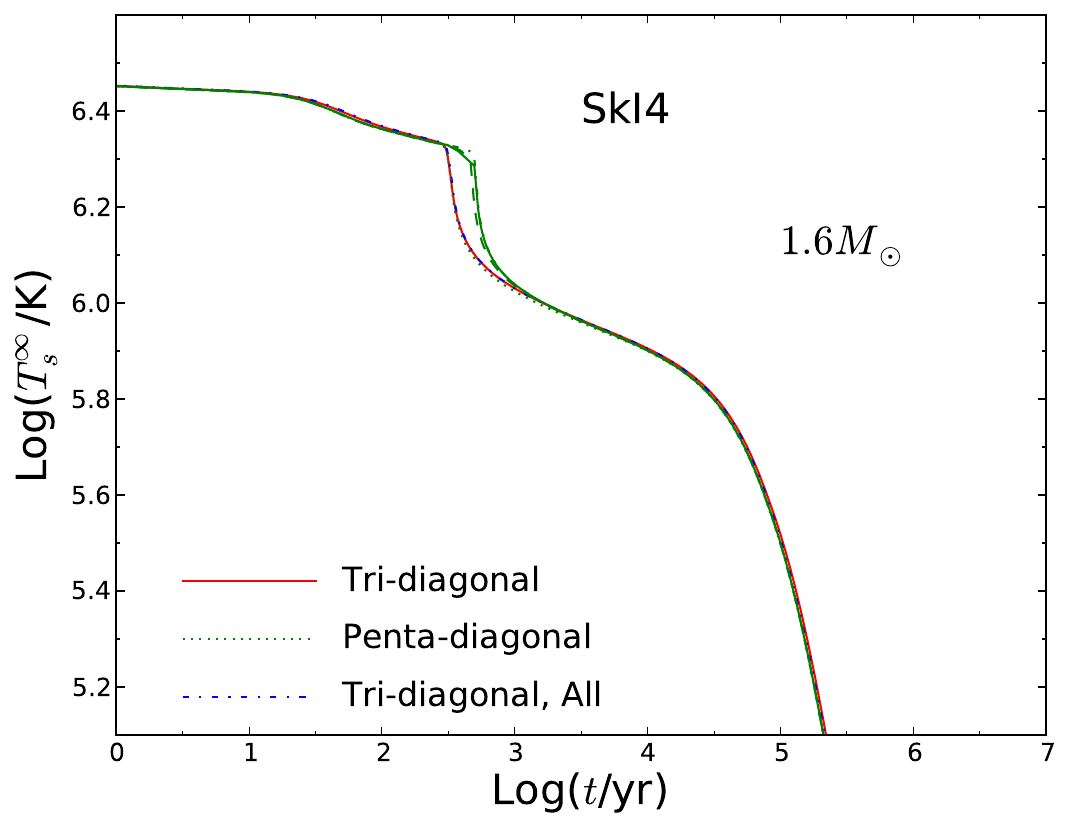} 
\includegraphics[scale=0.4]{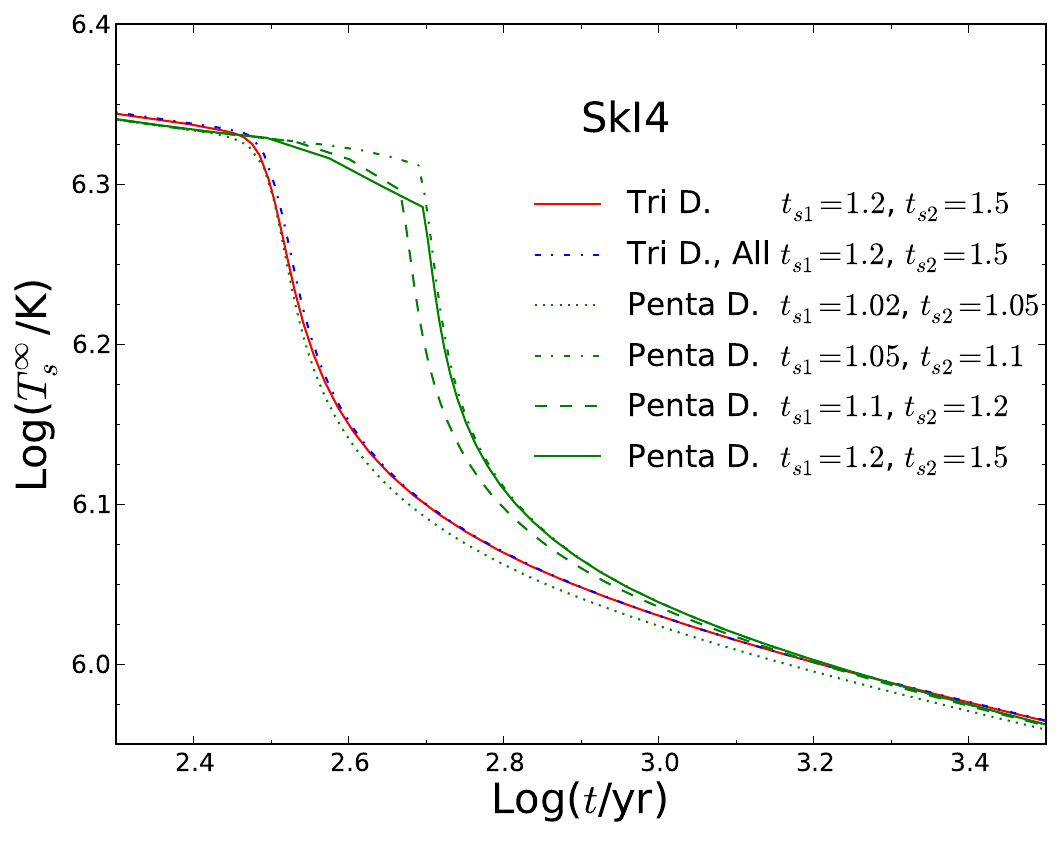}
}
\caption{(Color online) Left figure shows the large scale cooling curve. Right figure shows
the cooling curve near the critical temperature for superfluidity. 
Each curve shows different behavior near the critical temperature.}
\label{fig:time_step}
\end{figure}

In Fig.~\ref{fig:time_step}, we compare results from three different numerical methods.
If superfluidity occurs,
penta-diagonal method needs a smaller size of time step to make the result similar 
with the ones from both the tri-diagonal methods (even $L_r$ and odd $T$)
and the tri-diagonal methods in which $L_r$ and $T$ are defined in all grid points.

\section{Spatial zone and time step}

In neutron star cooling simulation, we make grids from the core to
outer boundary of crust ($\rho = 10^{10} \textrm{g}/\textrm{cm}^3) $
and connect the temperature $T_{b}$ with $T_{s}$ using uniform
luminosity approximation and $T_s-T_b$ relation \cite{gpe1983, pcy1997}.
The density of the core is around 
$\rho \simeq 10^{14} \sim 10^{15} \textrm{g}/\textrm{cm}^3$
and the crust has the density in the range of $10^{10}$ to $10^{14} \textrm{g}/\textrm{cm}^3$.
Even though, the size of crust is only $\sim$ $1 \textrm{km}$,
the nuclear phase changes from heavy nuclei with neutron and electron gas
to heavy nuclei with electron gas. Since the different equations of state
give different central, core-crust boundary, and neutron drip density,
it is reasonable to make mesh point,
\begin{eqnarray}
N_1 &=& W_1 \log_{10}\left( \frac{\rho_{c}}{\rho_{core}} \right)\,, \\
N_2 &=& W_2 \log_{10}\left( \frac{\rho_{core}}{\rho_{drip}} \right)\,, \\
N_3 &=& W_3 \log_{10}\left( \frac{\rho_{drip}}{\rho_{env}} \right)\,,
\end{eqnarray}
where $\rho_c$ is the central density, $\rho_{core}$ is the density for
core-crust boundary, $\rho_{drip}$ is the neutron drip denisty,
and $\rho_{env} = 10^{10} \textrm{g}/\textrm{cm}^3$ for density of
boundary of crust and envelope.
\begin{figure}
\centerline{
\includegraphics[scale=0.4]{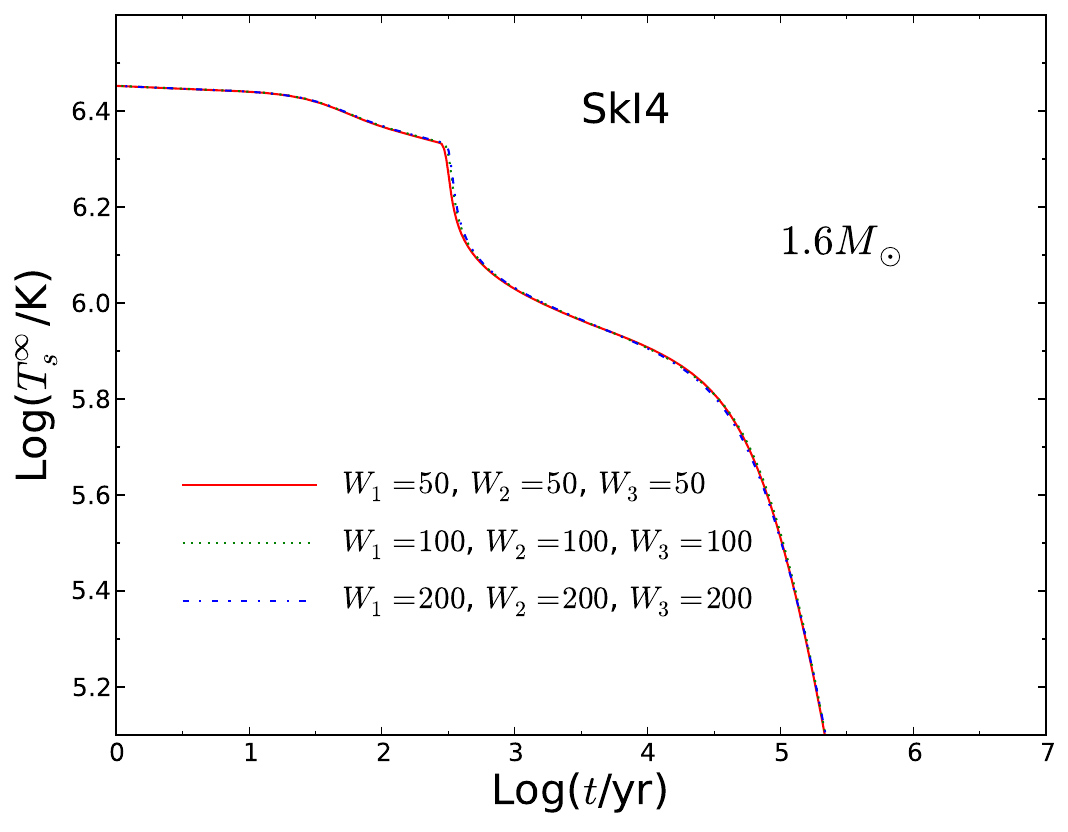} 
\includegraphics[scale=0.4]{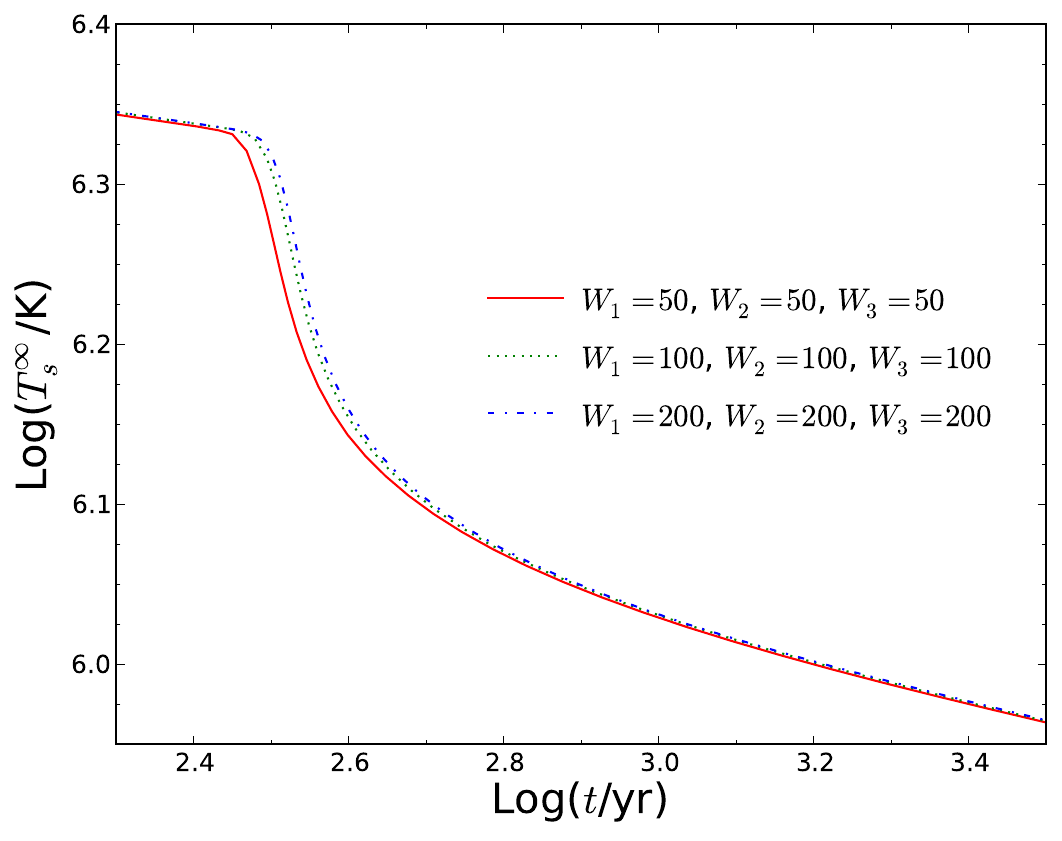}
}
\caption{(Color online) Different number of zone for core and crust region. Large scale figure
(left) shows that three curves give the same behavior. Enlarged figure near the
critical temperature shows that $W=100$ is enough for the numerical simulation.}
\label{fig:cool_zone}
\end{figure}
Fig.~\ref{fig:cool_zone} shows the cooling curves depends on the number of grid zones
for the same density interval. In large scale, the curves are not much different.
However, it is necessary to use enough number of grids per the density interval 
to make the cooling curves converge. We found that $W=100$ is enough for cooling
simulation.

Several constraints for time step $\Delta t$ are used. In general, as time goes, 
the numerical solution is more stabilized so that we can use larger time step. 
Here we use the $T_{eff}$ to
determine the next time step. We choose different $t_{scale}$,
$\Delta t^{n+1} = t_{scale} \Delta t^{n}$ for different conditions of
$T_{eff}^{n}$ and $T_{eff}^{n-1}$.
\begin{equation}\label{eq:normal_reg}
\begin{cases}
\Big \vert \frac{T_{eff}^{n} - T_{eff}^{n-1}}
{T_{eff}^{n-1}} \Big \vert > 0.1 \,, &  t_{scale} = 1.02 \,, \\
0.05 <\Big \vert \frac{T_{eff}^{n} - T_{eff}^{n-1}}
{T_{eff}^{n-1}} \Big \vert \le 0.1 \,, & t_{scale} = 1.1 \,, \\
0.01 < \Big \vert \frac{T_{eff}^{n} - T_{eff}^{n-1}}
{T_{eff}^{n-1}} \Big \vert \le 0.05 \,, & t_{scale} = 1.2 \,, \\
\Big \vert \frac{T_{eff}^{n} - T_{eff}^{n-1}}
{ T_{eff}^{n-1}}
\Big \vert \le 0.01 \,, & t_{scale} = 1.5\,.
\end{cases}
\end{equation}
Another constraint for the time step comes from total time.
In our simulation the next time step is always less than one tenth of total time,
\begin{equation}
\Delta t^{n+1} = \text{min}(t_{scale}\Delta t^{n}, \frac{1}{10}t)\,.
\end{equation}
If superfluidity occurs, a neutron star experiences drastic changes 
in specific heat, thermal conductivity, and neutrino emission rate.
Thus, when the internal temperature drops below the critical temperature for
superfluidity, we use adaptive time step method. The $T_{eff}^{n+1}$ should
change within maximum $5 \%$ of $T_{eff}^{n}$. 
For instance, if the numerical solution gives $T_{eff}^{n+1} < 0.95 \, T_{eff}^{n}$,
we solve the diffusion equations again 
with  the new time step 
$\Delta t^{n+1, i+1} = t_{reduce}\Delta t^{n+1, i} $ 
(where index $i$ indicates the $i$th trial time step)
until ${\cal T}^{n+1}_{eff} > 0.95 {\cal T}^{n}_{eff}$. 
In our simulation $t_{reduce} = 0.75$ to reduce the time step.
Once we find the solution, according to the temperature differences between
$t^n$ and $t^{n+1}$, we use the adaptive $t_{scale}$ for the next time
$t^{n+2}$.
In superfluid case, we use
\begin{equation}\label{eq:super_reg}
\begin{cases}
0.01 <\Big \vert \frac{T_{eff}^{n} - T_{eff}^{n-1}}
{T_{eff}^{n-1}} \Big \vert \le 0.05 \,, & t_{s_1} = 1.2 \,, \\
\Big \vert \frac{T_{eff}^{n} - T_{eff}^{n-1}}
{ T_{eff}^{n-1}}
\Big \vert \le 0.01 \,, & t_{s_2} = 1.5\,.
\end{cases}
\end{equation}

\end{document}